\documentclass[apj]{emulateapj}

\usepackage{amsmath}
\usepackage{txfonts}

\input{epsf}

\begin{document}
\def\lax    {\ifmmode{_<\atop^{\sim}}\else{${_<\atop^{\sim}}$}\fi}
\def\gax    {\ifmmode{_>\atop^{\sim}}\else{${_>\atop^{\sim}}$}\fi}
\def\gtorder{\mathrel{\raise.3ex\hbox{$>$}\mkern-14mu
             \lower0.6ex\hbox{$\sim$}}}
\def\ltorder{\mathrel{\raise.3ex\hbox{$<$}\mkern-14mu
             \lower0.6ex\hbox{$\sim$}}}
 

 
\title{XMM-Newton survey of the central region of M31. Spectral properties and variability of bright X-ray sources and source classification.}

\author{Sergey~P.~Trudolyubov}
\affil{Institute of Geophysics and Planetary Physics, University of California, Riverside, CA 92521,}
\affil{Space Research Institute, Profsoyuznaya 84/32, Moscow, 117997, Russia}
\author{William~C.~Priedhorsky}
\affil{Los Alamos National Laboratory, Los Alamos, NM 87545}
\author{and France~A.~C\'ordova}
\affil{Institute of Geophysics and Planetary Physics,} 
\affil{Department of Physics and Astronomy, University of California, Riverside, CA 92521}





\begin{abstract}
We present the results of the systematic survey of X-ray sources in the central region of M31 using the data from 
{\em XMM-Newton} observations performed in the years 2000-2004. The spectral properties and variability of 123 bright 
X-ray sources with apparent luminosities between $\sim 10^{36}$ and $5\times 10^{38}$ ergs s$^{-1}$ were studied in 
detail. The spectral properties of majority of sources in our sample have been found to be consistent with that of 
the low-mass X-ray binary systems. 20 sources in our sample coinside with globular clusters in M31, four sources are 
supernova remnant candidates and another six sources can be classified as background AGN candidates. The spectral 
distribution of M31 X-ray sources, based on the spectral fitting with a power law model is clearly bimodal with a 
main peak corresponding to a photon index $\Gamma\sim 1.75$ and a shoulder at $\Gamma\sim 2.0-2.2$ extending to the 
soft spectral region. The spectral distribution shows clear evolution with source luminosity, characterized by its 
narrowing and shift of its main peak to the harder spectral region for luminosities above $\sim(3-5)\times10^{37}$ 
ergs s$^{-1}$. The spectral properties of the high-luminosity sources in our sample closely resemble those of the 
bright LMXB in nearby galaxies of different morphological type. The distribution of absorbing columns towards M31 
sources derived from spectral analysis has a peak at N$_{\rm H}\sim1.2\times10^{21}$ cm$^{-2}$ extending up to 
$1.3\times10^{22}$ cm$^{-2}$, with an average value of $(1.52\pm0.02)\times10^{21}$ cm$^{-2}$. More than 80\% of 
sources observed in two or more observations show significant variability on the time scales of days to years. The 
fraction of variable sources in our survey is much higher than previously reported from {\em Chandra} survey of M31, 
and is remarkably close to the fraction of variable sources found in the M31 globular cluster X-ray source population. 
Timing analysis of several of the brightest sources reveals significant aperiodic variability in the $2\times10^{-5}-0.01$ 
Hz frequency range, resembling the very-low frequency variability of the bright Galactic LMXB. About 50\% of the sources 
in our sample are spectrally variable. The spectral evolution of a number of sources is correlated with the level of 
their X-ray flux, while some sources demonstrate complex patterns of evolution on the hardness-intensity diagram. 

Based on the similarity of the properties of M31 X-ray sources and their Galactic counterparts, we expect most of the 
X-ray sources in our sample to be accreting binary systems with neutron star and black hole primaries. A total of 44 
X-ray sources can be identified as probable X-ray binaries. We show that X-ray hardness-luminosity (spectral photon 
index vs. luminosity) diagrams can be an effective tool for characterising X-ray binary populations in external 
galaxies and classification of individual sources as neutron star and black hole candidates. Combining the results of 
X-ray analysis (detailed X-ray spectra, hardness-luminosity diagrams and variability) with available data at other 
wavelengths, we classify 7\% and 24\% of sources in our sample as, respectively, probable black hole and neutron star 
candidates.
\end{abstract}


\section{INTRODUCTION}
X-ray surveys of external galaxies are important for understanding populations of X-ray binaries and supernova 
remnants in these galaxies and our own Milky Way. The statistical study of the spatial and spectral distribution 
of these sources can help us to distinguish between different physical classes of X-ray emitters. X-ray binaries 
and supernova remnants comprise a fossil record of the stellar population, and may be used as a probe of galactic 
dynamics and star formation history. A detailed study of spectral properties and variability of X-ray binaries, 
combined with results of observations at other wavelengths, can help to distinguish between systems with black 
hole and neutron star primaries, providing invaluable information on the binary stellar evolution. 

The nearby, giant spiral M31 presents an excellent opportunity to study the global properties of a galaxy that 
is in many respects similar to our own. The proximity, well-established distance and favorable inclination of 
M31 galactic disk allow effective study and comparison of its X-ray source populations and their association 
with the bulge, spiral arms, or halo components. M31 has been observed extensively by {\em Einstein}, {\em ROSAT}, 
{\em Chandra} and {\em XMM-Newton} missions (Trinchieri \& Fabbiano 1991; Primini et al. 1993; Supper et al. 
2001; Shirey et al. 2001; Kong et al. 2002a; Williams et al. 2004a; Pietsch et al. 2005). Recent X-ray observations 
with {\em Chandra} and {\em XMM-Newton} has started to reveal important similarities and differences between X-ray 
source populations and hot interstellar medium in the bulge and disk of M31, and has allowed direct comparisons to 
the Milky Way galaxy (Shirey et al. 2001; Trudolyubov et al. 2002a; DiStefano \& Kong 2003; Trudolyubov et al. 2005b). 
The advanced capabilities of {\em XMM-Newton} and {\em Chandra} have allowed a detailed study of the variability and 
spectral properties of a large number of individual sources belonging to different classes of X-ray emitting objects 
like globular cluster sources, supersoft sources and supernova remnants (DiStefano et al. 2002; Kong et al. 2002b; 
Trudolyubov \& Priedhorsky 2004; Di Stefano et al. 2004). The spectral and timing analysis of individual bright M31
sources has allowed for the identification of some of them as accreting X-ray binaries and has revealed striking 
similarities between them and their Galactic counterparts. Timing studies of M31 X-ray sources led to discovery of 
X-ray pulsations (Osborne et al. 2001; Trudolyubov et al. 2005b), periodic dipping (Trudolyubov et al. 2002b; Mangano 
et al. 2004) and short bursts (Pietsch \& Haberl 2005). The spectral properties and evolution of a number of M31 
sources have led to their classification as probable black hole, neutron star or white dwarf systems. Regular 
monitoring observations of M31 have revealed several dozens of transient and recurrent X-ray sources ranging from 
black hole candidate X-ray novae to supersoft transient systems (Trudolyubov et al. 2001; Williams et al. 2006). 

Observations of the central part of M31 provide a unique opportunity to study stellar populations of the bulge 
and inner disk of the galaxy. This resion contains a significant fraction of the stellar mass of M31 and is the 
most densely populated region in X-rays with nearly 300 individual sources detected to date (Kong et al. 2002a; 
Pietsch et al. 2005). In this work we present the results of our survey of spectral properties and variability of 
123 brightest sources detected in the the central part of M31 ($15\arcmin$ radius), using data from {\em XMM-Newton} 
observations. 

\section{OBSERVATIONS AND DATA ANALYSIS}
\begin{figure}
\epsscale{.7}
\plotone{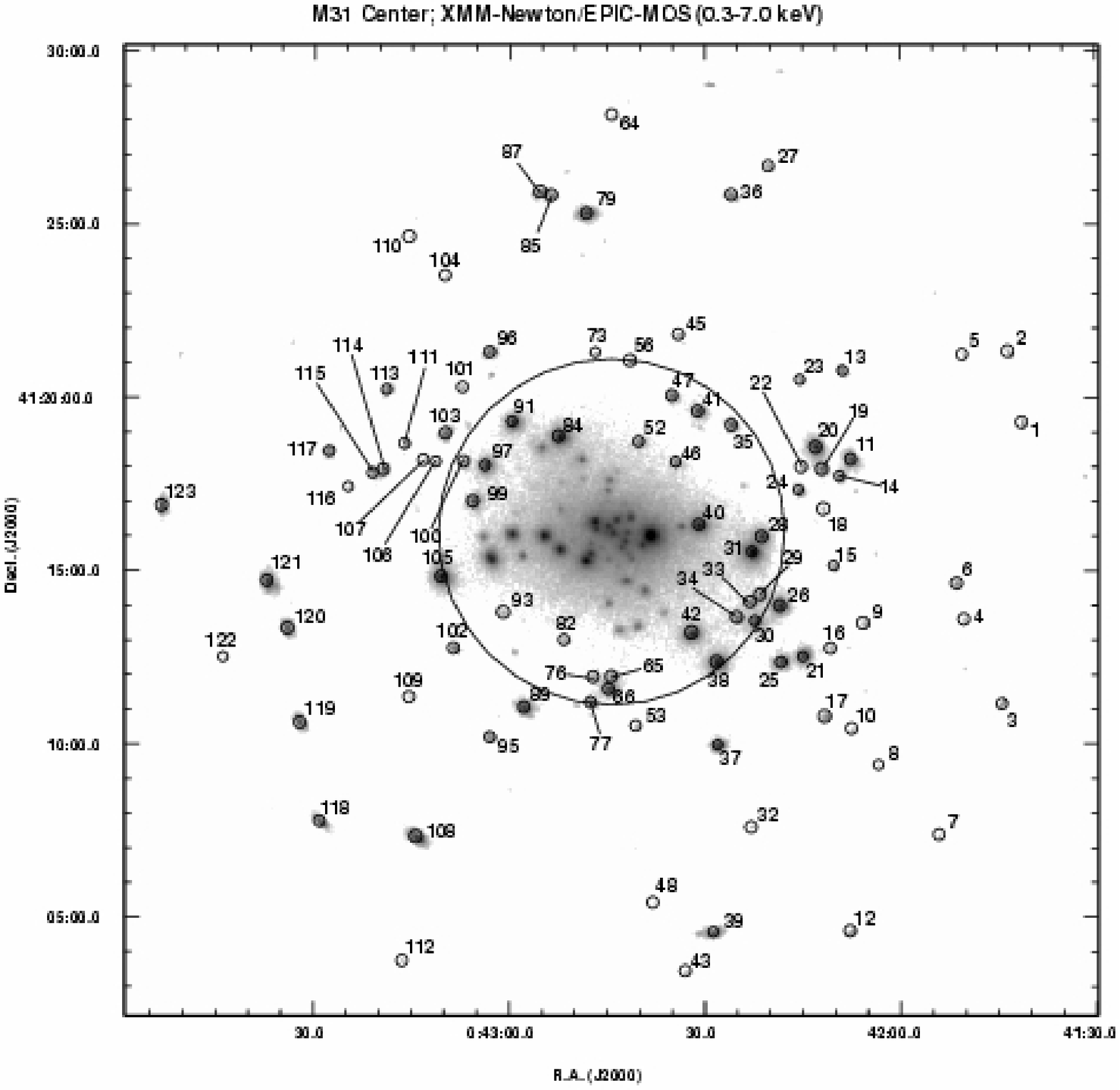}
\plotone{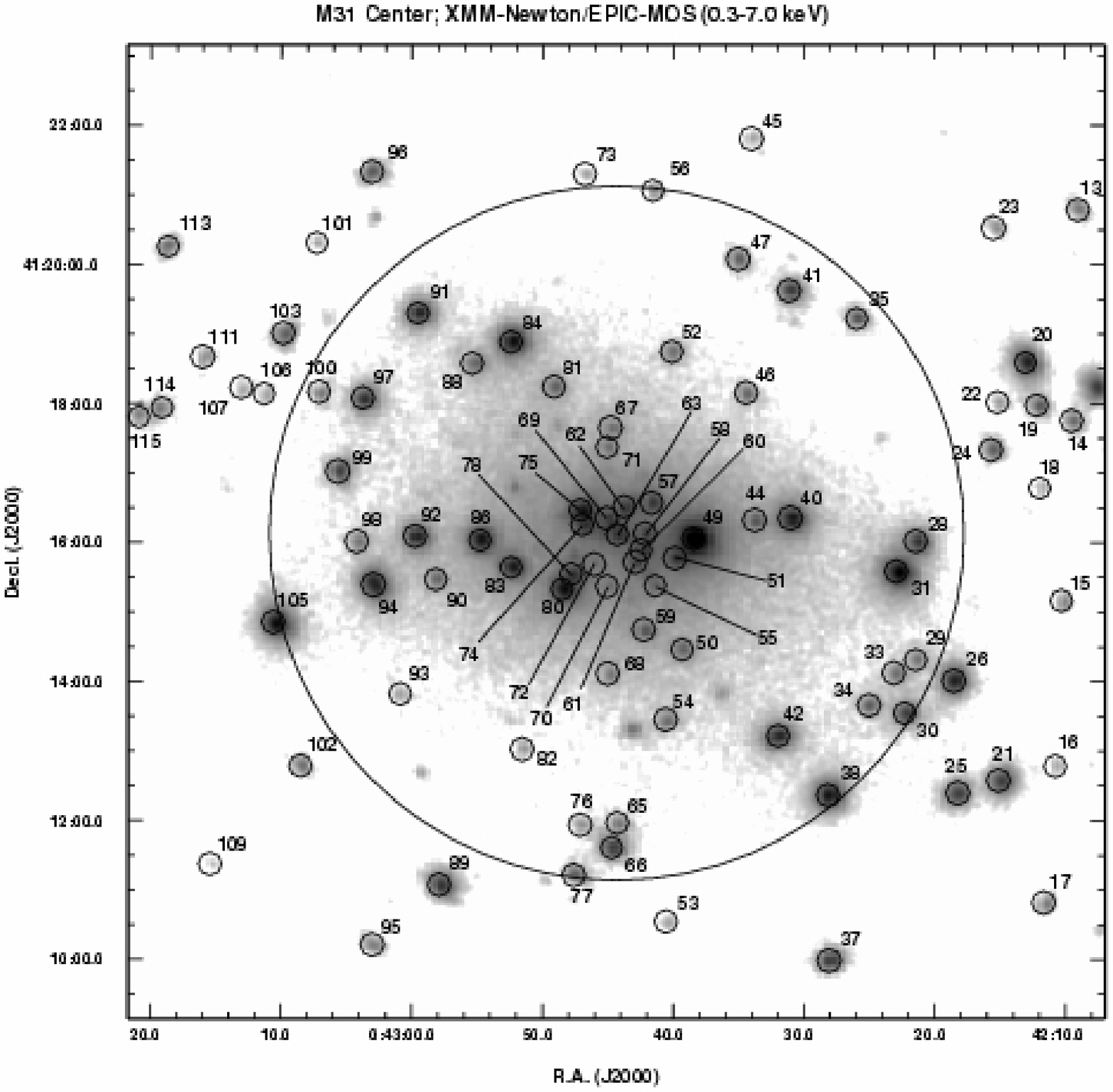}
\caption{{\em Upper panel:} Combined $0.3 - 7$ keV EPIC-MOS image of the central region of M31. The sources in our 
sample are marked with with circles and arrows. {\em Bottom panel:} Enlargement of the region within $7\arcmin$ of 
the nucleus. Source labels correspond to the numbering in Table 2. A $5\arcmin$ radius circle centered on the nucleus 
is shown for comparison. \label{image_general}}
\end{figure}

The central region of M31 was observed with {\em XMM-Newton} on 8 occasions in years 2000 -- 2004 
(Table \ref{obslog}; Fig. \ref{image_general}). In the following analysis we use the data from 
three European Photon Imaging Camera (EPIC) instruments: two EPIC MOS detectors \cite{Turner01} 
and the EPIC-pn detector \cite{Strueder01}. In all observations EPIC instruments were operated in 
the {\em full window} mode ($30\arcmin$ FOV) with the {\em medium} and {\em thin} optical blocking 
filters.

We reduced EPIC data with the latest version of {\em XMM-Newton} Science Analysis System (SAS v 
6.5.0)\footnote{See http://xmm.vilspa.esa.es/user}. Each of the original event files were screened 
for periods of high background. The remaining exposure times for each observation are listed in 
Table \ref{obslog}. The 2004 July 17 observation is affected by high background, so we excluded it 
from the analysis.

We generated EPIC-pn and MOS images of the central region of M31 (Fig. \ref{image_general}) in the 
0.3 -- 7.0 keV energy band, and used the SAS standard maximum likelihood (ML) source detection script 
{\em edetect\_chain} to detect and localize point sources. We used bright X-ray sources with known 
optical counterparts from USNO-B \cite{Monet03} and 2MASS catalogs \cite{Cutri03} to correct EPIC 
image astrometry. After applying the astrometric correction, we estimate residual systematic error 
in the source positions to be of the order $0.5 - 1\arcsec$. 

To generate lightcurves and spectra of X-ray sources, we used elliptical extraction regions 
with semi-axes size of $\sim 15 - 50 \arcsec$ (depending on the distance of the source from 
the telescope axis) and subtracted as background the spectrum of adjacent source-free regions, 
with subsequent normalization by ratio of the detector areas. For spectral analysis, we used 
data in the $0.3 - 10$ keV energy band. For the sources with soft (SNR candidates) and supersoft 
X-ray spectra we considered only the $0.3 - 3$ and $0.3 - 1.5$ keV spectral ranges, since their 
flux was negligible above 3.0 and 1.5 keV respectively. We used spectral response files generated 
by XMM SAS tasks. Spectra were grouped to contain a minimum of 20 counts per spectral bin in order 
to allow $\chi^{2}$ statistics and fit to analytic models using the 
XSPEC v.11\footnote{http://heasarc.gsfc.nasa.gov/docs/xanadu/xspec/index.html} fitting package 
\cite{arnaud96}. EPIC-pn, MOS1 and MOS2 data were fitted simultaneously, but with normalizations 
varying independently. 

We performed timing studies of the sources in our sample, using standard XANADU/XRONOS 
v.5\footnote{http://heasarc.gsfc.nasa.gov/docs/xanadu/xronos/xronos.html} tasks. We generated 
source and background X-ray lightcurves in the $0.3 - 7$ keV energy band with 5.2s time resolution 
using the data of individual EPIC detectors. For SNR candidates and supersoft sources the lightcurves 
in the $0.3 - 2$ keV energy range were extracted. To improve statistics, properly synchronized 
lightcurves from individual detectors were added to produce combined EPIC lightcurves for each source. 
To synchronize source and background lightcurves from individual detectors, we used the same time 
filtering criteria based on Mission Relative Time (MRT) (see also Barnard et al. 2006b). Fourier power 
density spectra (PDS) in the $2\times 10^{-5} - 0.1$ Hz frequency range were produced using both 
individual detector and combined EPIC background-subtracted lightcurves. The spectra were logarithmically 
rebinned when necessary to reduce scatter at higher frequencies and normalized such that the white noise 
level expected from the data errors corresponds to a power of 2 (Leahy-normalized)\cite{Leahy83}. In 
several cases, when significant aperiodic variability was detected, the PDS were also normalized to the 
square of fractional r.m.s. variability, with the expected white noise level subtracted. 

In the following analysis we assume M31 distance of 760 kpc (van den Bergh 2000). All parameter 
errors quoted are 68\% ($1\sigma$) confidence limits.

\section{Source Identification}
We study a sample of 123 X-ray sources in the central region of M31 selected on the basis of their 
brightness; each source was required to have more than 300 counts in the EPIC during at least one 
observation. The information on the positions, spectral properties and identifications of the X-ray 
sources is shown in Table \ref{source_ID_1}. The combined EPIC-MOS image of the central region of M31 
with source positions marked with circles is shown in Fig. \ref{image_general}.   

We searched for the X-ray, optical, infrared and radio counterparts to the sources in our sample using 
the existing catalogs and images from the CTIO/KPNO Local Group Survey (LGS) \cite{Massey01} and the 
Second Generation Digitized Sky Survey. We varied the search radius based on both the accuracy of the 
catalogs and localization errors of {\em XMM} sources. We used the following catalogs:

\noindent i) {\em X-ray sources:} the {\em Einstein} (Trinchieri \& Fabbiano 1991), {\em ROSAT}/HRI 
(Primini, Forman \& Jones 1993), {\em Chandra}/ACIS (Kong et al. 2002), {\em Chandra}/HRC (Williams 
et al. 2004; Kaaret 2002) and {\em XMM-Newton} (Pietsch et al. 2005) catalogs of X-ray sources in the 
field of M31.

\noindent ii) {\em Globular cluster candidates:} the Bologna catalog \cite{Bo87,Galleti04}, the 
catalog by Magnier (1993), and the HST globular cluster candidate catalog \cite{Barmby01}. The positions 
of 20 sources coinside with optically identified globular cluster candidates. 

\noindent iii) {\em Supernova remnant candidates:} the lists by Braun \& Walterbos (1993), Magnier et al. 
(1995) and Ford \& Jacoby (1978). 

\noindent iv) {\em Stellar objects (Galactic foreground stars/background AGN):} 
the catalogs of stellar objects: USNO-B \cite{Monet03}, 2MASS \cite{Cutri03}, catalogs 
by Magnier et al. (1992) and Haiman et al. (1994) and SIMBAD.

\noindent v) {\em Radio sources:} VLA All-sky Survey Catalog
\footnote{http://www.cv.nrao.edu/nvss/} \cite{nvss} and catalogs of radio sources in the field of 
M31 (Walterbos, Brinks, \& Shane 1985; Braun 1990).

All but three new transient sources (\#3, 104, 110) in our sample have been previously detected in X-rays. 
The positions of 20 sources coincide with optically identified globular cluster candidates. The X-ray source 
\#29 has a bright optical counterpart, and has been proposed by Pietsch \& Haberl (2005) as possible globular 
cluster candidate. Four sources (\#70, 85, 117, 122) coincide with supernova remnant (SNR) candidates in M31. 
Six sources (\#4, 6, 17, 27, 69, 103), not identified with SNR, have bright radio counterparts, and can be 
tentatively identified with background AGN. Two of these sources (\#6 and 103) also have point-like optical 
counterparts listed in Magnier et al. (1992) (MLV92 242628 and MLV92 267602) and seen in LGS images. The 
remaining four sources with radio counterparts have very faint or no detectable optical counterparts. 

\section{Note on the Effect of Background Sources and Multiple Unresolved X-ray Sources}
A notable fraction of the X-ray sources detected in the central region of M31 must be background objects. 
To estimate the expected contribution of the background sources in our sample, we used the results of the 
combined X-ray source counts from {\em ROSAT}, {\em Chandra} and {\em XMM-Newton} surveys \cite{Moretti03}. 
Taking into account spatial coverage of our survey, $\sim 7 - 10$ out of 87 sources with apparent 
luminosities above $5 \times 10^{36}$ ergs s$^{-1}$ (approximate completeness limit of our sample) should 
be background objects. 

Ten sources in our sample are unresolved composites of the two or more {\em Chandra} sources 
(Table \ref{source_ID_1}). In addition, the spatial resolution of both {\em Chandra} and {\em XMM-Newton} 
is not sufficient to resolve most possible multiple X-ray sources within globular clusters in M31 
\cite{DiStefano01,TP04}. Therefore, the effects of spectral blending and superposition of variability 
may affect the properties of several sources in our sample, and complicate direct comparisons with 
Galactic X-ray sources.

\section{Spectra}
The spectra of M31 X-ray sources were fitted with a variety of spectral models using XSPEC v11. We first 
considered a simple one-component spectral model: an absorbed simple power law. The results of fitting this 
model to the source spectra are given in Table \ref{source_ID_1}. The spectra of 115 sources in our sample 
(excluding SNR candidates and supersoft sources) can be generally described by an absorbed simple power 
law model with a photon index of $\sim 0.6 - 3.7$ and an equivalent absorbing column of 
$(0.1 - 13) \times 10^{21}$ cm$^{-2}$. The apparent absorbed luminosities of the X-ray sources in our sample 
differ by nearly three orders of magnitude and fall between $\sim 10^{36}$ and $\sim 5\times 10^{38}$ ergs 
s$^{-1}$ in the $0.3 - 10$ keV energy band, assuming a distance of 760 kpc.

\begin{figure}
\epsfxsize=18cm
\epsffile{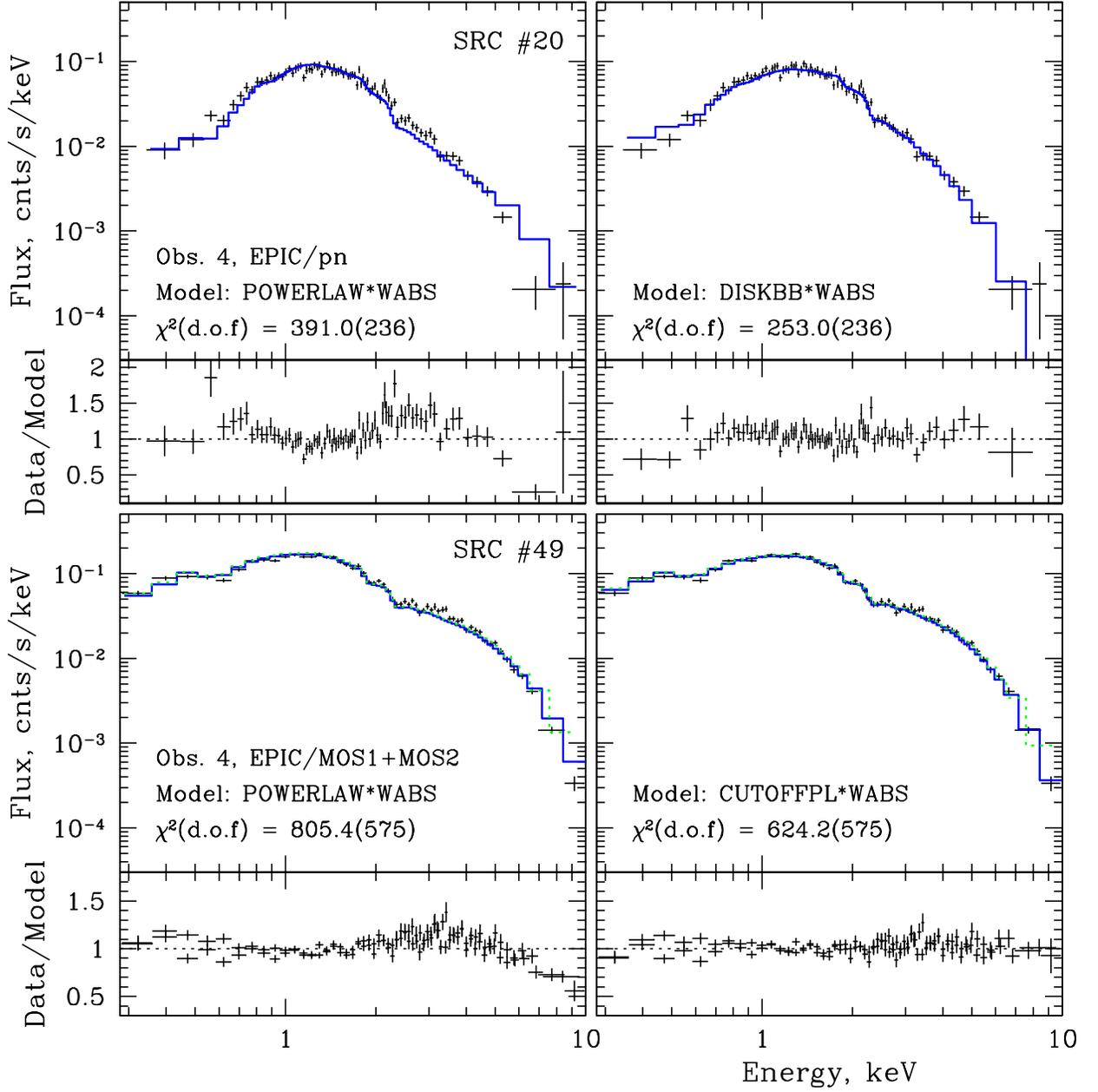}
\caption{\small Representative EPIC count spectra and model ratios of two sources (\#20 and \#49 in Table 2) 
showing spectral cutoff. The absorbed simple power law model approximation of the data for both sources is shown 
in {\em left} panels. Best-fit cut-off spectral model fits (Src. \#20 -- Multicolor disk black body (DISKBB) 
and Src. \#49 -- Cut-off power law (CUTOFFPL)) are shown in {\em right} panels. Note significant improvement 
in the $\chi^{2}$ fit statistics and data/model ratios for cut-off model approximation as compared to the power 
law. \label{spec_CUTOFF_fig}}
\end{figure}

In most cases, we obtained acceptable fits using a power law spectral model. However, for many brighter 
sources, a complex spectral models are required. For several sources with high luminosities (Table 
\ref{spec_par_CUTOFF},\ref{spec_par_two_comp},\ref{spec_par_TR}), the models with quasi-exponential cut-off 
at $\sim 0.6 - 8$ keV or two-component models describe the energy spectra significantly better than a simple 
power law (Fig. \ref{spec_CUTOFF_fig}). To approximate the spectra of these sources, we used an absorbed power 
law model with exponential cut-off (XSPEC CUTOFFPL model), a Comptonization model, a multicolor disk black body 
(DISKBB) model and a two types of two-component models described below. 

For the Comptonization model approximation (Table \ref{spec_par_two_comp}), we used the XSPEC model COMPTT 
\cite{ST80,T94}. This model includes a self-consistent calculation of the spectrum produced by the Comptonization 
of the soft photons in a hot plasma. It contains as free parameters the temperature of the Comptonizing electrons, 
$kT_{e}$, the plasma optical depth with respect to the electron scattering, $\tau$ and the temperature of the 
input Wien soft photon distribution, $kT_{0}$. In our spectral analysis a spherical geometry of the Comptonizing 
region was assumed. 

The spectra of many luminous neutron star LMXB are well fit with a two-component model consisting of a soft 
black body-like component which might represent emission from an optically thick accretion disk or from the 
neutron star surface, together with hard component which may be interpreted as emission from a corona-like 
structure or a boundary layer between the disk and a neutron star (White, Stella \& Parmar 1988). For the soft 
component we used a multicolor disk-blackbody (XSPEC DISKBB)\cite{Mitsuda84} model. The hard spectral component 
can be adequately described by various phenomenological and physical models involving a break in the slope of 
the spectrum or quasi-exponential spectral cut-off at higher energies. For the sake of easier comparison with 
the results for the Galactic LMXB, we use a simple black body (BBODYRAD) in combination with DISKBB as a soft 
component to approximate the hard component in the spectra of several bright sources in our sample (probable 
neutron star candidates).

The other two-component model, used in our analysis of several bright M31 sources (Fig. \ref{spec_TR_fig}; 
Table \ref{spec_par_TR}), is a combination of absorbed simple power law and DISKBB models, a standard model 
describing the spectra of Galactic black hole candidates \cite{MR04}.

\subsection{Transient Sources}
\begin{figure}
\epsscale{.7}
\plotone{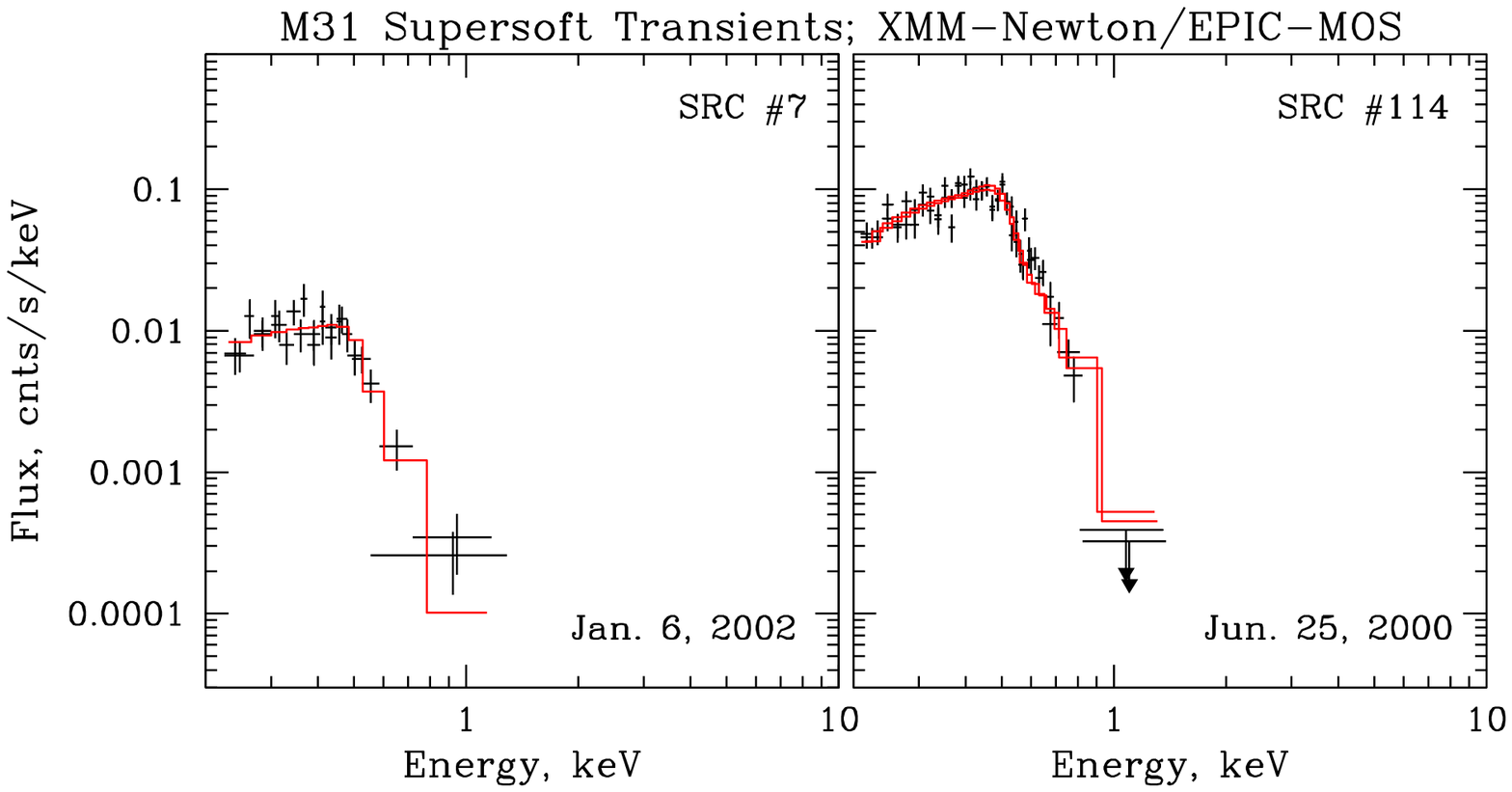}
\plotone{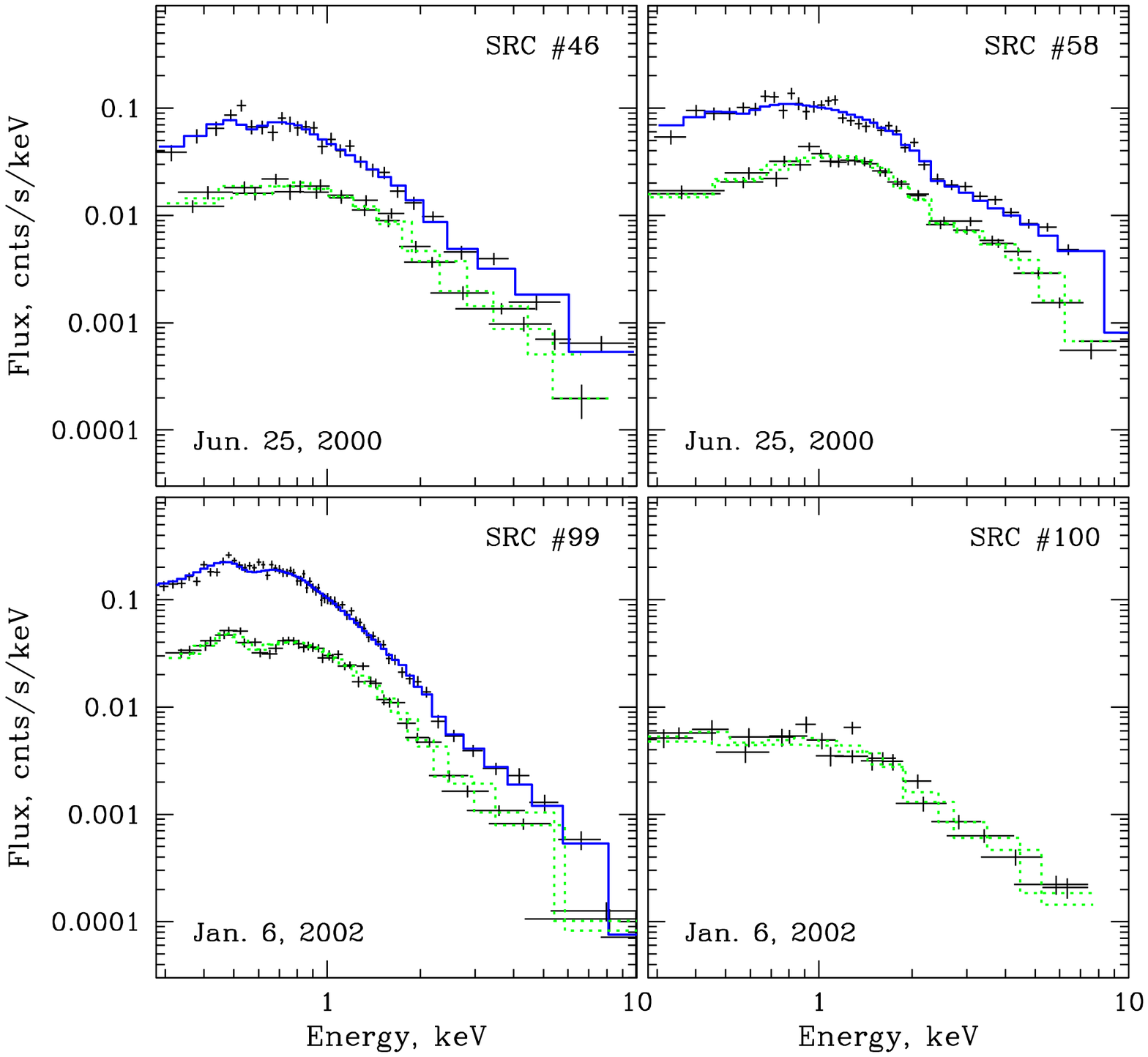}
\caption{\small EPIC count spectra of six bright X-ray transient sources detected in 2000 - 2002 {\em XMM} 
observation of the central part of M31. The best-fit models are shown with thick histograms for EPIC-pn 
data and dotted histograms for EPIC-MOS data. \label{spec_TR_fig}}
\end{figure}

A total of 10 bright transient sources with 0.3 - 10 keV luminosities above $10^{36}$ ergs s$^{-1}$ have been 
detected in seven {\em XMM-Newton} observations of the central part of M31 (Osborne et al. 2001; Trudolyubov, 
Borozdin \& Priedhorsky 2001; Pietsch et al. 2005; Trudolyubov, Priedhorsky \& Cordova 2006). All of them are 
included in our source sample (Table \ref{source_ID_1}). Two {\em XMM} transients (\#7 and 114) have supersoft 
spectra, five sources (\#3, 57, 99, 104, 110) have soft spectra with spectral photon indices above 2.4, and 
three sources (\#46, 58, 100) have harder spectra (Fig. \ref{spec_TR_fig}; Table \ref{spec_par_TR}). 

The energy spectra of the two supersoft transient sources (\#7 and 114) obtained with {\em XMM}/EPIC MOS1 
and MOS2 detectors are shown in Figure \ref{spec_TR_fig}. The EPIC energy spectra of the supersoft transients 
can be satisfactorily described by the absorbed blackbody radiation models (Table \ref{spec_par_TR}). The 
characteristic spectral temperatures and luminosities of both sources are typical for a supersoft X-ray sources 
found in our Galaxy and M31 \cite{O01,DiStefano04}.

The energy spectra of the two transient sources (\#3 and 57) are soft, and show curvature around 0.6 and 
1 keV. Due to the curvature in the source spectra, the DISKBB model provides a significantly better description 
to the data than a power law (Table \ref{spec_par_TR}). The observed spectra and luminosities of the sources bear 
clear resemblance to the Galactic black hole transients in the high/thermal-dominant state \cite{MR04}. Both 
sources have been proposed to be black hole candidates, based on their overall X-ray properties and additional 
constraints on the optical counterpart (Trudolyubov, Priedhorsky \& Cordova 2006; Williams et al. 2005b).  

The two-component model including absorbed low-temperature thermal component (DISKBB) and a hard power law tail 
gives the best description to the EPIC spectra of the transient sources \#46 and \#99 (Table \ref{spec_par_TR}). 
The X-ray spectra, long-term variability and luminosity of these sources are remarkably similar to the Galactic 
black hole transients in the low/hard state or during transition from the high/thermal-dominant to the low/hard 
state (Tomsick \& Kaaret 2000; McClintock \& Remillard 2006). Both objects have been previously classified as 
probable black hole candidates based on their X-ray properties and evolution observed with {\em XMM-Newton} and 
{\em Chandra} (Trudolyubov, Borozdin \& Priedhorsky 2001; Williams et al. 2004a).

The EPIC spectra of the transient source \#58 (Garcia et al. 2000, Trudolyubov, Borozdin \& Priedhorsky 2001) 
are relatively hard and show evidence of a high-energy cutoff. The absorbed power law model with exponential cut-off 
at 4-6 keV provides a significantly better approximation than a simple power law model (Table \ref{spec_par_TR}).

The spectrum of the transient source \#100 is hard and can be adequately fit by absorbed simple power law with 
photon index of $\sim 1.6$ (Table \ref{spec_par_TR}). 

The energy spectra of the remaining two transient sources \#104 and 110 are soft and can be fit with both 
steep power law or thermal models with characteristic temperatures of 0.3-0.4 keV (Table \ref{spec_par_TR}). 
The discussion of the detailed spectral analysis of these sources can be found in Trudolyubov, Priedhorsky \& 
Cordova (2006).

\subsection{Globular Cluster Sources}
The positions of 20 bright X-ray sources are consistent with globular cluster (GC) candidates in M31 (Table 
\ref{source_ID_1}). The results of the detailed analysis of the properties of 19 globular cluster candidates
\footnote{The recurrent X-ray source \#77 associated with globular cluster (Primini, Forman \& Jones 1993) 
candidate Bo 128 was not detected during the first four {\em XMM} observations} in our sample detected in the 
first four {\em XMM-Newton} observations can be found in Trudolyubov \& Priedhorsky (2004). In this work, we 
combine results of the previous study with results of the three subsequent 2004 {\em XMM-Newton} observations.

\subsection{Supersoft Sources}
Besides two transients, two other sources in our sample, \#83 and \#113 have supersoft spectra. The EPIC energy 
spectra of the supersoft transients can be generally described by the absorbed blackbody radiation models with 
characteristic temperature of 50-60 eV (Table \ref{spec_par_SSS}). Both sources are variable in X-rays, with 
source \#113 remaining below the EPIC detection threshold in some observations and flaring up to 
$\sim 1.3\times 10^{37}$ ergs s$^{-1}$ luminosity level (Table \ref{source_ID_1},\ref{spec_par_SSS}).

\subsection{Supernova Remnants}
\begin{figure}
\epsfxsize=18cm
\epsffile{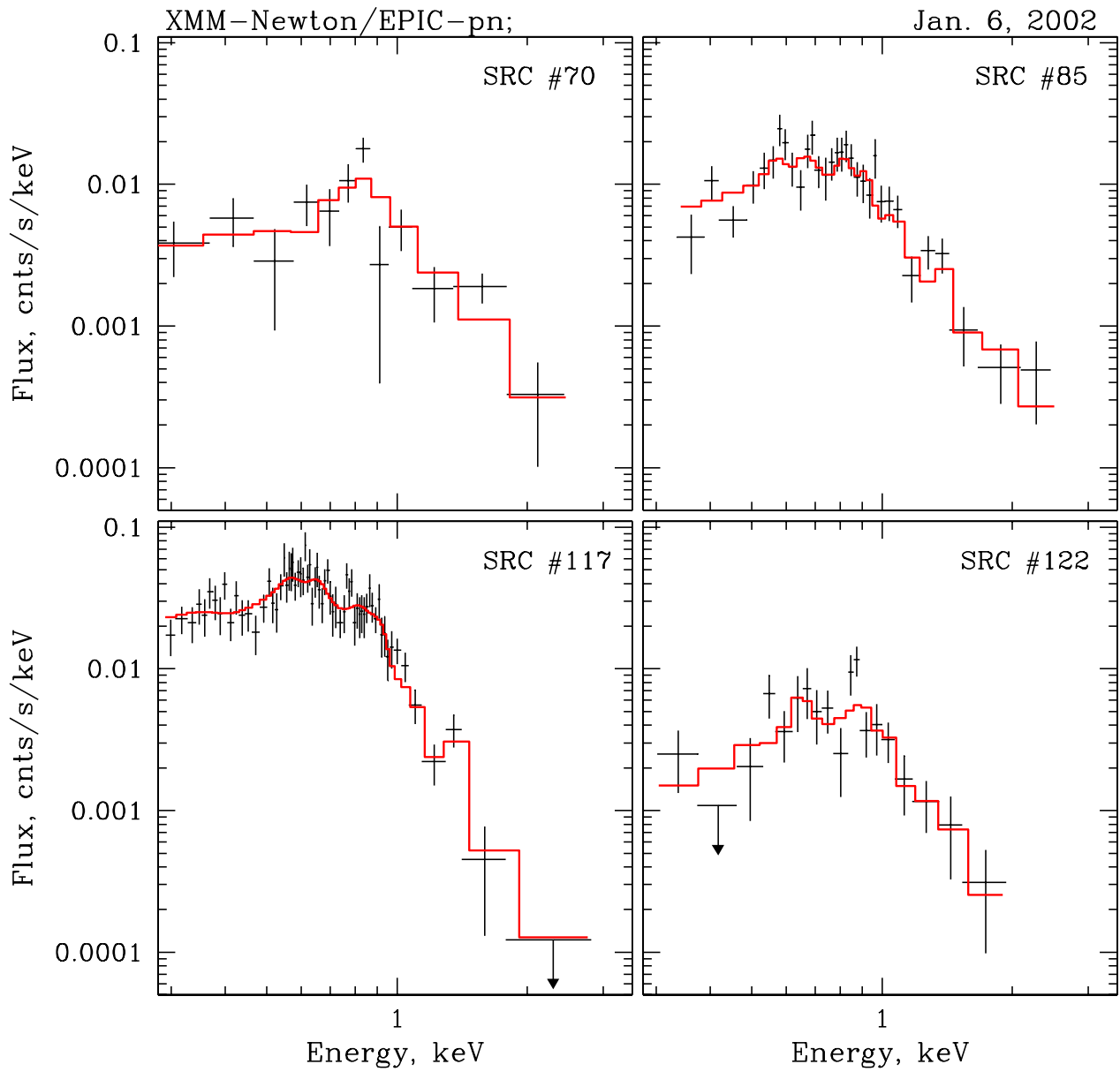}
\caption{\small EPIC-pn count spectra of four SNR candidates in our sample measured in the 2002, Jan. 6 observation. 
For each source the best-fit absorbed NEI (sources \#85 and 117) and MEKAL (sources \#70 and 122) models (Table 
\ref{spec_par_SNR}) are shown with solid histograms.\label{spec_SNR_fig}}
\end{figure}

The positions of three X-ray sources in our sample are consistent with supernova remnant (SNR) candidates 
from various optical and radio surveys (Baade \& Arp 1964; Braun 1990; Braun \& Walterbos 1993)(Table 
\ref{source_ID_1}). In addition, the X-ray source \#70 is coincident with planetary nebula candidate from 
Ford \& Jacoby (1978) and has been classified as SNR candidate based on its spectrum and luminosity (Pietsch 
et al. 2005).The EPIC spectra of all SNR candidates are soft and show clear presence of the emission lines 
(Fig. \ref{spec_SNR_fig}). 

We fitted the spectra of SNR candidates with XSPEC collisional equilibrium thermal plasma (MEKAL) (Mewe et 
al. 1985; Liedahl et al. 1995), and non-equilibrium ionization collisional plasma (NEI) models with 
interstellar absorption. The results of the analytical approximation of the {\em XMM}/EPIC data for SNR 
candidates are shown in Table \ref{spec_par_SNR}.

{\em SNR candidates \#85(BA 521) and \#117(BA 23).} To improve statistical quality of data, we combined the 
data of three observations (Obs. 1,3,4) of these two SNR candidates. For the MEKAL and NEI models, we first 
fixed the abundances at solar values \cite{AG89}. This set of parameters, however, left bump-like residuals 
indicating that we had overestimated the contribution from O-K, Ne-K and Fe-L shell emission lines. Then we 
fixed the abundances based on the values determined from optical spectroscopy of the optical counterpart to 
the sources \cite{Blair82} with Fe abundance fixed at solar value. Finally, we allowed the O, Ne and Fe 
abundances to be free parameters. In both cases the quality of the fit improved significantly compared to the 
same models with fixed solar abundances (Table \ref{spec_par_SNR}). 

The average absorbed $0.3 - 3$ keV luminosities of the sources \#85 and \#117 corresponding to the best-fit 
MEKAL and NEI models are $\sim 4.5\times 10^{36}$ and $\sim 5.3\times 10^{36}$ ergs s$^{-1}$ (Table 
\ref{spec_par_SNR}), which makes them two brightest thermal SNR candidates detected in M31 to date 
\cite{TP05,Pietsch05}. The plasma temperatures inferred from the best-fit spectral models for the sources \#85 
and \#117 were found to be typical for the SNR candidates in M31 (0.2-0.5 keV), in general agreement with 
previous results \cite{Kong02_SNR,Williams04_2,TP05,T05}.   

In their recent paper, based on the {\em XMM} and {\em Chandra} observations of the source \#117(BA 23), 
Williams et al. (2005{\em d}) report on the detection of the high-energy excess in the spectrum of the source, 
which they attribute to the contamination from nearby X-ray binary. Using the power law approximation, they 
estimate the contribution of the hard spectral component to be 26\% of the total X-ray luminosity. The 
results of our analysis, based on the larger number of observations, do not confirm their findings. After careful 
background screening and subtraction, we were not able to detect any statistically significant high-energy excess 
in the spectra of the source \#117 (Fig. \ref{spec_SNR_fig}). To estimate the possible contribution of the hard 
component, we fit source spectra from individual observations with a two-component model including MEKAL model with 
varying O, Ne and Fe abundances and a power law with two fixed values of the photon index ($\Gamma_{1} = 1.7$ and 
$\Gamma_{2} = 3.2$). The resulting $2\sigma$ upper limit on the contribution of the hard component to the total 
unabsorbed X-ray luminosity is $\sim$ 7\% for the power law model with photon index $\Gamma_{1} = 1.7$ and 
$\sim$ 16\% for $\Gamma_{2} = 3.2$.  

{\em SNR candidates \#70 and 122.} Because of the limited statistics, we used only the MEKAL model to approximate 
the spectra of the two fainter SNR candidates (sources \#70 and 122). We first fixed the abundances at solar values, 
and then allowed the O, Ne and Fe abundances to be free parameters. The models with free O, Ne and Fe abundances 
give marginal improvement of the quality of the fit as compared to the fixed solar abundances (Table 
\ref{spec_par_SNR}). The best-fit models give plasma temperatures of $\sim 0.50$ keV for the source \#70 and 
$\sim 0.33$ keV for the source \#122, typical for thermal SNR in M31 \cite{TP05,Kong02_SNR}, and an absorbed 
$0.3 - 3$ keV luminosities of $\sim 1.3\times 10^{36}$ and $\sim 1.4\times 10^{36}$ ergs s$^{-1}$ respectively.

\section{Spectral Distribution}
\begin{figure}
\epsfxsize=18cm
\epsffile{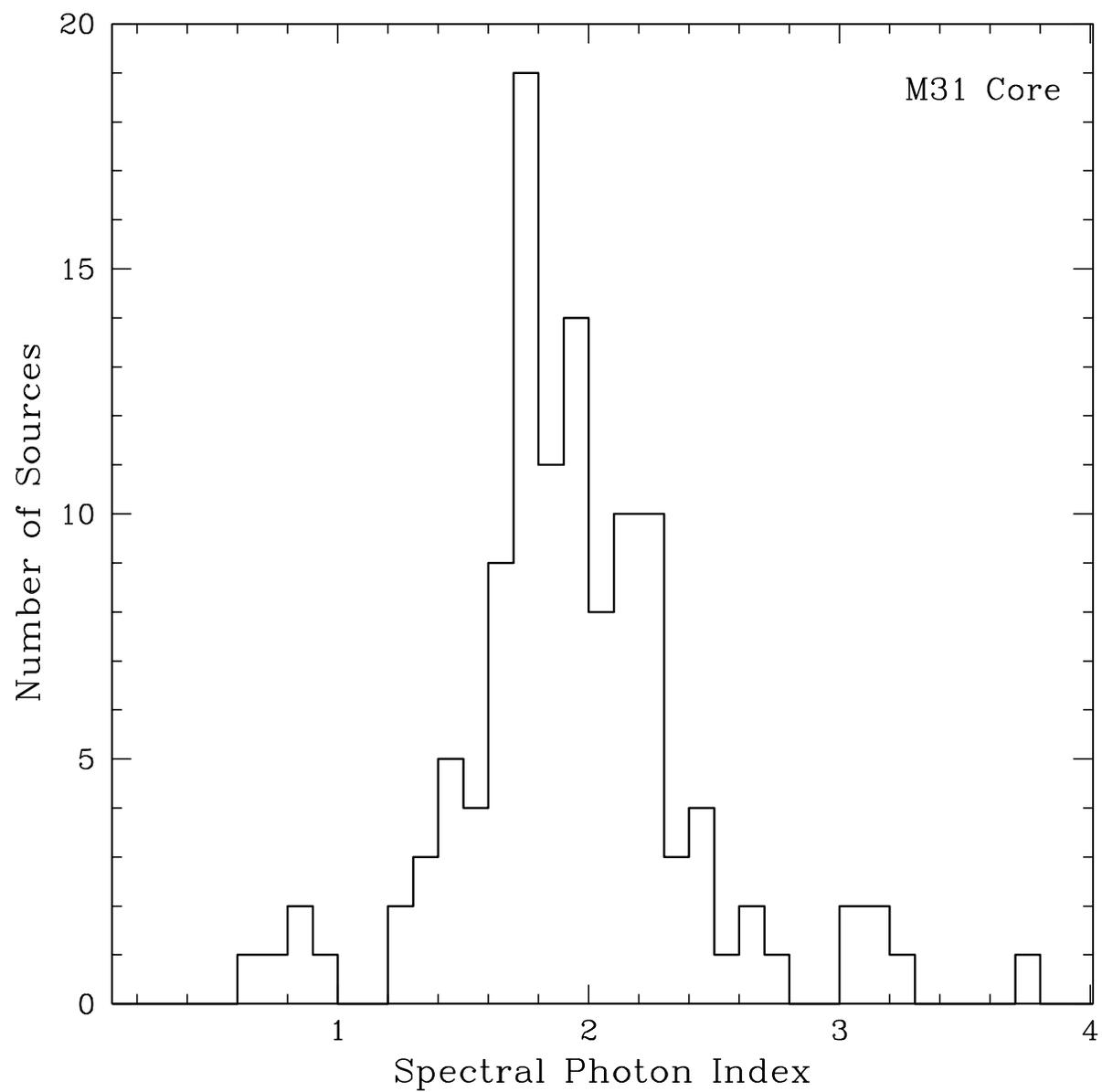}
\caption{The distribution of the spectral slopes derived from spectral analysis of bright X-ray sources with the 
exception of very soft sources. The histogram bins are 0.1 wide. \label{hardness_distribution}}
\end{figure}

To characterize the overall spectral properties of the bright M31 X-ray sources, we constructed 
a distribution of their spectral indices in the $0.3 - 10.0$ keV energy range using the model fits to 
{\em XMM-Newton}/EPIC data with an absorbed simple power law (Fig. \ref{hardness_distribution}). We chose 
this spectral model because it gives adequate representation to the spectra of most sources in our sample 
and allows easier comparison to the results of studies of X-ray binary populations in other galaxies 
(Irwin et al. 2003; Fabbiano \& White 2006; Fabbiano 2006 and references therein). The sources with extremely 
soft spectra (supersoft sources and thermal supernova remnants) were excluded from the sample. For the sources 
with multiple spectral measurements, we used weighted average values of the photon index. The spectral hardness 
distribution including 115 sources spans a wide range of photon indices between $\sim 0.6$ and $\sim 3.7$, and 
has an asymmetric shape with main concentration between $\Gamma \sim 1.2$ and $\Gamma \sim 2.8$ and additional 
groups of sources with soft and hard spectra (Fig. \ref{hardness_distribution}). The overall properties of the 
spectral distribution of the sources in our sample are generally consistent with results of earlier observations 
(Shirey et al. 2001). The main part of the distribution has a dominant peak at $\Gamma \sim 1.75$ and an additional 
shoulder at $\Gamma \sim 2.0-2.2$. The model approximation with a sum of two Gaussian functions provides a 
significantly better fit to the main part of hardness distribution of M31 X-ray sources in the $1.0<\Gamma<3.0$ 
range when compared to a single Gaussian model fit. The model fit with a single Gaussian function is not adequate 
with reduced $\chi^{2}_{r}$=9.4. The fit with sum of two Gaussians with centroids at $\Gamma_{1} = 1.78^{+0.04}_{-0.05}$ 
and $\Gamma_{2} = 1.95^{+0.05}_{-0.03}$, and widths of $0.07\pm0.03$ and $0.36\pm0.04$, accounting for $\sim 18\%$ and 
$\sim 82\%$ of the total number of sources, yields significantly lower $\chi^{2}_{r}$=1.3. This is the first case 
in which we clearly see bimodality of the spectral distribution of X-ray sources in an external normal galaxy.

\subsection{Evolution of Spectral Distribution with Source Luminosity}
\begin{figure}
\epsfxsize=18cm
\epsffile{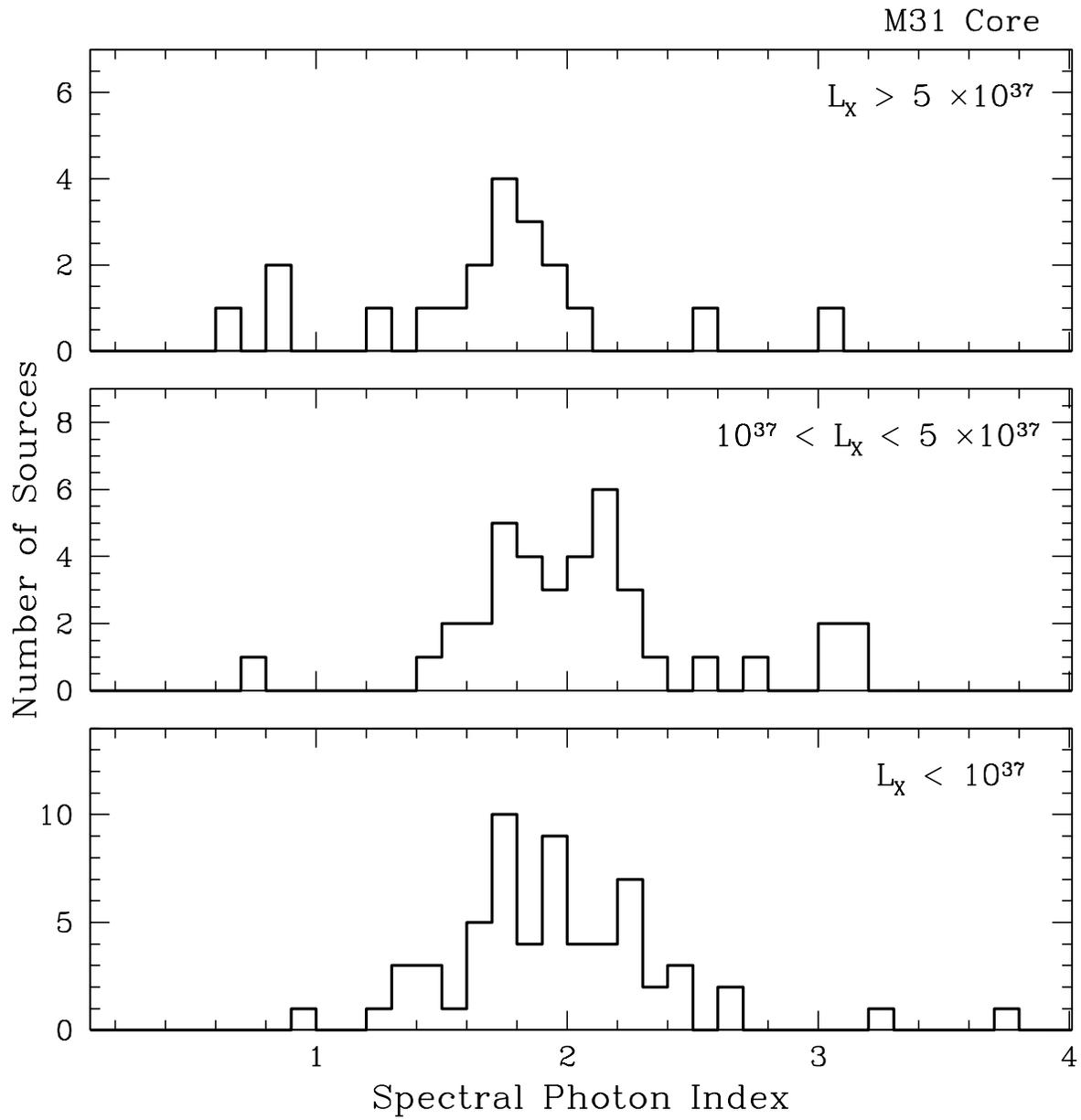}
\caption{Spectral hardness distributions of M31 X-ray sources for different ranges of source luminosity. The 
histogram bins are 0.1 wide. \label{hardness_distr_lum}}
\end{figure}

In order to study the dependence of the spectral hardness distribution on the source luminosity, we constructed 
three separate hardness distributions for sources with apparent luminosities below $10^{37}$ ergs s$^{-1}$ 
(Fig. \ref{hardness_distr_lum}, {\em lower panel})(low luminosity subsample), between $10^{37}$ and 
$5 \times 10^{37}$ ergs s$^{-1}$ (Fig. \ref{hardness_distr_lum}, {\em middle panel}) (intermediate luminosity 
subsample), and brighter than $5 \times 10^{37}$ ergs s$^{-1}$ (Fig. \ref{hardness_distr_lum}, {\em upper panel})
(high luminosity subsample). Each of the three distributions have a distinct central concentration and outlying 
parts representing soft and hard sources. Visual inspection of Fig. \ref{hardness_distr_lum} suggests that the 
spectra of the most luminous sources (L$_{\rm X}>5\times10^{37}$ ergs s$^{-1}$) are generally harder than that of 
the fainter sources. This can be confirmed with a statistical analysis. We used a Kolmogorov-Smirnov (KS) test to 
determine whether three hardness distributions shown in Fig. \ref{hardness_distr_lum} were drawn from the same 
distribution. The results for the low and intermediate luminosity subsamples are consistent with the same parent 
population. The hypotheses that the high luminosity distribution is extracted from the same parent distribution as 
the intermediate and low luminosity distributions can be rejected at the confidence levels of 98.9\% and 99.2\% 
respectively. 

\section{X-ray Hardness-Luminosity Diagram}
\begin{figure}
\epsfxsize=18cm
\epsffile{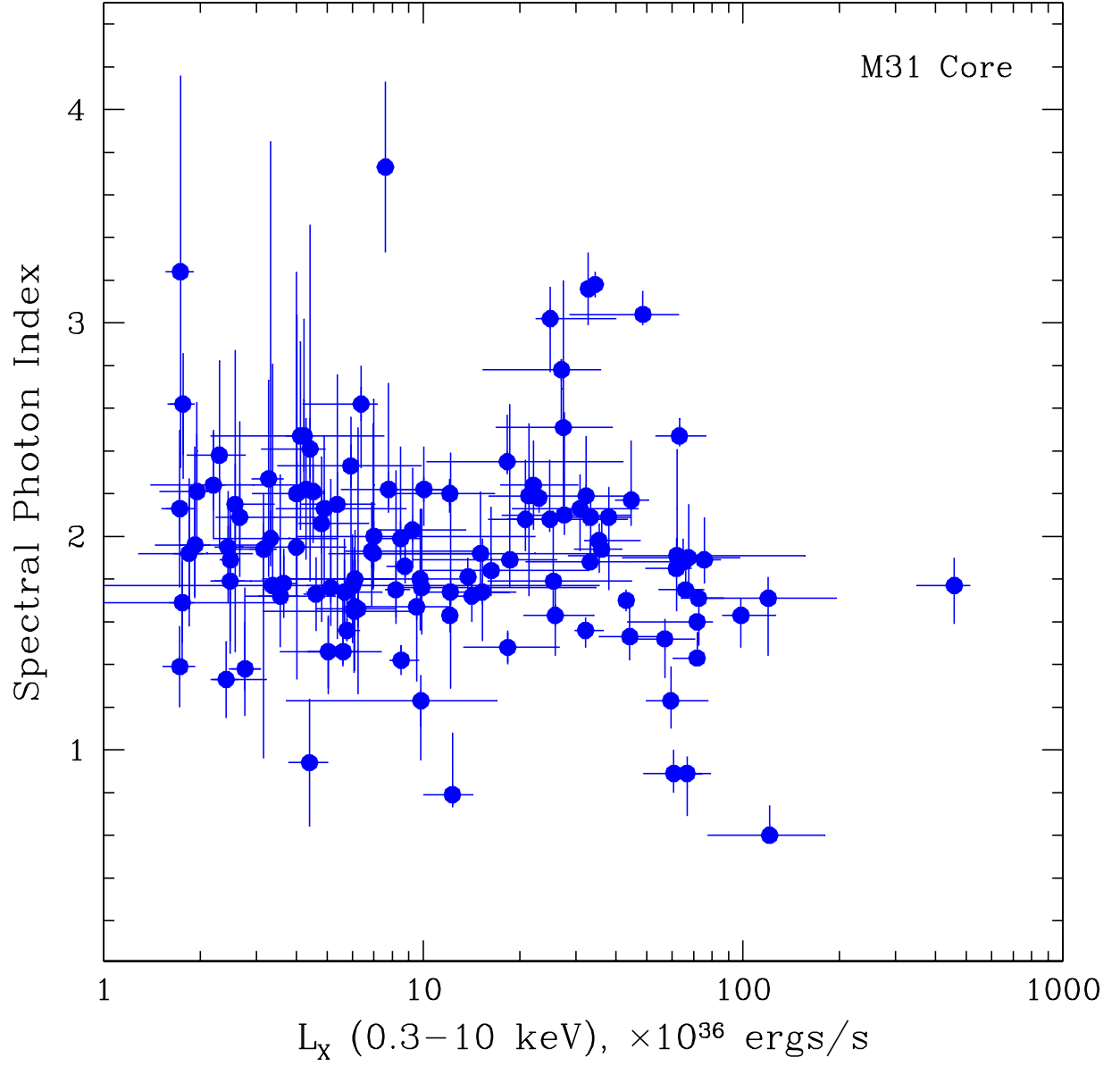}
\caption{Spectral photon index of the bright X-ray sources detected in the M31 core field 
vs. their absorbed X-ray luminosity in the $0.3 - 10$ keV energy band. The error bars in 
X-axis reflect statistical uncertainty of the source flux determination and in some cases 
the range of source X-ray luminosities observed with {\em XMM-Newton}. 
\label{hardness_luminosity}}
\end{figure}

We studied the relation between the hardness of the spectrum and luminosity of X-ray sources in our sample 
using the results of spectral model fitting. In Fig. \ref{hardness_luminosity} the hardness of the spectrum 
of X-ray sources expressed in terms of spectral photon index is shown as a function of their absorbed X-ray 
luminosity in the $0.3 - 10$ keV energy band calculated assuming the distance of 760 kpc. The sources with 
extremely soft spectra (supersoft sources and thermal supernova remnants) have been excluded from the sample 
and are not shown in this plot. The sources in the low-luminosity ($L_{\rm X} < 10^{37}$ ergs s$^{-1}$) part 
of the diagram show a wide scatter of photon indices ($\Gamma \sim 0.9 - 3.8$). At higher luminosities 
($L_{\rm X} > 10^{37}$ ergs s$^{-1}$), one can identify tree main source concentrations/branches in the 
hardness-intensity diagram. The first densely populated group comprises sources with the spectral photon 
indices $\Gamma \sim 1.4 - 2.3$ with the most luminous sources having narrower range of photon indices 
$\Gamma \sim 1.5 - 1.9$. The two other groups include sources with soft spectra ($\Gamma > 2.3$) and sources 
with extremely hard spectra ($\Gamma < 1.4$). 

\section{Low-Energy Absorption}
\begin{figure}
\epsfxsize=18cm
\epsffile{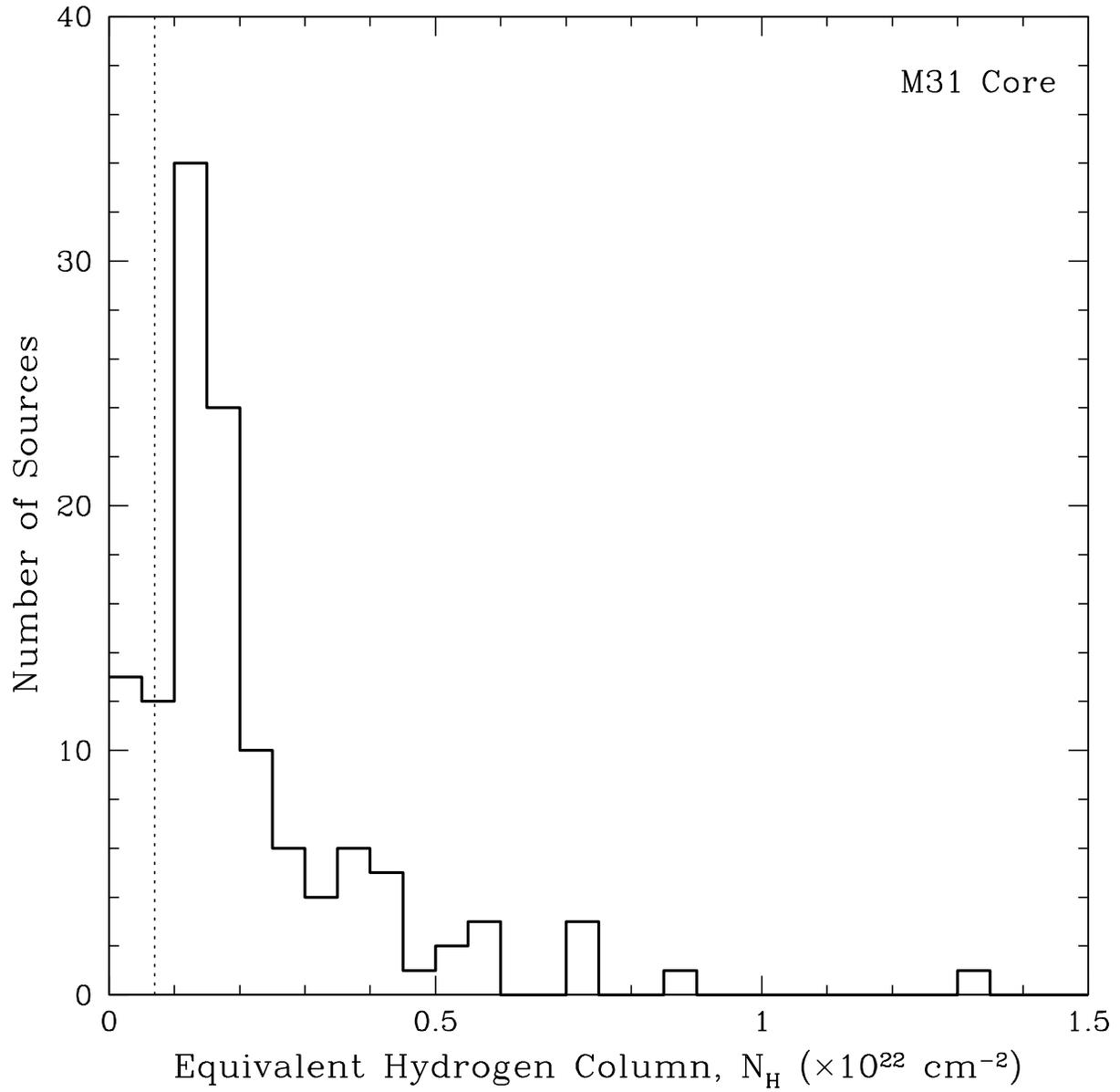}
\caption{The distribution of the absorbing columns derived from spectral analysis of bright 
X-ray sources. Each bin along X-axis has a width of $5\times 10^{20}$ cm$^{-2}$. The Galactic 
foreground absorbing column in the direction of M31 ($7 \times 10^{20}$ cm$^{-2}$) is marked 
with dotted line. \label{N_H_distr}}
\end{figure}

The X-ray absorption in the source spectra could provide information on their location and help to map 
the structure of ISM inside M31 \cite{TP04}. We estimated the value of equivalent hydrogen absorbing 
column, $N_{\rm H}$ for each source in our sample from the best-fit model approximation of 
their spectra (Tables \ref{source_ID_1},\ref{spec_par_CUTOFF},\ref{spec_par_two_comp},\ref{spec_par_TR},
\ref{spec_par_SSS})\footnote{It should be noted, that the amount of low-energy absorption derived 
from spectral fitting is usually sensitive to the type of continuum model used to approximate X-ray spectrum. 
The choice of a particular spectral model introduces additional uncertainty in the value of $N_{\rm H}$ 
\cite{TP04}}. For the sources with multiple spectral measurements, we used weighted average values of 
$N_{\rm H}$. The resulting distribution of absorbing columns towards X-ray sources in the central M31 field 
is shown in Fig. 
\ref{N_H_distr}. For the majority of sources in our sample the derived value of $N_{\rm H}$ is either in 
excess or consistent with Galactic hydrogen column $N_{\rm H}^{\rm Gal} \sim 7 \times 10^{20}$ cm$^{-2}$ 
in the direction of M31 \cite{DL90}. The distribution of absorbing columns has a prominent peak centered 
at $\sim 1.2 \times 10^{21}$ cm$^{-2}$ and a tail structure extending to higher columns with most absorbed 
source having $N_{\rm H} \sim 1.3 \times 10^{22}$ cm$^{-2}$ (Fig. \ref{N_H_distr}). The average value of 
$N_{\rm H}$ derived from our source sample is $(1.52\pm0.02)\times 10^{21}$ cm$^{-2}$ or approximately 
twice the Galactic foreground value. Given a considerably large size and completeness of our source sample, 
it can be used as a measure of the average level of X-ray absorption towards sources in M31.


\section{Variability of M31 sources}

\subsection{Long-term flux variability}
Combining the data of multiple {\em XMM-Newton} observations, we searched for long-term flux variability of 
sources in our sample. We found that more than 80\% of sources observed in two or more observations show 
significant variability (at the level of $3\sigma$ and higher). Ten sources in our sample are transients with 
outburst-to-quiescent luminosity ratios of $\sim 100 - 1000$. Three other sources show recurrent outbursts on 
the time scales of years (Table \ref{source_ID_1}). The fraction of variable sources in our survey is much 
higher than 50\% reported from earlier {\em Chandra}/ACIS survey of the central region of M31 (Kong et al. 
2002a). On the other hand, it is remarkably close to the fraction of variable sources ($>80\%$) found in M31 
GC X-ray source population \cite{TP04} and in the smaller sample of bright M31 X-ray sources (90\%)(Barnard 
et al. 2006a). 

\subsection{Short-term aperiodic variability}
The {\em XMM-Newton} observations of the central part of M31 have led to the detection of a periodic variability 
of three sources included in our sample: X-ray pulsations in the supersoft transient source \#115 (Osborne et al. 
2001) and periodic dipping in the GC X-ray source \#109 (Trudolyubov et al. 2002b) and in the source \#103 
(Mangano et al. 2004). A detection of strong aperiodic variability in a number of M31 sources has been also reported 
(Barnard et al. 2003a; Barnard et al. 2004; Williams et al. 2005a; Barnard et al. 2006a). The reported variability 
has been characterized by broken power law PDS with a break at $0.01 - 0.1$ Hz, and interpreted as a signature of 
the disk-accreting X-ray binaries. However, recently it has been discovered that this aperiodic variability has 
artificial origin, resulting from improper addition of lightcurves from individual EPIC detectors (Barnard et al. 
2006b).

\begin{figure}
\epsfxsize=18cm
\epsffile{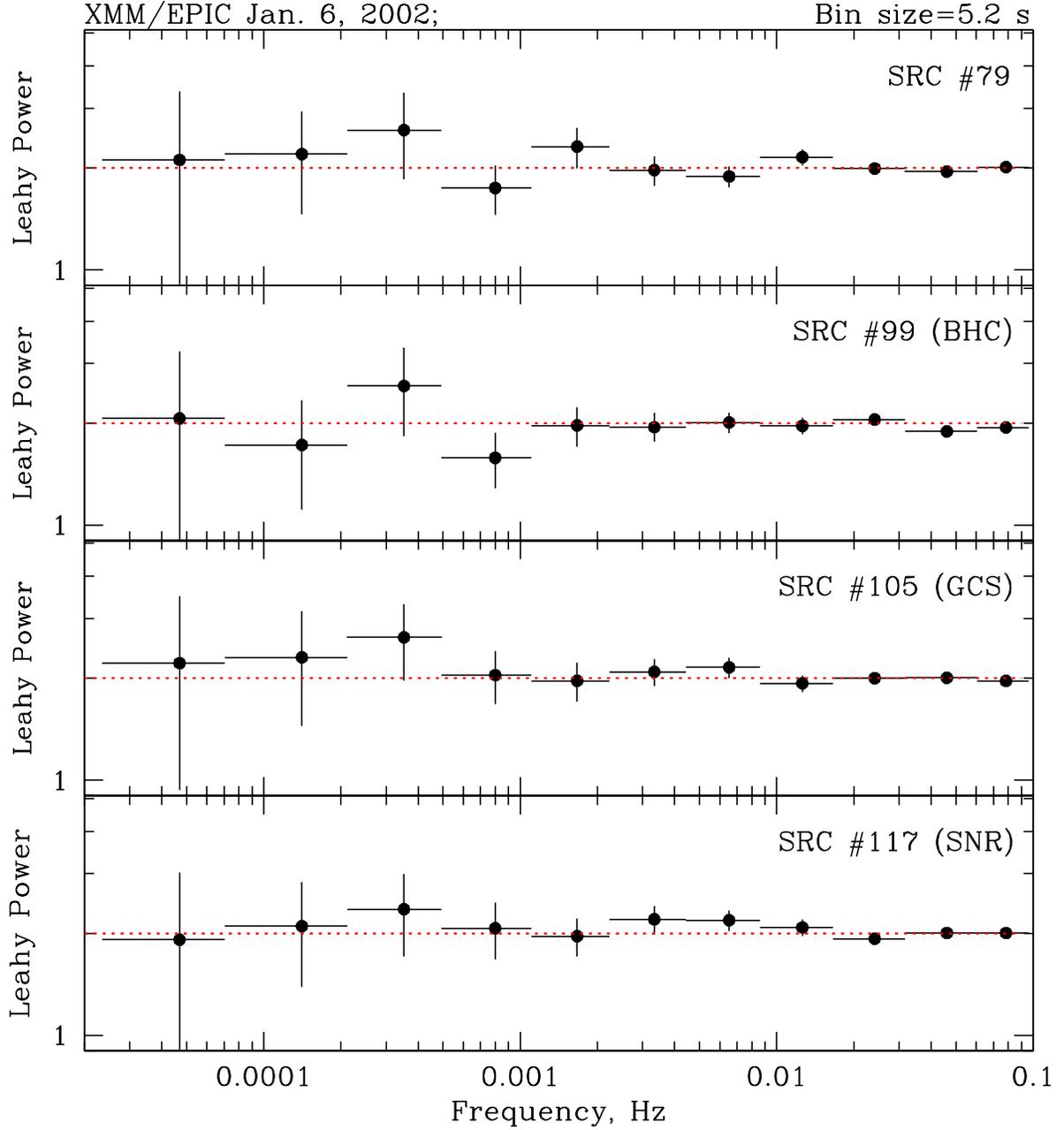}
\caption{Leahy-normalized power density spectra of four sources in our sample with no significant aperiodic 
variability detected: src. 79, a black hole candidate X-ray transient src. 99, a globular cluster src. 105 
and a supernova remnant candidate src. 117. Combined EPIC-pn, MOS1 and MOS2 data in the $0.3-7$ keV (Src. 79, 
99, 105) and $0.3-2$ keV (Src. 117) energy ranges, $2\times 10^{-5}-0.1$ Hz frequency range. The PDS 
were rebinned logarithmically, to reduce scatter at higher frequencies. Note that all PDS are consistent 
with pure counting noise (expected Poisson noise level is shown with dotted lines). \label{pds_no_var}}
\end{figure}

\begin{figure}
\epsscale{.6}
\plotone{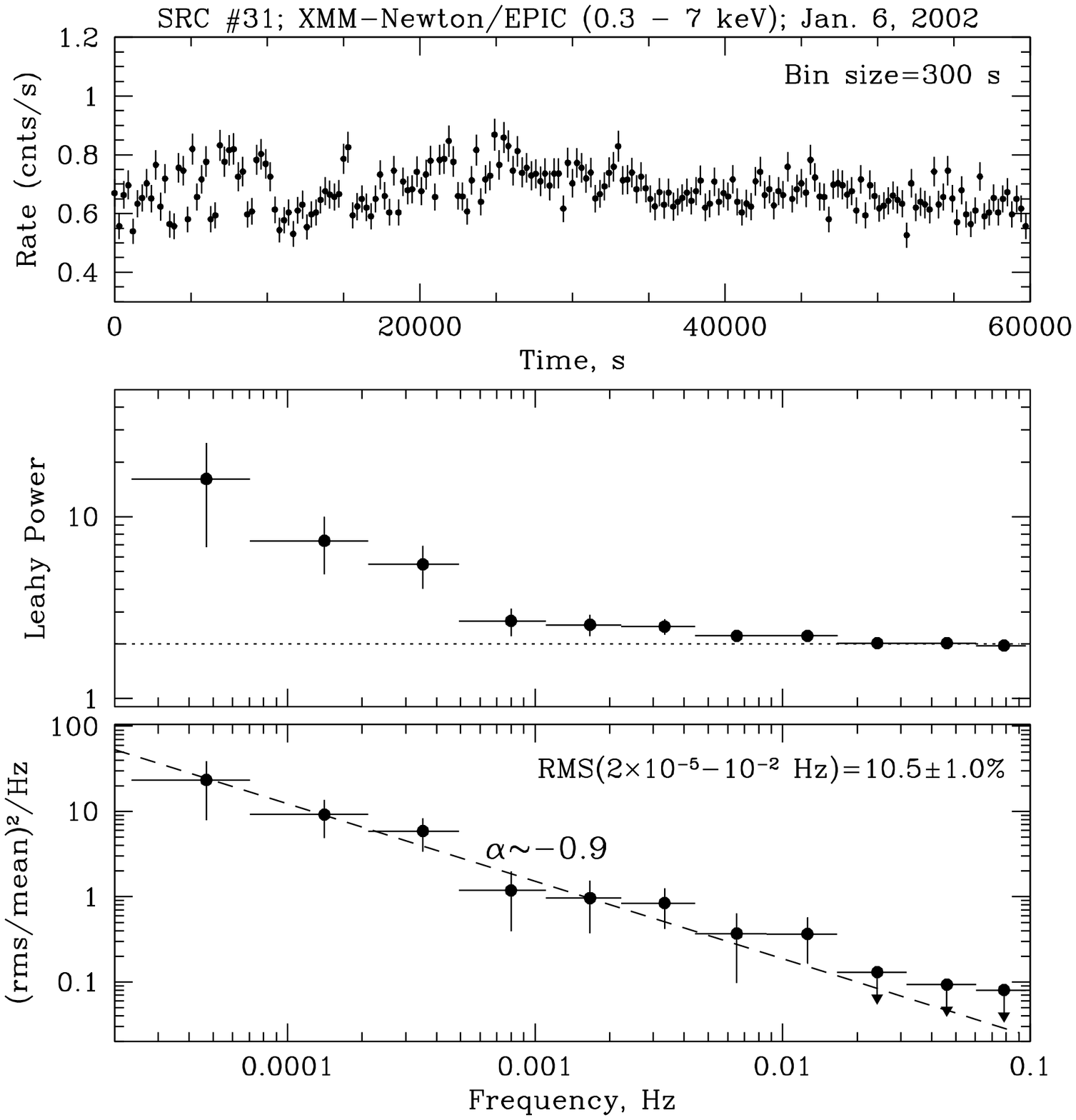}
\plotone{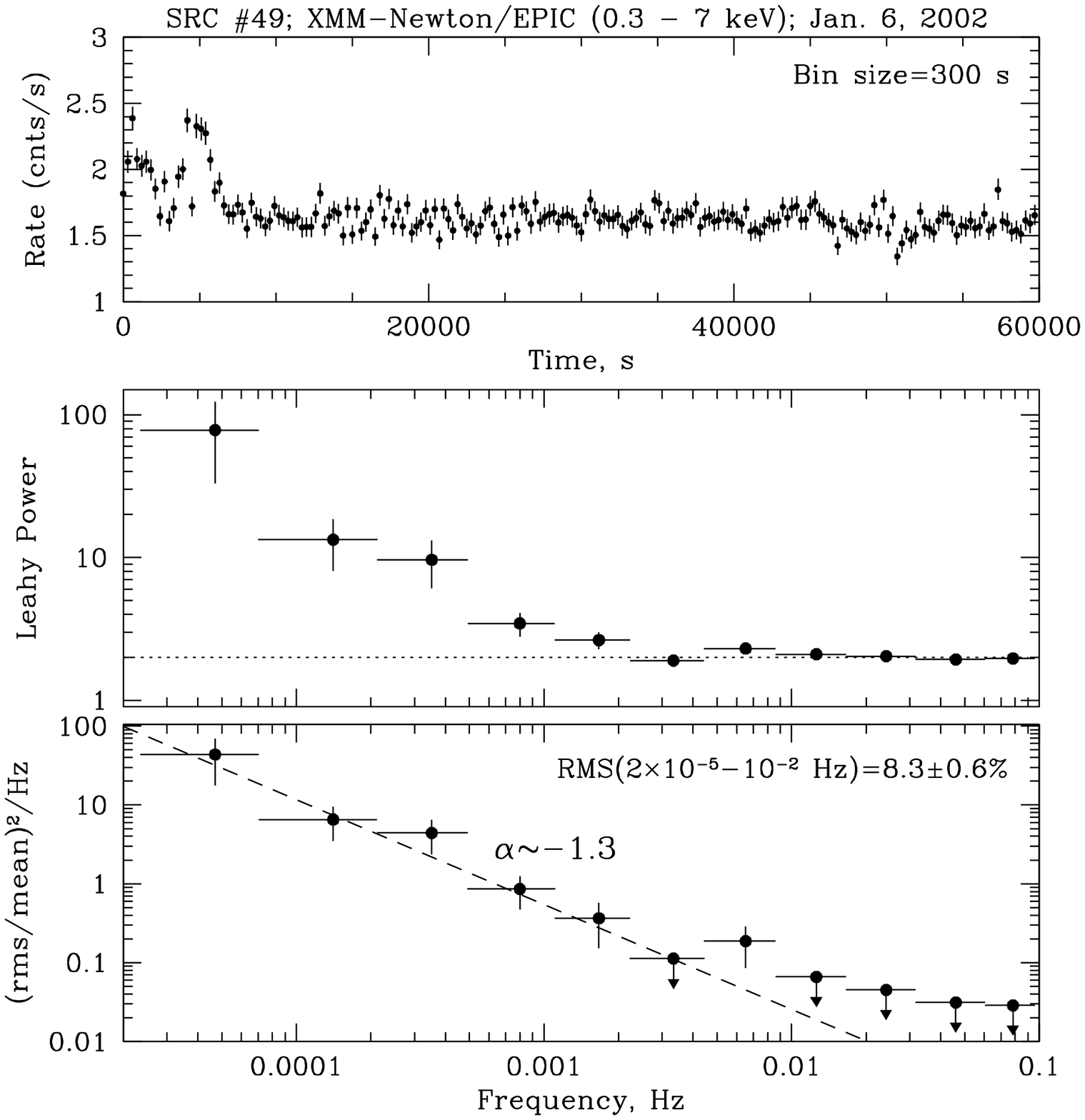}
\caption{Lightcurves (upper panels) and PDS (middle and lower panels) of the two sources (Src. 31 and 
49) showing significant levels of aperiodic variability. \label{pds_var}}
\end{figure}

We re-examined the data of {\em XMM-Newton} observations in a search for short-term aperiodic variability of X-ray 
sources in our sample. For each observation, we constructed power density spectra of X-ray sources in the 
$2\times 10^{-5}-0.1$ Hz frequency range using the lightcurves from individual and combined EPIC detectors. Because 
of the lack of statistics, none of the sources in our sample shows detectable aperiodic variability at frequencies 
above $10^{-2}$ Hz. The PDS of majority of sources do not indicate significant source variability over the wide 
range of frequencies, being consistent with white noise expected from data errors. Typical examples of the 
Leahy-normalized broad-band power density spectra of M31 sources with no detectable variability are shown in Fig. 
\ref{pds_no_var}. However, several brighter sources show signs of very-low frequency variability with two brightest 
sources, \#31 and \#49, being the most variable (Fig. \ref{pds_var}). 

The Leahy-normalized power density spectra of sources \#31 and \#49 measured in the Jan. 6, 2002 {\em XMM} observation 
clearly exceed the expected white noise level at frequencies below $\sim 10^{-2}$ Hz (Fig \ref{pds_var}, middle panels) 
with source fractional variability amplitudes of $10.5\pm1.0\%$ and $8.3\pm0.6\%$ in the $2\times 10^{-5}-10^{-2}$ Hz 
frequency band. The PDS of the sources \#31 and \#50 normalized to the square of fractional r.m.s. variability with 
expected white noise level subtracted are shown in lower panels of Fig. \ref{pds_var}. Both PDS can be adequately 
approximated by power law models (${\rm P}_{\nu} \sim \nu^{-\alpha}$) with exponent $\alpha = 0.9^{+0.2}_{-0.1}$ (source 
\#31) and $\alpha = 1.3\pm0.2$ (source \#49). 

The very-low frequency PDS of the M31 X-ray sources \#31 and \#49 resemble that of the bright Galactic low-mass 
X-ray binaries (LMXB) \cite{GA06}. In their recent survey of the very-low frequency X-ray variability of the persistent 
Galactic LMXB, Gilfanov \& Arefiev (2006) show that PDS of the sources can be generally approximated by broken power law 
which is nearly flat at frequencies below the break and has a slope of $0.6 \sim 1.5$ at higher frequencies. The break 
frequency is found to be correlated with binary orbital frequency for a broad range of binary periods. The observed 
very-low frequency variability of LMXB has been explained as a result of the slow variations in the mass accretion rate 
being generated in the outer parts of the accretion flow and propagated to the region of the main energy release on a 
viscous time scale of the accretion flow. The observed slopes of the PDS are in general agreement with the slope 
$\alpha\sim 1$ predicted from the model involving local fluctuations in the mass accretion rate in the viscous accretion 
disk (Lyubarskii 1997). The break in the PDS slope is a signature of the finite size of the accretion disk related to the 
viscous time on the outer boundary of the disk. If sources \#31 and \#49 are indeed accretion-powered X-ray binaries in 
M31, as suggested by their overall X-ray properties, they can be seen as further support for Lyubarskii's model.

\subsection{Long-term spectral variability}
We searched for long-term spectral variability of M31 X-ray sources combining the data of multiple {\em XMM-Newton} 
observations. Since many of the sources in our sample have not enough counts for establishing spectral variability 
from model fits, we used their hardness ratios to search for spectral variability. These hardness ratios were 
calculated using the source counts in three energy bands of MOS and pn detectors: $0.3 - 1$ keV (soft band, S), 
$1 - 2$ keV (medium band, M) and $2 - 7$ keV (hard band, H), and defined as: HR1=(M-S)/(M+S) and HR2=(H-M)/(H+M). 
In order to study long-term spectral variability, we computed a set of two spectral variability parameters for 
each observation and each of the EPIC detectors following Kong et al. (2002a) : 
\begin{equation}
\delta({\rm HR1}) = \frac{|{\rm HR1}_{\rm max}-{\rm HR1}_{\rm min}|}{\sqrt{\sigma_{{\rm HR1}_{\rm max}}^{2}+\sigma_{{\rm HR1}_{\rm min}}^{2}}}
\end{equation}

\begin{equation}
\delta({\rm HR2}) = \frac{|{\rm HR2}_{\rm max}-{\rm HR2}_{\rm min}|}{\sqrt{\sigma_{{\rm HR2}_{\rm max}}^{2}+\sigma_{{\rm HR2}_{\rm min}}^{2}}}
\end{equation}

\noindent where ${\rm HR1}_{\rm min}$, ${\rm HR2}_{\rm min}$ and ${\rm HR1}_{\rm max}$, ${\rm HR2}_{\rm max}$ are the 
minimum and maximum values of the source hardness ratios during the 4 years of observations and $\sigma$ denote 
corresponding errors. We define the source to be spectrally variable if $\delta({\rm HR1})$ or $\delta({\rm HR2})$ exceeds 
level of 3 in both EPIC-MOS (combined MOS1 and MOS2) and pn detectors. Applying this criterion, we found that about $50\%$ 
of sources in our sample with multiple flux measurements available show significant spectral variability (Table 
\ref{source_ID_1}). This estimate should be considered as a lower limit, since we have not enough sensitivity to detect 
possible small changes in the spectra of a number of fainter sources and because of the sparseness of 
{\em XMM-Newton} observations. We found the fraction of spectrally variable sources in our sample to be much higher 
than the 5\% previously reported from {\em Chandra} survey (Kong et al. 2002a), which demonstrates a significant 
improvement in the sensitivity and temporal coverage in comparison to previous studies of X-ray sources in the 
center of M31. 

\begin{figure}
\epsfxsize=18cm
\epsffile{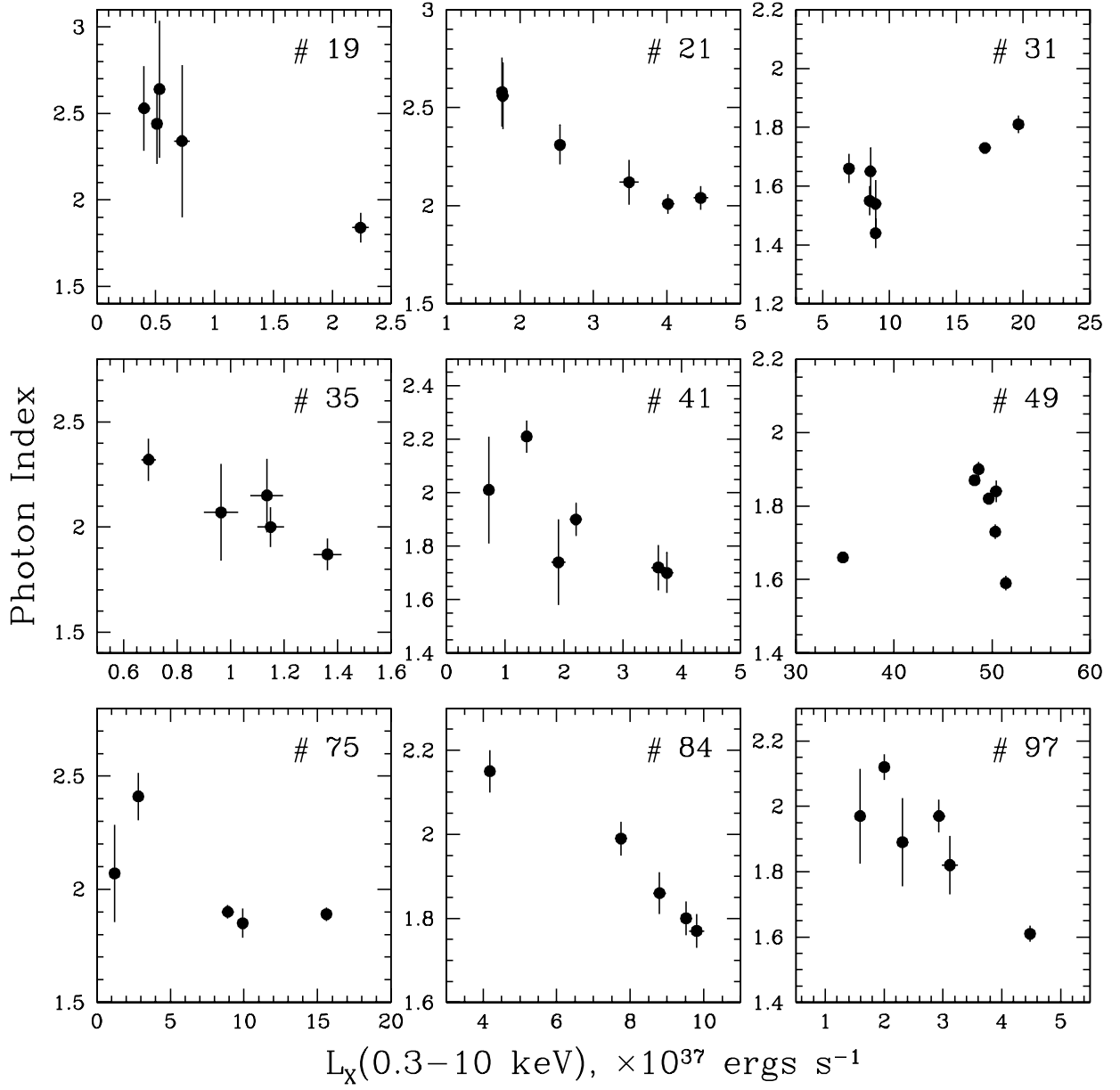}
\caption{Spectral variability of 9 bright X-ray sources. The hardness of the spectrum expressed 
in terms of the spectral photon index (y-axis) is plotted against source X-ray luminosity in the 
$0.3 - 10$ keV energy band (x-axis). \label{spec_var}}
\end{figure}

A significant fraction of the sources in our sample provide sufficient number of counts in individual EPIC observations 
to permit a detailed study of their spectral variability using spectral model fits.  In Fig. \ref{spec_var} the long-term 
evolution of the spectral hardness is shown for 9 brighter X-ray sources, showing the highest levels of spectral variability. 
In this figure the hardness of the X-ray spectrum, expressed in terms of the best-fit photon index, is shown as a function 
of the source luminosity. The majority of spectrally variable sources with luminosities above $\sim 10^{37}$ and below 
$\sim 10^{38}$ ergs s$^{-1}$ demonstrate a correlation between the level of their X-ray flux and the hardness of their 
spectrum: as the source flux increases, the spectrum becomes harder (Fig. \ref{spec_var}). The brightest sources with 
luminosities above $10^{38}$ ergs s$^{-1}$ (sources \#31 and \#49 in Fig. \ref{spec_var}) show a more complex relation 
between the spectral slope and flux. The spectral evolution of these brightest sources is generally consistent with that of 
the Galactic Z sources (Hasinger \& van der Klis 1989).     

\subsection{Short-term spectral variability}
\begin{figure}
\epsscale{.6}
\plotone{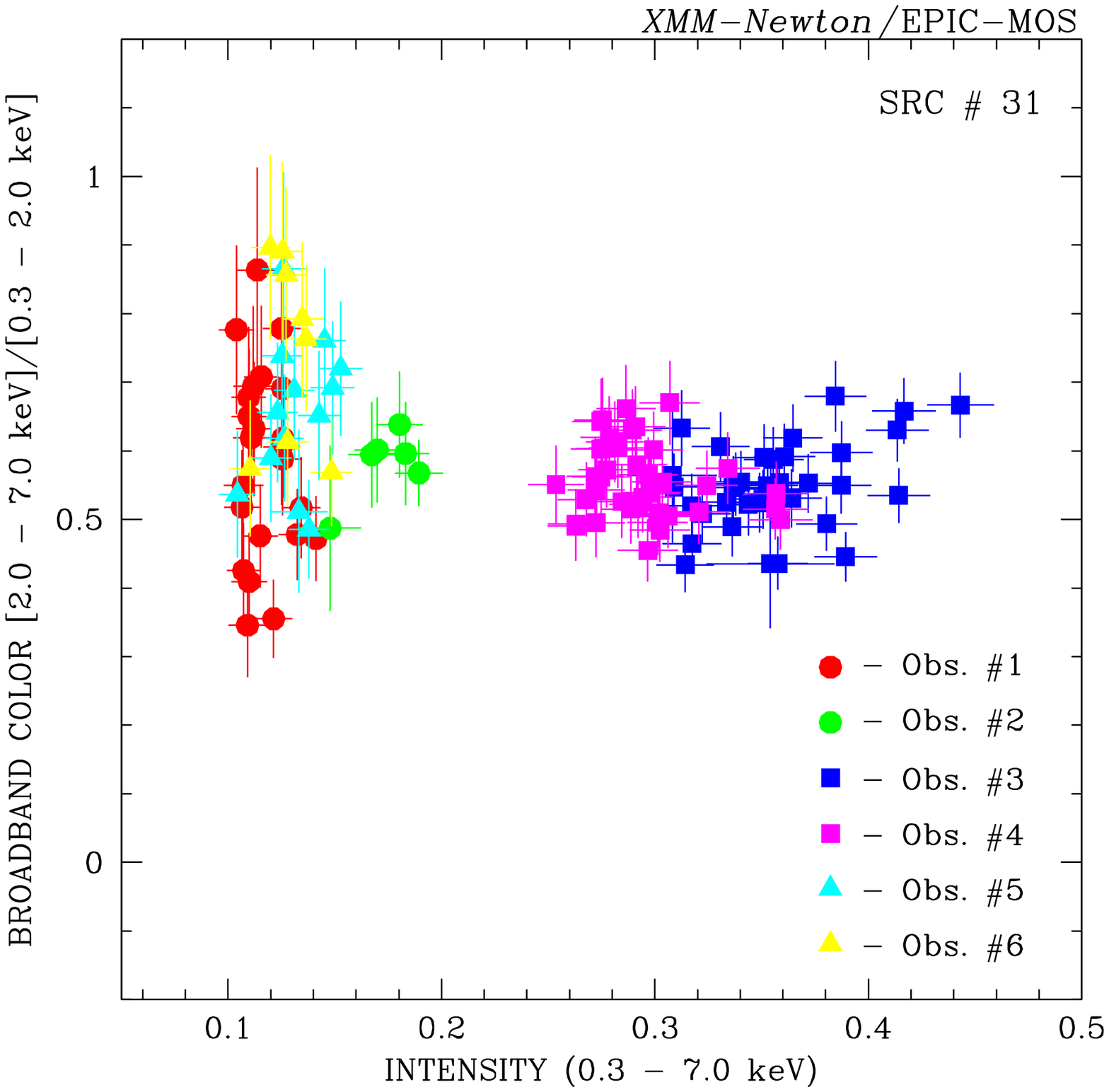}
\plotone{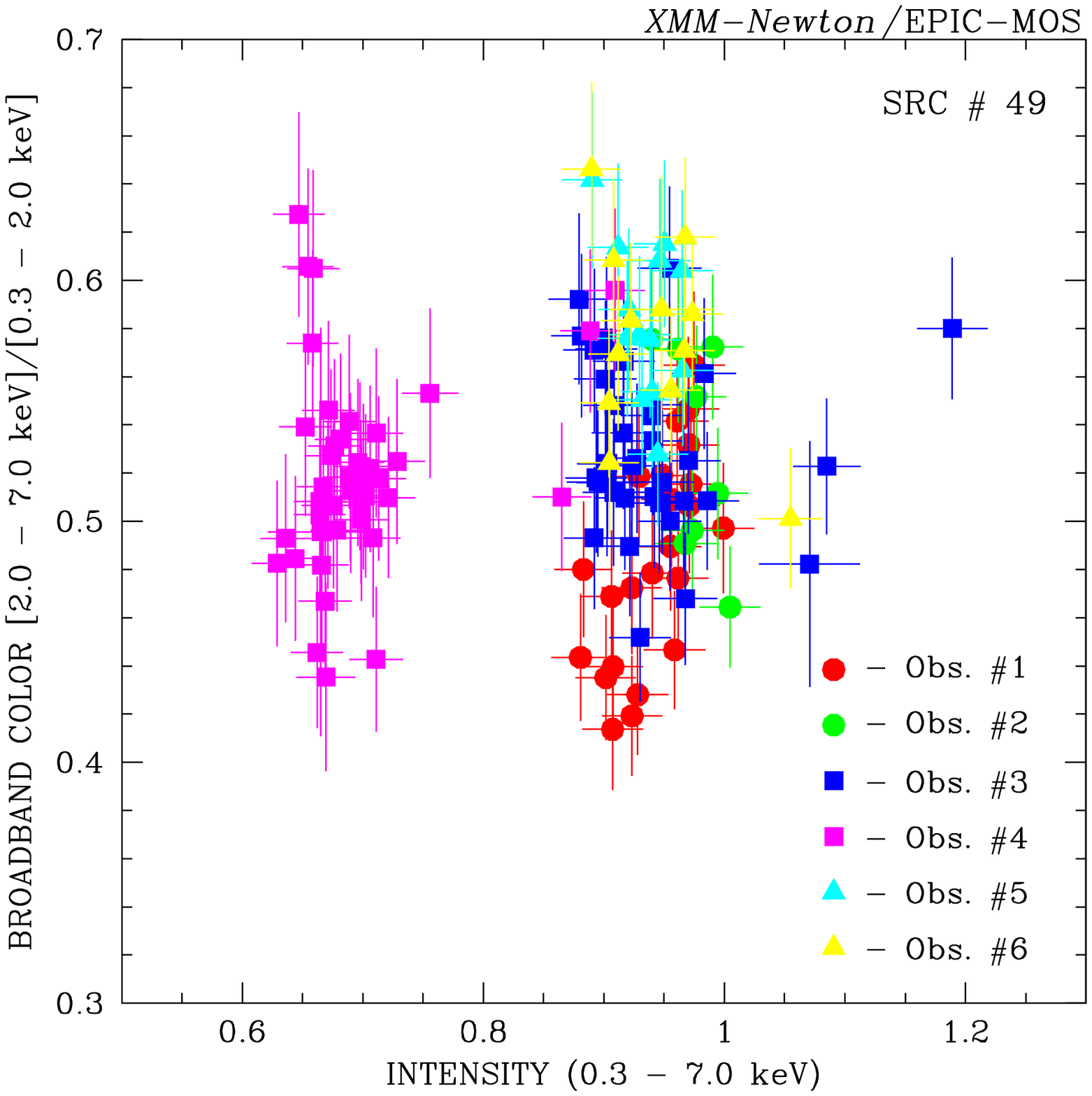}
\caption{{\em Upper panel:} Broadband color (hardness) vs. intensity for six {\em XMM} observations of X-ray 
source \#31. The broadband color is defined as ratio of source intensities in the $2 - 7$ and 
$0.3 - 2$ keV energy bands. The EPIC-MOS data is binned to 1500s. The corresponding source luminosity 
changes between $\sim 7 \times 10^{37}$ and $\sim 2 \times 10^{38}$ ergs s$^{-1}$ in the $0.3 - 10$ 
keV energy band. {\em Lower panel:} Broadband color vs. intensity for the source \#49. The EPIC-MOS data 
is binned to 1500s. The corresponding source luminosity changes between $\sim 3 \times 10^{38}$ and 
$\sim 5.5 \times 10^{38}$ ergs s$^{-1}$ in the $0.3 - 10$ keV energy band.\label{hardness_intensity}}
\end{figure}

The unprecedented sensitivity of {\em XMM-Newton} provides a unique opportunity to study spectral evolution of 
the brightest X-ray sources in M31 on a time scale of hundreds and thousands of seconds. To study short-term 
spectral variability of the bright sources, we constructed their X-ray hardness-intensity diagrams using the 
data of {\em XMM-Newton}/EPIC observations. The X-ray hardness was defined as the ratio of the source intensities 
in the $2.0 - 7.0$ keV and $0.3 - 2$ keV energy bands with data integration times of $1000 - 3000$ s depending on 
the source intensity. Several sources show a complex patterns of evolution on the hardness-intensity diagram 
somewhat reminiscent of the Galactic Z and atoll sources (Hasinger \& van der Klis 1989; Trudolyubov \& Priedhorsky 
2004). In Fig. \ref{hardness_intensity} we show the hardness-intensity diagrams of two of these sources (sources 
\#31 and \#49) based on the EPIC-MOS data. The hardness-intensity diagrams of both sources clearly demonstrate 
hysteretic behavior with the same values of the broad-band color corresponding to the different levels of the source 
luminosity. The X-ray source \#49 has been previously classified as a Z source candidate (Barnard et al. 2003b). 
The results of our analysis are generally consistent with Z source classification of this source. The X-ray source 
\#31 has been previously proposed by Barnard et al. (2003a) as a stellar-mass black hole candidate in M31 mainly 
on the basis of its short-term aperiodic variability. Recently, it has been shown that the aperiodic variability of 
the source reported in Barnard et al. (2003a) has artificial origin (Barnard et al. 2006b), making its black hole 
classification questionable. Moreover, the spectrum, luminosity and the pattern of the short-term spectral variability 
of the source \#31 (Fig. \ref{hardness_intensity}, {\em upper panel}) clearly resemble those of the high-luminosity 
neutron star systems rather than stellar-mass black hole candidates. 

\section{Source classification}

\subsection{Hardness-Luminosity diagrams}
The results of recent {\em XMM-Newton} and {\em Chandra} observations of individual X-ray sources in M31 suggest 
close similarity to their Galactic counterparts (Trudolyubov, Borozdin \& Priedhorsky 2001; DiStefano et al. 2002; 
Trudolyubov \& Priedhorsky 2004; Williams et al. 2006). Based on their similarity to Galactic sources, the 
majority of sources in our sample should be accreting X-ray binaries with neutron star and black hole primaries. 
The problem of distinguishing between these two types of systems among M31 sources is of fundamental importance. 
It is critical for understanding the processes of binary evolution and star formation history of M31. The study 
of the short-term X-ray variability can provide a definitive answer on the nature of the source, if Type I X-ray 
bursts or X-ray pulsations are observed. Unfortunately, the observed source count rates for most M31 sources are 
too low to allow us to probe their fast ($t < 10 - 20$ s) variability and search for typical Type I X-ray bursts. 
Another attractive possibility is to try to classify M31 sources based on the comparisons of their spectral 
properties and variability with that of the canonical Galactic neutron star and black hole X-ray binary systems. 

X-ray colors can be a sensitive discriminator of the source type, when applied to study of X-ray source populations 
in nearby galaxies (DiStefano \& Kong 2003; Prestwich et al. 2003). Spectral analysis of the high-quality data has
proven to be even more effective, because it can account for the low-energy absorption and describe detailed spectral 
shape (i.e two-component spectra, spectral lines). Nonetheless, this analysis often needs to be combined with 
information on X-ray variability and counterparts at other wavelengths for secure source classification. 

The spectral hardness-luminosity diagrams can be very useful for classification of X-ray sources in nearby galaxies, 
because of the good source count statistics in {\em XMM} and a small relative uncertainty in the source distances. 
In this work we made use of the hardness (spectral photon index, $\Gamma$)-luminosity diagram of M31 sources (Fig. 
\ref{hardness_luminosity}) combined with detailed spectral model fits and spectral variability information to 
identify a number of sources as black-hole and neutron star binary candidates. 

As a first step, we tried to define the regions in the hardness-luminosity diagram corresponding to the established 
spectral states of X-ray binaries, using spectral model fits and an extensive set of published spectral model 
parameters for a number of selected neutron star and black hole binaries with well established distances. In most 
cases, we used spectral data of {\em ASCA}, {\em BeppoSAX} and {\em EXOSAT}, because of their common energy range 
with {\em XMM-Newton}. We used standard models to fit the spectra of the established black hole and neutron star 
X-ray sources in a variety of spectral states (White, Stella \& Parmar 1988; McClintock \& Remillard 2006 and 
references therein). The resulting spectral model parameters were then used to simulate {\em XMM-Newton} spectra of 
these objects as if they were observed in M31. When simulating the spectra of Galactic binaries, an 
absorbing column $N_{\rm H} = 1.2\times 10^{21}$ cm$^{-2}$, corresponding to the peak of the absorbing column 
distribution for M31 sources was assumed. The spectral photon indices and source luminosities were derived from 
fitting of the simulated spectra by the absorbed power law model in the $0.3 - 10$ keV energy band. 

Typically, the energy spectrum of the stellar-mass black hole candidates in the high/soft state consists of the 
dominant thermal disk component with $kT \lesssim 1.5$ keV and a power law tail with a photon index 
$\Gamma \sim 2.0 - 5.0$. Some sources also show the very-high state characterized by dominant power law component 
with $\Gamma > 2.4$ and a soft thermal component with temperature of up to $\sim 1.8 - 2$ keV. The intrinsic energy 
spectrum of black hole candidates in the hard/low state is usually described as a power law with a photon index 
$\Gamma = 1.4 \sim 2.1$ and sometimes a low-temperature disk component with $kT \sim 0.2 - 0.4$ keV. The spectrum 
in the intermediate state corresponding to the transition between low and high/very high states bears resemblance 
to both low and high states. The estimated $0.3 - 10$ keV luminosities of the stellar-mass black hole candidates 
in the high/very high are usually well above $2 \times 10^{37}$ ergs s$^{-1}$, while in the low/hard state they tend 
to be below that level. The luminosities in the intermediate state fall in the range $(1 - 3)\times 10^{37}$ ergs 
s$^{-1}$ The majority of simulated $0.3 - 10$ keV EPIC spectra of the black hole candidates in the high and very-high 
states, scaled to the M31 distance, can be described by simple power law models with photon index 
$\Gamma = 2.1 \sim 5.0$, with the exact value of $\Gamma$ depending on the parameters of the input spectrum: the 
temperature of the soft component, slope of the hard tail and relative normalization of the soft and hard components. 
The characteristic values of the photon index in the simulated low state spectra lie between $\sim 1.4$ and $\sim 2.2$, 
while for the intermediate state they are close to that of the high state. 

The energy spectra of the neutron star binaries at high luminosities $\gtrsim (3 - 5)\times 10^{37}$ ergs s$^{-1}$ 
are usually well fit with a two-component models consisting of a soft black body-like component with characteristic 
temperature of $\sim 0.5 - 1.5$ keV, together with harder component with $kT \sim 1.5 - 2.5$ keV. The resulting 
simulated EPIC spectra of high-luminosity neutron star systems can be approximated by power law models with photon 
index $\Gamma = 1.4 \sim 1.9$, with gradual hardening and reduced scatter in photon index as luminosities approach and 
exceed $10^{38}$ ergs s$^{-1}$. The simulated spectra of neutron stars at low and intermediate luminosities show 
wider scatter of the photon indices: $\Gamma = 1.4 \sim 2.5$.   

\begin{figure}
\epsfxsize=18cm
\epsffile{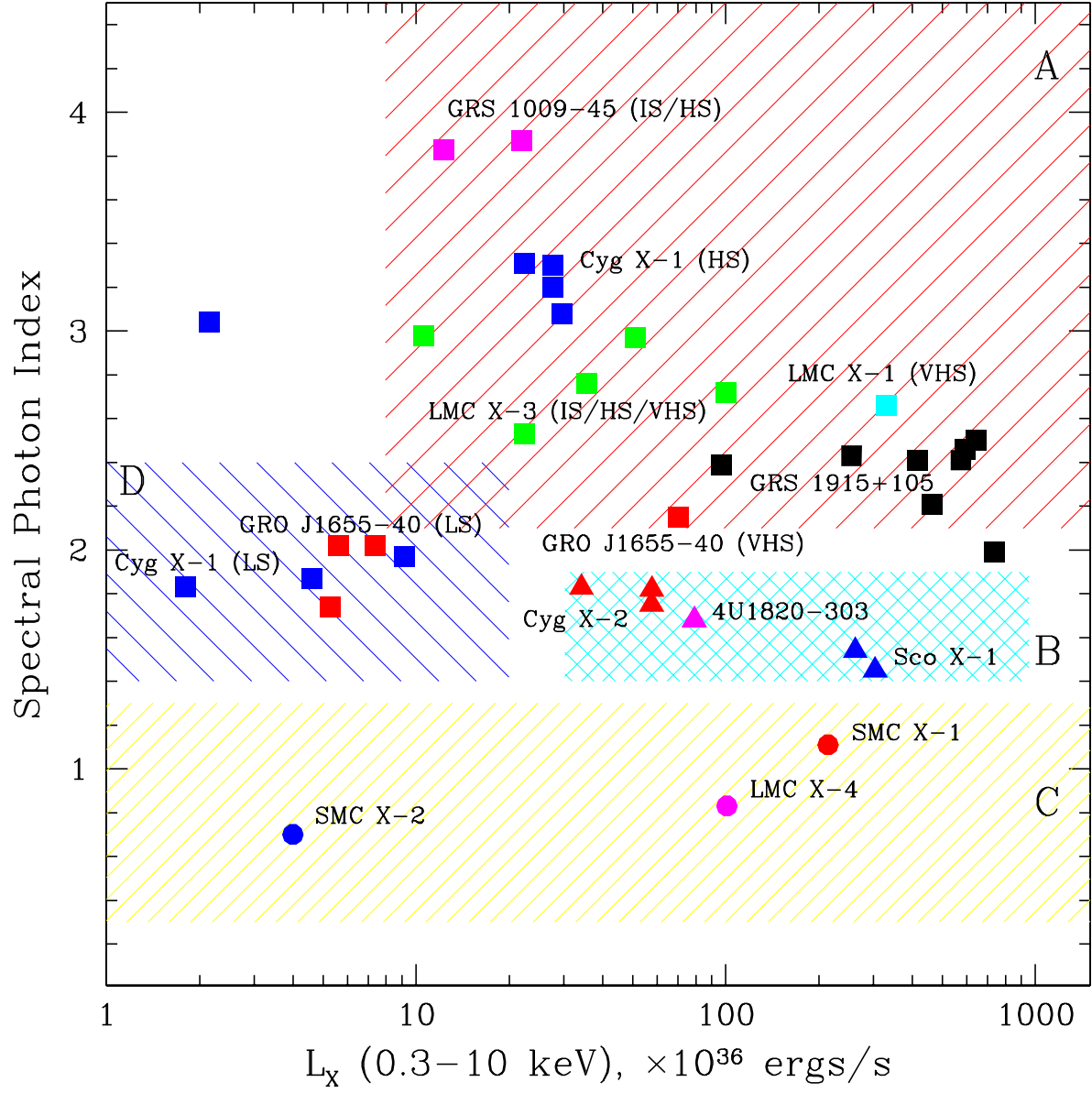}
\caption{Spectral hardness -- luminosity diagram (photon index vs. absorbed luminosity in the $0.3 - 10$ 
keV energy range) of X-ray sources based on simulated EPIC spectra with selected sources shown 
for comparison. The shaded regions identify four different classes of spectral states: black hole 
intermediate/high/very-high states (region A), high-luminosity states of neutron star X-ray binaries 
(region B), the region occupied by accretion-powered X-ray pulsars and high-inclination X-ray binaries 
(dippers, eclipsing systems and coronal sources)(region C) and the mix of neutron star and black hole 
systems in the low luminosity states (region D). \label{hardness_luminosity_class}}
\end{figure}

The spectral photon index vs. luminosity diagram of X-ray sources with shaded regions roughly representing the established 
states of neutron star and black hole binaries is shown in Fig. \ref{hardness_luminosity_class}. The simulated data for 
several black hole candidates (Cyg X-1, GRS1915+105, GRO J1655-40, GRS 1009-45, LMC X-1, LMC X-3), neutron star 
binaries (Sco X-1, Cyg X-2 and globular cluster source X1820-303) and accretion-powered X-ray pulsars (SMC X-1, SMC X-2 
and LMC X-4) are shown for comparison. The first region (A) with source luminosities above $\sim 10^{37}$ ergs s$^{-1}$ 
and photon indices higher than 2.1 represents canonical intermediate high/soft (thermal-dominated) and very-high (steep 
power law) spectral states of stellar-mass black hole candidates. The second region (B) (L$_{\rm X} > 3 \times 10^{37}$ 
ergs s$^{-1}$ and $1.4 \lesssim \Gamma \lesssim 1.9$) corresponds to high-luminosity states of neutron star systems. The 
third region (C) in the bottom of the figure ($\Gamma \lesssim 1.4$) covers typical accretion-powered X-ray pulsars 
(Nagase 1989; Yokogawa et al. 2003) and some of the obscured X-ray binaries and high-inclination systems (dippers, 
eclipsing and coronal sources). The fourth region (D) in the diagram is occupied by a mix of black hole and neutron star 
binaries in low luminosity states. 

As is clearly seen from Figure \ref{hardness_luminosity_class}, a hardness-luminosity diagram based on the EPIC spectral 
data can be a very effective tool for distinguishing between luminous neutron star and stellar-mass black hole candidate 
systems. The spectral photon index vs. luminosity relation can be also used to identify high-inclination systems and 
accreting pulsar candidates. On the other hand, it is still impossible to make a distinction between neutron star and 
black hole systems in the low luminosity states based on this diagram alone. 

The spectral photon index vs. luminosity plot of M31 X-ray sources in our sample is shown in Figure 
\ref{hardness_luminosity_class_M31}. The {\em XMM} sources that are unresolved composites of two or more 
{\em Chandra} sources have been excluded from the plot. The hardness-luminosity distribution of M31 globular 
cluster sources from Trudolyubov \& Priedhorsky (2004) is also shown for comparison. Approximately half of 
the sources in our sample have spectra and luminosities typical for both neutron star and black hole binaries 
in the low luminosity states (region D). A number of sources in the intermediate luminosity range 
$10^{37}<{\rm L}_{\rm X}<3\times 10^{37}$ ergs s$^{-1}$ with photon indices between 1.8 and 2.2 can be either 
neutron stars or black hole systems in the transition between low and high spectral states. All sources with 
soft spectra $\Gamma > 2.4$ and luminosities above $2\times 10^{37}$ ergs s$^{-1}$ have properties consistent 
with high/very high states of stellar-mass black hole candidates. Fourteen sources with luminosities above 
$3\times 10^{37}$ ergs s$^{-1}$ and photon indices between 1.4 and 1.9 fall into luminous neutron star region 
(region B). The remaining group of 6 hard sources with $\Gamma \lesssim 1.4$ have spectra and luminosities 
characteristic of X-ray pulsars and high-inclination systems (region C) with two sources (\#102 and 108) showing 
regular dips in their lightcurves \cite{dipper02,Mangano04}. 

It should be noted that energy spectra of the majority of background AGN resemble the spectra of neutron star and 
black hole systems in both low and high luminosity states. Therefore, some sources in our sample that are background 
objects can be misidentified as X-ray binary candidates in M31. We expect $\sim 7 - 10$ sources in our sample with 
apparent luminosities above $5 \times 10^{36}$ ergs s$^{-1}$ to be background objects. The expected number of 
background objects drops below 1 for source apparent luminosities above $3\times 10^{37}$ ergs s$^{-1}$. Another 
type of source that can be mistaken for X-ray binary systems are Crab-like supernova remnants, which have spectral 
properties similar to that of the neutron star and black hole binaries in the low luminosity states. Since typical 
Crab-like SNR are often associated with emission line regions and do not show significant variability in X-rays, 
additional information on the X-ray variability and/or optical counterpart can be used to distinguish between them 
and X-ray binaries. A hypothetical accreting intermediate-mass black hole with mass above $100 M_{\odot}$ 
in the hard spectral state could fall into the region in the hardness-luminosity diagram usually occupied by bright 
neutron star binaries with luminosities above $3\times 10^{37}$ ergs s$^{-1}$.  
 
We now briefly describe the properties of the X-ray binary candidates in our sample identified using hardness-luminosity 
diagram and detailed spectral and timing analysis.

\subsection{X-ray Binary Candidates}
{\em i) Stellar-Mass Black Hole Candidates.}
Based on the hardness-luminosity diagram, 7 sources in our sample can be classified as stellar-mass black hole 
candidates in the intermediate/high/very-high spectral state: sources \#3, 20, 25, 57, 99, 110, 123. The energy 
spectra of all these sources are soft with $\Gamma > 2.4$ and are best described by multicolor disk model or a 
two-component models including disk and power law tail (Tables \ref{spec_par_CUTOFF},\ref{spec_par_TR}), typical 
for canonical black hole systems (McClintock \& Remillard 2006 and references therein). Four of these sources are 
transients and has been previously classified as black hole candidates (Trudolyubov, Borozdin \& Priedhorsky 2001; 
Williams et al. 2006 and references therein; Trudolyubov, Priedhorsky \& Cordova 2006), with X-ray source \# 99 
showing the whole range of spectral evolution of bright black hole X-ray nova. Three black hole candidates 
(\# 20, 25, 123) are persistent sources showing significant variability on a time scales of years. There is a clear 
correlation between the spectral temperature derived in the DISKBB spectral fits to the spectra of these sources and 
their X-ray flux with temperature increasing with increase of the flux (Table \ref{spec_par_CUTOFF}). At the same 
time, the characteristic emitting radius, $R_{in} \sqrt{cos~i}$ remains essentially constant, despite a significant 
change in the X-ray flux, which is the effect also observed in the Galactic black hole candidates in the high spectral 
state \cite{TL95}. The X-ray source \#46 can be also tentatively classified as a black hole candidate, based on its 
transient behavior, two-component spectrum and luminosity (Trudolyubov, Borozdin \& Priedhorsky 2001). That brings 
the total number of the proposed black hole candidates to 8, or $\sim$7\% of the total number of sources in our sample. 
By analogy with Galactic X-ray binaries, we expect most of M31 black hole candidates to be transient systems. Therefore, 
the fraction of these objects is likely to become higher with more new transients detected in the future monitoring 
observations of the central part of M31.

{\em ii) Neutron Star Candidates.} 14 sources in our sample (sources \#26, 31, 37, 38, 40, 49, 58, 77, 80, 84, 86, 92, 
105, 121)(Fig. \ref{hardness_luminosity_class_M31}) have spectra, luminosities and variability similar to that of the 
bright (L$_{\rm X} > 3\times 10^{37}$ ergs s$^{-1}$) Galactic neutron star binaries. Seven of them coincide with M31 
globular clusters (Trudolyubov \& Priedhorsky 2004) and one source, \#49 shows spectral variability characteristic of 
Galactic Z-sources (Barnard et al. 2003b). The remaining 13 sources associated with globular clusters can be also added 
to the neutron star candidates list, based on the similarities with their Galactic counterparts. The X-ray source \# 29 
has been previously classified as a neutron star candidate, based on a short burst detected in one of {\em XMM-Newton} 
observations, interpreted as radius-expansion Type I X-ray burst (Pietsch \& Haberl 2005). Another X-ray source \#102 
(Mangano et al. 2004), which shows regular dipping with a period of 1.78 hours, has overall properties similar to the 
Galactic dipping sources with neutron star primaries. Therefore, 29 sources or $\sim24\%$ of the total number of sources 
in our sample can be identified as probable neutron star X-ray binary candidates. 

{\em iii) White Dwarf Candidates.}
Four sources in our sample, \# 7, 83, 113, 114 have supersoft spectra (not shown in hardness-luminosity diagram). Two 
sources are transients, and one source (\# 113) shows recurrent outbursts. The properties of all four sources are similar 
to the Galactic supersoft sources, identified with white dwarfs, sustaining a thermonuclear burning of the accreted matter 
on their surface (Kahabka \& van den Heuvel 2006). We can therefore propose these four sources as probable accreting white 
dwarf candidates.

The remaining 2 transient sources (\#100 and 104) have observed properties consistent with both black hole or neutron 
star X-ray binary interpretation. Finally, adding these two objects to the X-ray binary candidates identified above, 
brings the total number of high-confidence X-ray binary candidates with black hole, neutron star and white dwarf 
primaries to 44 or $\sim35\%$ of sources in our sample.

\section{Comparison with X-ray Binary Source Populations in Globular Clusters and Normal Galaxies}
The similarities between M31 and Milky Way GC X-ray sources suggest that M31 GC source population is likely to be 
dominated by LMXB with neutron star primaries, and in some sense can be regarded as a prototype LMXB population. 
Therefore, the results from the study of M31 GC sources provide an important benchmark for comparison with the results 
from study of galactic X-ray source populations of mixed nature, helping to constrain the neutron star X-ray binary 
contribution. Similarities between our {\em XMM-Newton} survey of the central region of M31 and a previous 
{\em XMM-Newton}/{\em Chandra} survey of its GC system \cite{TP04}(overlapping observations with identical observational 
setup and data analysis techniques etc.) make comparison of their results straightforward. 

\begin{figure}
\epsfxsize=18cm
\epsffile{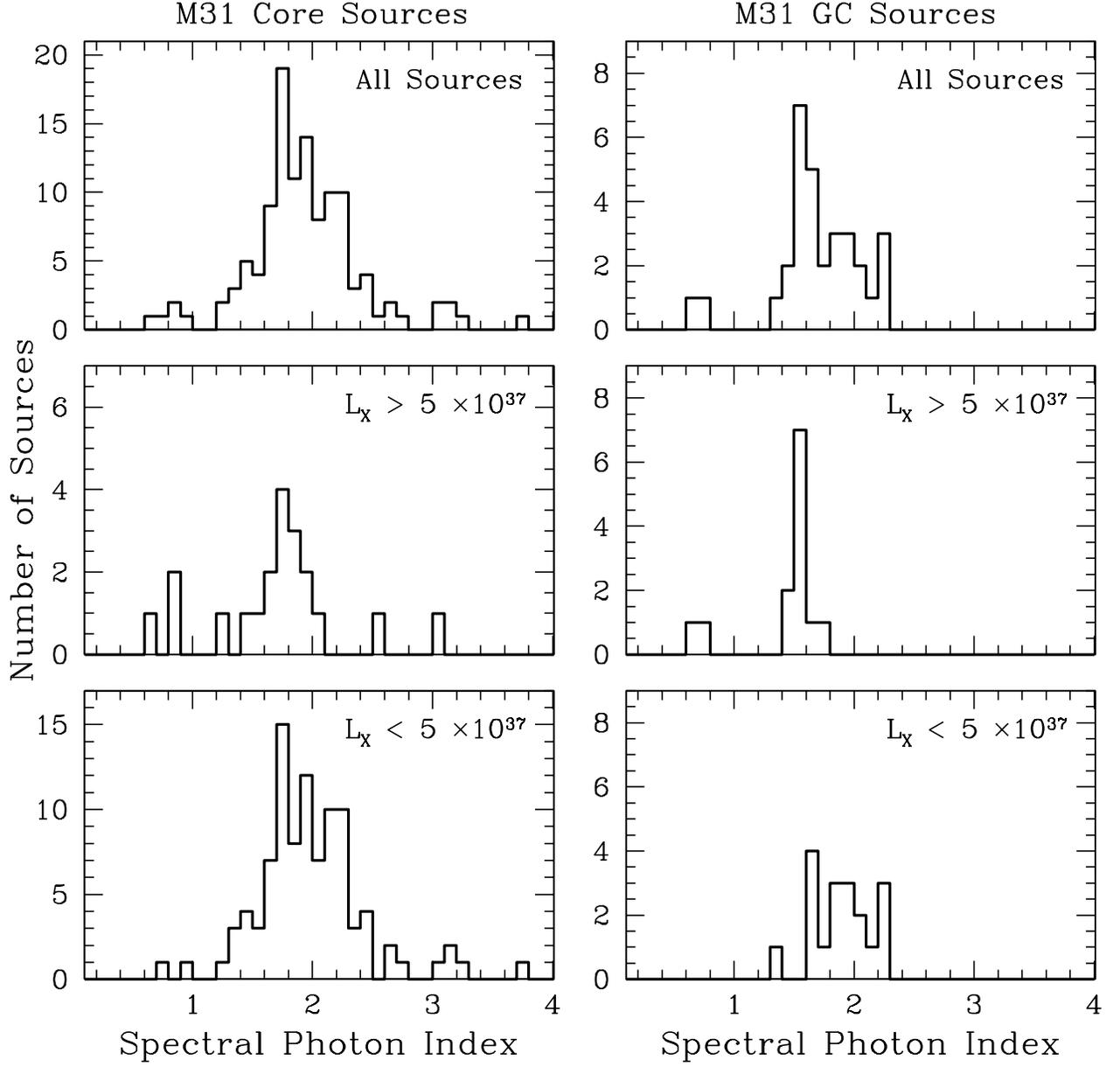}
\caption{A comparison of power law spectral index distributions of the bright X-ray sources in the central 
part of M31 (left panels) and M31 globular cluster X-ray sources from Trudolyubov \& Priedhorsky (2004)(right 
panels). \label{core_GCS_hardness_compare}}
\end{figure}

The power law spectral index distributions and their dependence on the source luminosity in our source sample and 
in the M31 GC source sample are shown in Fig. \ref{core_GCS_hardness_compare}. Although the total distributions 
(upper panels in Fig \ref{core_GCS_hardness_compare}) appear to be qualitatively similar (both distributions have 
well-defined maxima corresponding to harder spectra and a shoulder in the softer spectral region), there are 
significant quantitative differences between them: the maximum of the GC distribution corresponds to harder spectra 
than that of the central region of M31, and it lacks soft sources with $\Gamma > 2.3$, while central region of M31 
contains a significant number (17) of such sources. According to the results of K-S test, the hypothesis that these 
two distributions are drawn from the same parent distribution can be rejected at 99.6\% confidence level. The 
hardness distributions of the sources in the central region of M31 and M31 GC sources differ at both high and low 
source luminosities. The main peak of the spectral distribution of the brightest GC sources with 
L$_{\rm X}>5\times 10^{37}$ ergs s$^{-1}$ (middle panels in Fig. \ref{core_GCS_hardness_compare}) appears to be 
narrower and shifted towards harder spectra than that of the central region of M31. For lower luminosities (lower 
panels in Fig. \ref{core_GCS_hardness_compare}), the hardness distribution of sources in our sample tends to be more 
peaked in the hard spectral region and extends further into soft spectral region, when compared to the GC source 
distribution.    

\begin{figure}
\epsfxsize=18cm
\epsffile{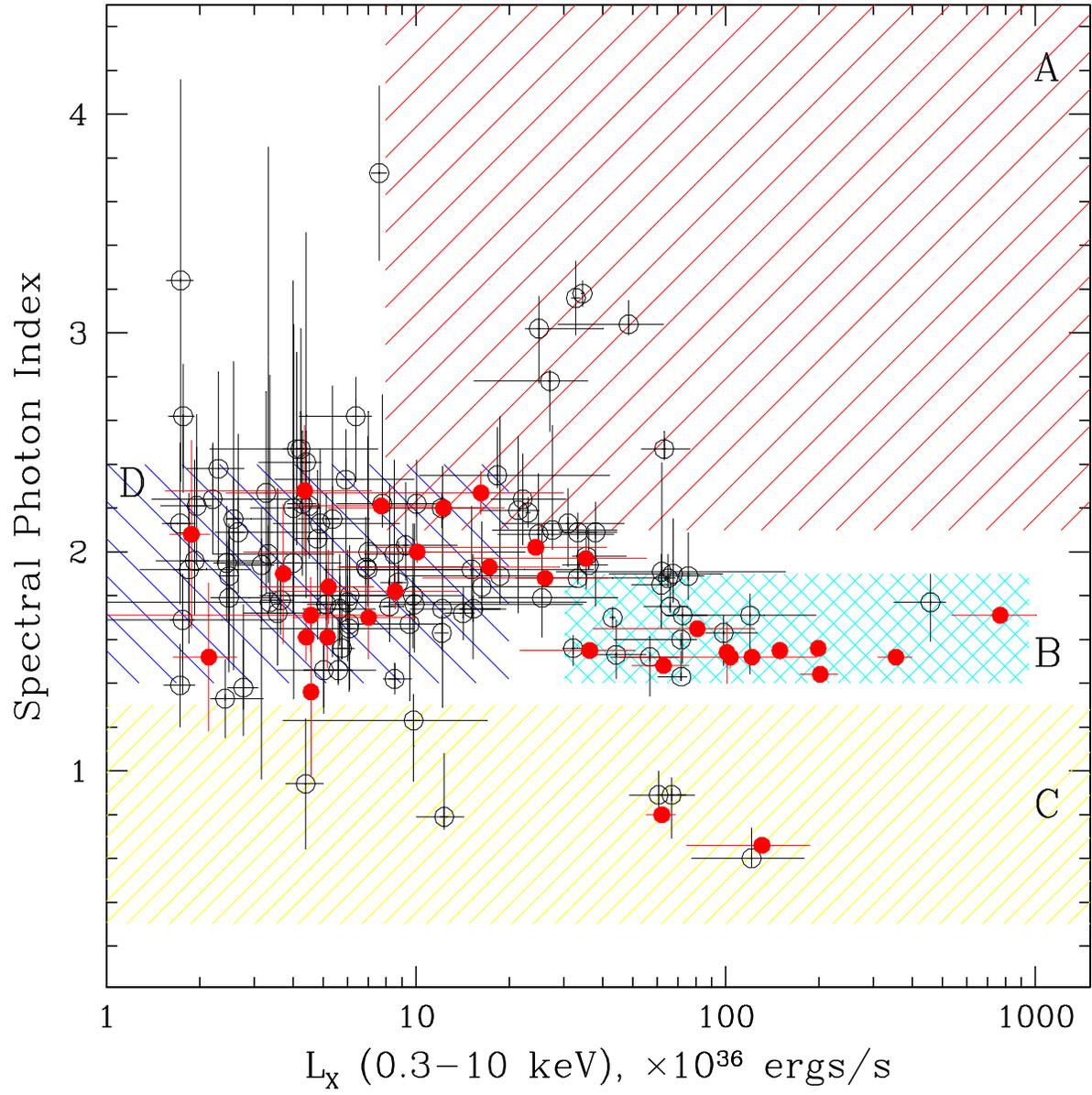}
\caption{Spectral hardness -- luminosity diagram (photon index vs. absorbed luminosity in the $0.3 - 10$ 
keV energy range) of M31 X-ray sources in our sample with M31 globular cluster candidates shown with red 
points for comparison. The definitions of shaded regions are the same as in Figure \ref{hardness_luminosity_class}. 
\label{hardness_luminosity_class_M31}}
\end{figure}

The spectral photon index-luminosity diagram (Fig. \ref{hardness_luminosity_class_M31}) is another useful tool for an 
effective comparison between the sources in the central part of M31 and M31 GC sources. The hardness-intensity data 
for M31 GC sources fits into the regions of the diagram, covering spectral states of neutron star X-ray binaries. This 
is consistent with similarities between M31 GC sources and their Galactic counterparts, all of which are confirmed 
neutron star systems. The X-ray sources in the central region of M31 follow essentially the same pattern in the 
hardness-luminosity diagram, except for the prominent group of spectrally soft sources with luminosities above 
$2\times 10^{37}$ ergs s$^{-1}$, most of them classified as black hole candidates. Therefore, it is natural to assume 
that majority of bright X-ray sources in the bulge and inner disk of M31 represents a mix of X-ray binaries with 
neutron star and black hole primaries, dominated by neutron star LMXBs, similar to our Milky Way galaxy \cite{Liu01}. 
The luminosity dependence of the spectral distribution of M31 sources 
(Fig. \ref{hardness_distr_lum},\ref{core_GCS_hardness_compare}) can be generally understood as a result of mixing a
neutron star and black hole LMXB populations. For high luminosities (L$_{\rm X} > (3 - 5) \times 10^{37}$ ergs s$^{-1}$), 
the spectra of neutron star systems in the $0.3-10$ keV energy band tend to be hard with typical photon indices of 
$\sim 1.4 - 1.8$, while most of the stellar-mass black hole systems are found in the high/very-high state with softer 
spectra ($\Gamma \gtrsim 2.1$). The resulting hardness distribution with the majority of sources being neutron star systems 
should have the main peak in the hard spectral region ($\Gamma \sim 1.4 - 1.8$) and a shoulder extending to the soft spectral 
region ($\Gamma \gtrsim 2$). The relative normalization of the main peak and the shoulder depends on the ratio of the 
number of neutron star and black hole systems. The overall form and centroid position of the main peak will be determined 
mainly by luminosity distribution of the neutron star systems in that source sample. For luminosities between $10^{37}$ 
and $5\times10^{37}$ ergs s$^{-1}$, the spectra of neutron star systems have wider range of photon index 
($\Gamma \sim 1.6 - 2.4$), and most of the stellar-mass black hole systems are still in the intermediate/high/very-high 
state. The luminosities below $10^{37}$ ergs s$^{-1}$ correspond to the low states of both neutron star and black hole 
candidates with typical range of spectral photon indices $\Gamma = 1.4\sim2.4$. The resulting spectral hardness 
distributions at intermediate and low luminosities should be broader and softer than that of the high luminosity group. 
As seen from Fig. \ref{hardness_distr_lum}, \ref{hardness_luminosity_class_M31}, \ref{core_GCS_hardness_compare}, the 
hardness-luminosity data for both M31 core and GC sources seem to support this picture.      

More than 80\% of the sources in our sample show significant variability on the time scales of months to years, which 
is very close to the fraction of variable objects among the M31 GC X-ray sources found in the {\em XMM-Newton}/{\em Chandra} 
survey of comparable sensitivity \cite{TP04}. The similarity of the fraction of variable sources may suggest that in 
general, there is no significant difference between the long-term variability of the field LMXB and their counterparts 
in globular clusters.

The luminosity distribution of the bright X-ray sources in the central region of M31 deviates from that of the M31 GC 
sources \cite{TP04}. The fraction of the bright sources with luminosities above $5\times10^{37}$ ergs s$^{-1}$ is much 
higher for the GC population ($\sim$30\%) than for the bulge ($\sim16-18\%$, depending on the background source contribution). 
Moreover, the difference in the fraction of the sources with luminosities above $10^{38}$ ergs s$^{-1}$ is even higher, 
with 6 out of 43 GC sources steadily exceeding and 3 three more sources occasionally exceeding this limit ($\sim14-21\%$ 
of the total number) and only 3 sources in the bulge and inner disk of M31 being persistently and 2 sources occasionally 
brighter than $10^{38}$ ergs s$^{-1}$ ($\sim3-6\%$ of the total number). A total of 20 sources in our sample are associated 
with globular clusters (Table \ref{source_ID_1}). GC sources make a significant contribution to the bright source counts in 
the central region of M31: on average, $\sim20-25\%$ of the sources with apparent luminosities above $10^{37}$ ergs s$^{-1}$ 
reside in the globular clusters, with GC source contribution rising to $\sim33-50\%$ at luminosities above $10^{38}$ ergs 
s$^{-1}$. The fraction of bright GC sources in the central region of M31 appears to be much higher than observed in our own 
Galaxy, and stands closer to the fraction of GC sources found in early-type galaxies (Angelini, Lowenstein \& Mushotzky 
2001; Kundu, Maccarone \& Zepf 2002; Fabbiano 2006 and references therein). 

The central region of M31 covers most of the bulge and inner disk, containing a significant fraction of X-ray sources detected 
in M31, and is suitable for direct comparison with LMXB populations in other galaxies. In the recent study of a large sample of 
nearby early-type galaxies Irwin, Athey \& Bregman (2003) performed a spectroscopic survey of their X-ray binary populations. 
They found that the spectrum of the sum of the sources with luminosities below $10^{39}$ ergs s$^{-1}$ in a given galaxy is 
similar from galaxy to galaxy, and can be approximated with simple power law models with photon index $\Gamma=1.4\sim1.7$. 
The combined spectrum of LMXB in the sample of Irwin et al. (2003) can be fit with a simple power law model with photon index 
of $1.56\pm0.02$. Since the low luminosity limit of their sample for most galaxies is well above $3\times10^{37}$ ergs s$^{-1}$ 
and sometimes exceeds $10^{38}$ ergs s$^{-1}$, it is suitable for the comparison with high-luminosity subsample of M31 sources. 
The formal weighted average photon index for the high-luminosity M31 sources $<\Gamma>_{\rm bright}=1.54\pm0.01$ is consistent 
with photon index derived for the combined bright X-ray sources in early-type galaxies. The similarity between these two is not 
surprising since both populations are probably dominated by bright persistent LMXB with neutron star primaries. The weighted 
average spectral photon index of 115 sources in our sample $<\Gamma>=1.67\pm0.01$ also appears to be in good agreement with 
the photon index ($1.63\pm0.04$) of the composite spectrum of M31 bulge sources measured with {\em Chandra} (Irwin, Athey \& 
Bregman 2003). 

\begin{figure}
\epsfxsize=18cm
\epsffile{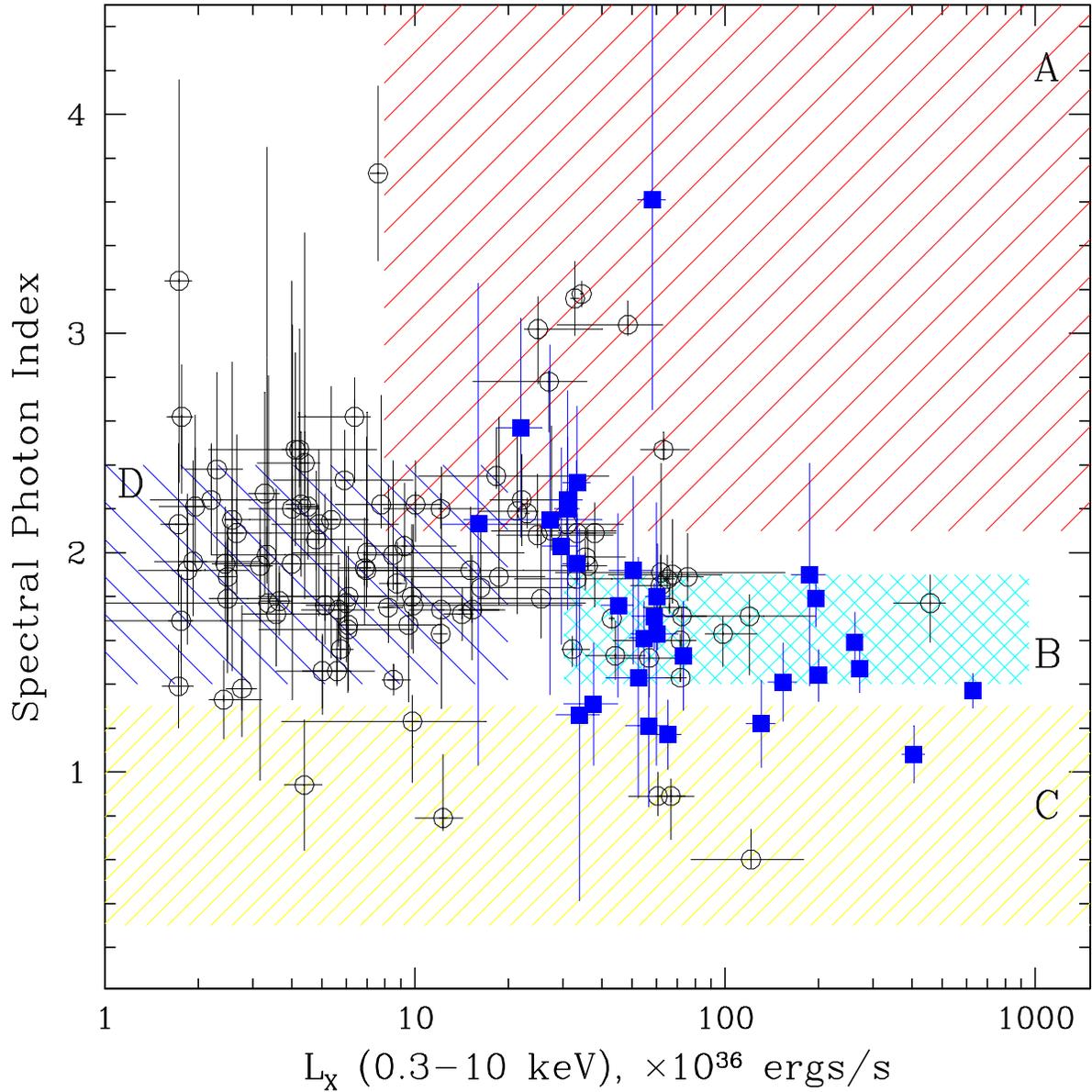}
\caption{A comparison of spectral hardness -- luminosity distributions (photon index vs. absorbed luminosity 
in the $0.3 - 10$ keV energy range) of X-ray sources in the central part of M31 (open circles) and bright 
X-ray sources detected in the {\em Chandra} survey of M81 \cite{Swartz03}(filled squares). The absorbed 
luminosities of M81 sources in the 0.3 -- 10 keV energy band were calculated using the original unabsorbed 
0.3 -- 8 keV luminosities and spectral indices from Swartz et al. (2003), assuming an absorbing column of 
$1.2\times10^{21}$ cm$^{-2}$. The definitions of shaded regions are the same as in Fig. 
\ref{hardness_luminosity_class}.\label{hardness_luminosity_M31_M81}}
\end{figure}

The study of detailed spectral properties and variability of individual sources in most large nearby galaxies is usually confined 
to a few brightest objects. However, recent deep {\em Chandra} and {\em XMM-Newton} observations of some of them allow us to 
obtain spectra of larger number of sources with apparent luminosities above $few\times10^{37}$ ergs s$^{-1}$. The observations of 
the spiral galaxy M81 with {\em Chandra} allowed for a study of the spectra of 30 X-ray sources with estimated unabsorbed luminosities 
above $2\times10^{37}$ ergs s$^{-1}$ (Swartz et al. 2003). In order to make a direct comparison of the spectral distributions of 
sources in M81 and M31, we used spectral fit parameters of bright M81 sources from Table 3 of Swartz et al. (2003) (spectral indices 
and unabsorbed luminosities in the 0.3--8 keV energy range) to calculate their absorbed luminosities in the 0.3--10 keV energy band, 
assuming an absorbing column of $1.2\times10^{21}$ cm$^{-2}$, typical for M31 sources. The resulting hardness-luminosity distribution 
(spectral photon index vs. luminosity diagram) of bright M81 X-ray sources (Fig. \ref{hardness_luminosity_M31_M81}) is similar to the 
distribution of the brighter sources in the center of M31 galaxy. The spectra and luminosities of the majority of M81 sources are 
close to those of the neutron star systems in the high luminosity states and during transitions between high and low luminosities. 
In addition, the properties of two M81 sources with steeper spectra ($\Gamma > 2.5$) can be consistent with black hole candidate 
interpretation. The similarity between spectral properties and luminosities of M31 and M81 X-ray sources provides another evidence 
for common spectral formation mechanism in X-ray binaries. It also demonstrates that hardness-luminosity diagrams can be an effective 
tool for characterising X-ray binary populations and classification of individual sources in external galaxies.     
        
\section{Summary and Conclusions}
We performed a detailed survey of spectral properties and variability of 123 bright X-ray sources detected in the central part 
of M31 (within $15\arcmin$ radius from the galactic nucleus) using the data of four years of observations with {\em XMM-Newton}. 
The observed luminosities of the sources, scaled to the M31 distance of 760 kpc, range from $\sim10^{36}$ to $\sim5\times10^{38}$ 
ergs s$^{-1}$ with luminosity of $\sim5\times10^{36}$ ergs s$^{-1}$ corresponding to a completeness limit of our sample. Based on 
the background source statistics, we expect a relatively small fraction ($\sim8-12\%$) of sources with luminosities above the 
completeness limit to be background objects. 20 sources in our sample are associated with optically-identified globular clusters, 
and 4 sources coincide with supernova remnant candidates in M31. The X-ray properties and radio/optical counterparts of 6 
sources have been found to be consistent with that of the background AGN. 
 
A large number of sources with a wide range of apparent luminosities allowed us to study a rich spectrum of states of different 
classes of X-ray emitting objects. The majority of sources in our sample have X-ray properties reminiscent of Galactic low-mass 
X-ray binaries, consistent with the older stellar population of the bulge. We fitted the energy spectra of individual sources with 
a variety of spectral models ranging from simple power law and blackbody radiation models to two-component models including soft 
and hard spectral components. In most cases, we obtained acceptable fits using an absorbed power law spectral model, but for many 
brighter sources a more complex spectral model is required. For these sources, a models with quasi-exponential cut-off at 
$\sim 0.6-8$ keV or two-component models describe the energy spectra significantly better than simple power law. Four sources 
have supersoft spectra that can be approximated with blackbody radiation models with temperatures of $\sim 50-60$ eV. Another four 
sources in our sample are SNR candidates with soft thermal spectra showing presence of spectral lines, that can be approximated 
by thermal plasma models with characteristic temperatures of $0.2\sim0.5$ keV. 

Using the results of spectral analysis, we studied a distribution of X-ray absorption towards M31 sources. The distribution of 
absorbing columns peaks at N$_{\rm H}\sim1.2\times10^{21}$ cm$^{-2}$, and extends up to $1.3\times10^{22}$ cm$^{-2}$. The 
average value of the absorbing column derived from our sample is $<{\rm N}>_{\rm H}=(1.52\pm0.02)\times10^{21}$ cm$^{-2}$, or 
nearly twice the Galactic foreground value. 

The spectral photon index distribution of the sources in the central part of M31 has a main peak at $\Gamma\sim1.75$ and a broad 
shoulder at $\Gamma\sim2.0-2.2$ extending to the softer spectral region. The main part of the photon index distribution can be 
adequately described by sum of two Gaussian functions with centroids at $\Gamma_{1}=1.78$ and $\Gamma_{2}=1.95$, accounting for 
$\sim 18\%$ and $\sim 82\%$ of the total number of sources. The spectral distribution shows clear evolution with source luminosity, 
characterized by narrowing and shift of its main peak to the harder spectral region for luminosities above $\sim(3-5)\times10^{37}$ 
ergs s$^{-1}$. The spectral properties of the brighter sources in our sample clearly resemble those of the bright LMXB observed in 
other nearby normal galaxies of different morphological type. The spectral distribution of the sources in the central region of 
M31 shows an apparent excess of soft sources when compared to the distribution of M31 globular cluster sources. This fact can be 
probably explained by the presence of black hole binaries in the intermediate/high/very-high state among the field sources in the 
central region of M31, and their absence among the globular cluster sources.

We found that more than 80\% of sources in our sample with two or more flux measurements available show significant variability 
on a time scales of days to years. Ten sources are transients with outburst-to-quiescent flux ratios of $\sim100-1000$, and 
three other sources show recurrent outbursts on a time scale of years. The fraction of variable sources in our survey is much 
higher than previously reported from {\em Chandra} surveys of the central region of M31. At the same time, it is very close to 
the fraction of variable sources among GC sources in M31. 

Timing analysis of {\em XMM-Newton}/EPIC data allowed us to study short-term aperiodic variability of M31 sources in the 
$2\times10^{-5}-0.1$ Hz frequency range. Because of the lack of statistics, for most sources in our sample it was not possible 
to detect any significant aperiodic variability. However, several of the brightest sources showed characteristic very-low 
frequency variability at frequencies below 0.01 Hz, resembling the noise variability seen in the bright Galactic LMXB.  

About 50\% of the sources in our sample show spectral variability between individual {\em XMM-Newton} observations. The 
spectral evolution of a number of sources is correlated with level of X-ray flux: the spectrum becomes harder as the flux 
increases. Several bright sources demonstrate complex patterns of evolution on the hardness-intensity diagram, somewhat 
reminiscent of the Galactic Z and atoll sources.     
   
The main features of the hardness-luminosity diagram of M31 sources can be interpreted as resulting from a mixed population 
dominated by neutron star and black hole X-ray binaries. Based on observational data, one can identify four main regions in 
the hardness-luminosity diagram of M31 sources, with properties corresponding to distinct spectral states of X-ray binaries. 
Combining the position in the hardness-luminosity diagram with detailed spectral and variability information, we classify 
$\sim7\%$ of sources in our sample as probable black hole candidates in the intermediate/high/very-high spectral state and 
$\sim24\%$ sources as neutron star binary systems. The properties of four sources with very soft thermal spectra are close to 
that of the Galactic supersoft sources, and can be classified as probable white dwarf candidates. In summary, a total of 44 
X-ray sources in our sample can be identified as probable X-ray binary candidates, based on their spectral properties and 
variability. 

The central region of M31 is one of the best places to study X-ray binary populations, and provides an important benchmark for 
comparison with our own Galaxy and other external galaxies. The observed similarities between the properties of bright X-ray 
sources in the central region of M31, its GC component, and other normal galaxies of different morphological type, help to 
identify universal properties of extragalactic LMXB populations. A complete X-ray survey of M31 and new high-sensitivity 
observations of nearby galaxies should allow a more detailed comparison of their X-ray emitting populations, putting better 
constraints on the fraction of black hole and neutron star binaries and ultimately help to improve our understanding of 
fundamental processes of binary evolution and star formation history.  

\section{Acknowledgments}
Support for this work was provided through NASA Grant NAG5-12390. Part of this work was initiated during a 2005 summer 
workshop ``Revealing Black Holes'' at the Aspen Center for Physics, S. T. is grateful to the Center for their hospitality. 
XMM-Newton is an ESA Science Mission with instruments and contributions directly funded by ESA Member states and the USA 
(NASA). This research has made use of data obtained through the High Energy Astrophysics Science Archive Research Center 
Online Service, provided by the NASA/Goddard Space Flight Center.

\clearpage

\begin{table}
\small
\caption{{\em XMM-Newton} observations of the central region of M31 used in the analysis. 
\label{obslog}}
\small
\begin{tabular}{ccccccc}
\hline
\hline
Obs. $\#$ &Date, UT & Obs. ID  & RA (J2000)$^{a}$ & Dec (J2000)$^{a}$ & Exp.(mos)$^{b}$&Exp.(pn)$^{b}$\\
& &&  (h:m:s)   &(d:m:s)&(ks)&(ks)\\             
\hline
$\#1$&2000 Jun 25    &0112570401&00:42:43.00&41:15:46.1&28.9&24.9\\
$\#2$&2000 Dec 28    &0112570601&00:42:43.00&41:15:46.1&12.1& 9.4\\
$\#3$&2001 Jun 29    &0109270101&00:42:43.00&41:15:46.1&29.0&24.9\\
$\#4$&2002 Jan 06    &0112570101&00:42:43.00&41:15:46.1&63.0&49.9\\
$\#5$&2004 Jul 16    &0202230201&00:42:42.12&41:16:57.1&19.6&16.4\\
$\#6$&2004 Jul 18    &0202230401&00:42:42.26&41:16:58.2&17.6&13.1\\
$\#7$&2004 Jul 19    &0202230501&00:42:42.23&41:16:57.6& 9.6& 6.5\\
\hline
\end{tabular}
\begin{list}{}{}
\item[$^{a}$] -- coordinates of the center of the field of view
\item[$^{b}$] -- instrument exposure used in the analysis
\end{list}
\end{table}

\clearpage

\begin{table}
\small
\caption{List of bright X-ray sources detected in the central part of M31. \label{source_ID_1}}
\begin{tabular}{cccccccl}
\hline
\hline
Source & R.A. & Decl.   & Photon                    & Flux$^{b}$ & $L_{\rm X}^{c}$ & Class$^{d}$ & Optical/Radio/X-ray \\
 ID    &      &         & Index$^{a}$               &            &                 &             &     ID$^{e,f}$      \\
\hline
\hline
  1 & 00 41 41.32&41 19 17.7& $2.15\pm0.15$ (1.52-2.76) & $0.60 - 1.02$  &    $42 - 71$  & SV          & P165\\
  2 & 00 41 43.48&41 21 20.9& $1.23\pm0.12$ (0.95-1.28) & $0.54 - 2.45$  &   $37 - 169$  &             & P169\\
  3 & 00 41 44.70&41 11 10.0& $3.16\pm0.17$ (2.74-3.36) & $3.84 - 4.71$  &   $<1 - 326$  & TR,BHC,SV   & \\
  4 & 00 41 50.27&41 13 36.8& $1.89\pm0.21$ (1.71-2.04) & $0.36\pm0.03$  &               & R,BKG       & B90 34,r3-110,P176\\
  5 & 00 41 50.51&41 21 15.5& $1.94\pm0.26$ (0.96-2.06) & $0.46\pm0.05$  &         $32$  &             & r3-109,P178\\
  6 & 00 41 51.59&41 14 39.0& $1.77\pm0.12$ (1.50-2.01) & $0.62 - 1.00$  &               & R,BKG       & B90 35,5C3.099,r3-81,P181\\
  7 & 00 41 54.12&41 07 23.9&                           & $<0.014 - 0.54$&  $<1-37$      & TR,SSS      & P191\\
  8 & 00 42 03.84&41 09 27.6& $0.94\pm0.30$             & $0.64\pm0.09$  &         $44$  &             & P200\\
  9 & 00 42 05.72&41 13 30.3& $1.72\pm0.12$ (1.70-1.79) & $1.79 - 2.24$  &  $123 - 155$  & REC         & PFJ3,P202\\
 10 & 00 42 07.52&41 10 28.0& $1.95\pm0.26$ (1.75-2.04) & $0.35\pm0.03$  &         $25$  &             & r3-94,P211\\
 11 & 00 42 07.73&41 18 15.2& $2.09\pm0.03$ (1.75-2.23) & $4.86 - 6.46$  &  $335 - 446$  &             & TF12,PFJ4,r3-61,P212\\
 12 & 00 42 07.77&41 04 37.8& $1.92\pm0.16$ (1.75-2.53) & $0.73 - 2.05$  &   $50 - 141$  &             & P213\\
 13 & 00 42 09.08&41 20 48.6& $1.80\pm0.08$ (1.56-2.13) & $0.53 - 2.07$  &   $37 - 143$  & SV          & r3-60,P218\\
 14 & 00 42 09.48&41 17 45.4& $2.22\pm0.08$ (2.11-2.72) & $0.80 - 1.67$  &   $55 - 115$  & GCS,SV      & MIT140,PFJ5,r3-59,TP8,P221\\
 15 & 00 42 10.28&41 15 10.0& $2.20\pm0.15$ (1.33-3.04) & $0.42 - 0.71$  &    $29 - 49$  &             & PFJ6,r3-58,P223\\
 16 & 00 42 10.98&41 12 48.3& $2.62\pm0.24$ (2.27-2.63) & $0.26\pm0.02$  &         $18$  & SV          & r3-57,P226\\
 17 & 00 42 11.66&41 10 49.3& $1.75\pm0.13$ (1.59-2.31) & $1.10 - 1.32$  &               & R,BKG       & B90 56,PFJ7,r3-56,P227\\
 18 & 00 42 11.93&41 16 49.5& $1.95\pm0.15$ (1.72-3.24) & $0.39 - 0.83$  &    $27 - 57$  &             & r3-55,P228\\
 19 & 00 42 12.16&41 17 58.7& $2.00\pm0.07$ (1.84-2.64) & $0.58 - 3.25$  &   $40 - 224$  & GCS,SV      & Bo78,PFJ8,r3-54,TP9,P230\\
 20 & 00 42 13.12&41 18 36.8& $3.04\pm0.03$ (2.99-3.15) & $4.15 - 9.16$  &  $286 - 632$  & BHC,SV      & TF14,PFJ9,r3-52,P233\\
 21 & 00 42 15.11&41 12 34.8& $2.10\pm0.03$ (2.01-2.58) & $2.56 - 6.47$  &  $176 - 446$  & SV          & PFJ10,r3-50,P235\\
 22 & 00 42 15.26&41 18 01.3& $2.24\pm0.25$ (2.12-2.50) & $0.20 - 0.65$  &    $14 - 45$  & SV          & r3-49,P236\\
 23 & 00 42 15.45&41 20 31.9& $1.99\pm0.13$ (1.93-3.85) & $0.32 - 0.79$  &    $22 - 54$  & SV          & PFJ11,r3-48,P237\\
 24 & 00 42 15.67&41 17 21.3& $1.74\pm0.06$ (1.29-2.39) & $1.39 - 2.83$  &   $96 - 195$  & SV          & PFJ12,r3-47,P238\\
 25 & 00 42 18.33&41 12 24.0& $3.02\pm0.05$ (2.77-3.17) & $3.26 - 5.82$  &  $225 - 401$  & BHC,SV      & TF19,PFJ14,r3-45,P244\\
 26 & 00 42 18.62&41 14 02.1& $1.52\pm0.02$ (1.34-1.61) & $6.15 - 10.3$  &  $424 - 710$  & GCS,SV      & Bo86,TF20,PFJ15,r3-44,TP11,P246\\
 27 & 00 42 20.47&41 26 42.0& $2.21\pm0.24$ (2.13-2.34) & $0.65\pm0.05$  &               & R,BKG       & B90 64,PFJ16,K22,P251\\
 28 & 00 42 21.43&41 16 01.5& $2.09\pm0.03$ (1.92-2.18) & $4.18 - 5.46$  &  $289 - 377$  &             & TF21,PFJ17,r3-42,P252\\
 29 & 00 42 21.55&41 14 20.1& $2.47\pm0.17$ (2.25-3.02) & $0.31 - 1.09$  &    $22 - 76$  & SV,Burst    & r3-41,P253\\
 30 & 00 42 22.40&41 13 34.3& $1.89\pm0.03$ (1.76-2.62) & $1.34 - 3.81$  &   $93 - 263$  &             & TF22,PFJ18,r3-40,P255\\
 31 & 00 42 22.93&41 15 35.5& $1.71\pm0.01$ (1.44-1.81) & $10.1 - 28.5$  &  $699 - 1964$ & SV          & TF23,PFJ20,r3-39,P257\\
 32 & 00 42 22.95&41 07 39.5& $2.13\pm0.37$ (1.98-2.25) & $0.25\pm0.03$  &         $17$  &             & r3-73,P258\\
 33 & 00 42 23.12&41 14 07.8& $2.06\pm0.14$ (1.60-2.37) & $0.59 - 0.98$  &    $41 - 67$  &             & TF24,PFJ21,r3-38,P259\\
 34 & 00 42 25.12&41 13 40.9& $2.33\pm0.10$ (1.68-2.56) & $0.51 - 1.43$  &    $35 - 99$  & SV          & PFJ22,r2-45,P262\\
 35 & 00 42 26.04&41 19 14.8& $2.03\pm0.05$ (1.87-2.32) & $1.01 - 1.98$  &   $69 - 136$  & GCS,SV      & Bo96,TF26,PFJ23,r2-36,TP14,P263\\
 36 & 00 42 26.16&41 25 52.6& $1.74\pm0.06$ (1.51-1.99) & $2.00 - 2.70$  &  $138 - 186$  & SV          & PFJ24,r3-87,P264\\
 37 & 00 42 28.18&41 10 00.8& $1.56\pm0.03$ (1.48-1.62) & $4.30 - 5.32$  &  $297 - 367$  &             & r3-36,P269\\
 38 & 00 42 28.26&41 12 23.4& $1.71\pm0.01$ (1.49-1.75) & $9.24 - 12.6$  &  $637 - 872$  & SV          & TF27,PFJ25,r2-35,P270\\
 39 & 00 42 28.98&41 04 36.4& $2.35\pm0.05$ (2.29-2.57) & $1.48 - 6.13$  &  $102 - 423$  & SV          & TF28,PFJ26,r3-111,P271\\
 40 & 00 42 31.14&41 16 21.6& $1.75\pm0.02$ (1.72-1.84) & $7.87 - 11.0$  &  $543 - 757$  & SV          & TF32,PFJ27,r2-34,P275\\
 41 & 00 42 31.24&41 19 39.0& $1.92\pm0.03$ (1.70-2.21) & $1.05 - 5.44$  &   $73 - 375$  & GCS,SV      & Bo107,TF33,PFJ28,r2-33,TP16,P277\\
 42 & 00 42 32.06&41 13 14.6& $1.23\pm0.02$ (1.10-1.39) & $7.19 - 11.3$  &  $496 - 780$  & SV          & TF34,PFJ29,r2-32,P280\\
 43 & 00 42 33.03&41 03 28.5& $1.76\pm0.18$ (1.69-2.27) & $0.64 - 0.86$  &    $45 - 59$  & GCS,SV      & Bo110,PFJ30,TP17,P282\\
 44 & 00 42 33.87&41 16 19.8& $1.99\pm0.08$ (1.83-2.42) & $0.99 - 1.42$  &    $68 - 98$  &             & TF36,PFJ32,r2-30,P285\\
 45 & 00 42 34.11&41 21 50.2& $1.46\pm0.17$ (1.26-1.62) & $0.63 - 0.90$  &    $43 - 62$  &             & r3-35,P286\\
 46 & 00 42 34.41&41 18 09.7& $2.18\pm0.07$             & $3.33\pm0.06$  &        $230$  & TR,BHC      & r2-29,P287\\
 47 & 00 42 35.23&41 20 06.3& $1.76\pm0.08$ (1.54-2.13) & $1.18 - 1.67$  &   $81 - 115$  &             & PFJ34,r2-27,P290\\
 48 & 00 42 37.98&41 05 26.9& $1.79\pm0.34$ (1.51-2.06) & $0.31 - 0.52$  &    $22 - 36$  &             & r3-100,P296\\
 49 & 00 42 38.56&41 16 03.8& $1.77\pm0.01$ (1.59-1.90) & $50.5 - 74.6$  & $3481 - 5144$ & Z,SV        & TF41,PFJ35,r2-26,P297\\
 50 & 00 42 39.53&41 14 29.2& $2.62\pm0.08$ (2.32-2.80) & $0.61 - 1.04$  &    $42 - 72$  & SV          & PFJ36,r2-25,P298\\
\hline
\end{tabular}
\end{table}

\clearpage

\begin{table}
\small
\caption{List of bright X-ray sources deteted in the central part of M31 (continued). \label{source_ID_2}}
\begin{tabular}{cccccccl}
\hline
\hline
Source & R.A. & Decl.   & Photon                    & Flux$^{b}$ & $L_{\rm X}^{c}$ & Class$^{d}$ & Optical/Radio/X-ray \\
 ID    &      &         & Index$^{a}$               &            &                 &             &     ID$^{e,f}$      \\
\hline
\hline
 51 & 00 42 39.98&41 15 48.0& $2.13\pm0.03$ (1.93-2.29) & $2.74 - 6.83$  &  $189 - 471$  & SV          & TF44,PFJ37,r1-15,P299\\
 52 & 00 42 40.22&41 18 45.2& $1.67\pm0.06$ (1.32-1.73) & $0.83 - 1.76$  &   $58 - 122$  &             & r2-24,P301\\
 53 & 00 42 40.65&41 10 33.2& $1.69\pm0.19$ (1.35-1.90) & $0.14 - 0.54$  &    $10 - 37$  & GCS         & MIT212,r3-34,P303\\
 54 & 00 42 40.68&41 13 27.6& $1.67\pm0.08$ (1.37-1.95) & $0.82 - 0.99$  &    $56 - 68$  & SV          & PFJ38,r2-22,P302\\
 55 & 00 42 41.48&41 15 24.3& $1.81\pm0.09$             & $2.00\pm0.08$  &        $138$  & GCS         & MIT213,r1-32,TP20,P307\\
 56 & 00 42 41.66&41 21 05.5& $1.78\pm0.16$ (1.74-1.91) & $0.46 - 0.58$  &    $32 - 40$  &             & r3-31,P308\\
 57 & 00 42 41.81&41 16 35.9& $2.47\pm0.05$ (2.33-2.55) & $7.85 - 10.7$  &  $543 - 742$  & TR,REC,BHC,SV & TF47\\
 58 & 00 42 42.16&41 16 08.1& $1.60\pm0.03$ (1.56-1.60) & $6.30 - 12.5$  &  $435 - 863$  & TR          & r1-5,P310\\
 59 & 00 42 42.31&41 14 45.8& $2.22\pm0.06$ (2.05-2.42) & $1.18 - 1.77$  &   $82 - 122$  & SV          & TF48,PFJ39,r2-21,P311\\
 60 & 00 42 42.48&41 15 53.9& $1.63\pm0.03$ (1.44-1.77) & $2.98 - 4.94$  &  $206 - 340$  & SV          & TF49,PFJ41,r1-14,P314\\
 61 & 00 42 42.91&41 15 43.7& $1.48\pm0.07$ (1.40-1.56) & $1.93 - 3.87$  &  $133 - 267$  &             & TF52,PFJ42,r1-13,P317\\
 62 & 00 42 43.75&41 16 31.5& $2.08\pm0.05$ (1.93-2.36) & $2.31 - 3.69$  &  $159 - 255$  & SV          & TF55,PFJ43,r1-12,r1-29,r1-11,P319\\
 63 & 00 42 44.33&41 16 08.4& $2.51\pm0.05$ (2.07-3.20) & $2.44 - 5.67$  &  $168 - 391$  &             & TF56,PFJ44,r1-10,r1-21,r1-9,P321\\
 64 & 00 42 44.33&41 28 10.8& $1.73\pm0.17$             & $0.67\pm0.06$  &         $46$  &             & P322\\
 65 & 00 42 44.38&41 11 58.3& $2.13\pm0.09$ (1.71-2.47) & $0.50 - 1.28$  &    $35 - 89$  & SV          & r3-30,P323\\
 66 & 00 42 44.80&41 11 38.1& $2.19\pm0.03$ (1.72-2.53) & $2.32 - 3.62$  &  $160 - 250$  & SV          & TF58,PFJ45,r3-29,P326\\
 67 & 00 42 44.90&41 17 40.0& $1.46\pm0.07$ (1.40-1.70) & $0.52 - 1.07$  &    $36 - 74$  &             & r2-18,P324\\
 68 & 00 42 45.01&41 14 07.3& $1.74\pm0.15$ (1.57-1.99) & $0.62 - 1.07$  &    $43 - 74$  &             & PFJ46,r2-17,P328\\
 69 & 00 42 45.05&41 16 21.4& $2.19\pm0.05$ (2.05-2.47) & $3.14 - 6.41$  &  $216 - 442$  & SV          & 37W135,TF57,PFJ47,r1-4,P325\\
 70 & 00 42 45.11&41 15 25.0&                           & $0.25\pm0.03$  &         $17$  & PN,SNR      & Ford21,r1-26,P327\\
 71 & 00 42 45.22&41 17 22.1& $1.93\pm0.08$ (1.67-2.16) & $0.82 - 2.14$  &   $57 - 147$  & REC,SV      & r2-16,P329\\ 
 72 & 00 42 46.17&41 15 44.0& $2.38\pm0.44$             & $0.33\pm0.07$  &         $23$  &             & PFJ48,r1-18,P332\\
 73 & 00 42 46.87&41 21 19.7& $1.38\pm0.22$ (1.28-1.76) & $0.40\pm0.05$  &         $28$  &             & r3-28,P333\\
 74 & 00 42 46.97&41 16 15.7& $1.88\pm0.03$ (1.87-1.99) & $3.04 - 9.47$  &  $210 - 653$  & SV          & TF59,PFJ49,r1-3,P334\\
 75 & 00 42 47.16&41 16 28.5& $1.91\pm0.02$ (1.85-2.41) & $1.74 - 22.6$  & $120 - 1560$  & SV          & TF60,PFJ50,r1-2,P335\\
 76 & 00 42 47.22&41 11 57.7& $2.09\pm0.16$ (1.81-2.54) & $0.32 - 0.67$  &    $22 - 46$  &             & r3-27,P337\\
 77 & 00 42 47.89&41 11 14.0& $1.70\pm0.05$             & $6.22\pm0.17$  &        $430$  & GCS,REC     & Bo 128,PFJ51 \\
 78 & 00 42 47.91&41 15 33.1& $2.08\pm0.03$ (2.02-2.36) & $2.65 - 6.33$  &  $183 - 437$  & SV          & TF61,PFJ52,r1-6,P338\\
 79 & 00 42 48.48&41 25 22.5& $1.85\pm0.03$ (1.65-1.99) & $7.18 - 10.0$  &  $495 - 690$  & SV          & TF62,PFJ53,r3-25,P340\\
 80 & 00 42 48.51&41 15 21.4& $1.89\pm0.02$ (1.78-2.09) & $8.80 - 12.4$  &  $607 - 853$  & SV          & TF63,PFJ54,r1-1,P341\\
 81 & 00 42 49.24&41 18 16.1& $1.42\pm0.07$ (1.38-1.49) & $1.13 - 1.41$  &    $78 - 97$  & SV          & TF64,PFJ55,r2-14,P343\\
 82 & 00 42 51.66&41 13 03.1& $1.96\pm0.25$ (1.72-2.42) & $0.21 - 0.46$  &    $14 - 32$  & SV          & r2-39,P350\\
 83 & 00 42 52.52&41 15 40.2&                           & $4.11 - 5.66$  &  $284 - 391$  & SSS         & TF69,PFJ58,r2-12,P352\\
 84 & 00 42 52.54&41 18 54.8& $1.90\pm0.02$ (1.77-2.15) & $6.07 - 14.2$  &  $419 - 981$  & SV          & TF68,PFJ57,r2-13,P353\\
 85 & 00 42 53.67&41 25 51.8&                           & $0.62\pm0.02$  &         $43$  & SNR         & BA 521,B90 106,PFJ59,r3-69,P354\\
 86 & 00 42 54.90&41 16 03.4& $1.88\pm0.02$ (1.84-1.99) & $9.26 - 9.96$  &  $639 - 687$  & SV          & TF71,PFJ60,r2-11,P357\\
 87 & 00 42 55.33&41 25 57.6& $1.84\pm0.06$ (1.72-2.14) & $1.44 - 4.79$  &   $99 - 331$  & SV          & TF70,PFJ61,r3-23,P358\\
 88 & 00 42 55.41&41 18 35.1& $1.86\pm0.08$ (1.81-2.05) & $1.11 - 1.35$  &    $77 - 93$  & GCS         & Bo138,PFJ62,r2-10/r2-10,TP23,P360\\
 89 & 00 42 57.90&41 11 05.0& $1.98\pm0.02$ (1.83-2.06) & $4.62 - 6.94$  &  $319 - 479$  & SV          & TF73,PFJ64,r3-22,P364\\
 90 & 00 42 58.28&41 15 29.6& $1.66\pm0.09$ (1.26-2.51) & $0.67 - 1.20$  &    $46 - 83$  & SV          & TF74,PFJ65,r2-7,P365\\
 91 & 00 42 59.67&41 19 19.6& $1.94\pm0.02$ (1.92-2.11) & $4.29 - 6.03$  &  $296 - 416$  & GCS,SV      & Bo143,TF76,PFJ66,r2-6,TP24,P372\\
 92 & 00 42 59.87&41 16 05.9& $1.53\pm0.01$ (1.42-1.69) & $5.12 - 8.04$  &  $353 - 554$  & GCS,SV      & Bo144,TF77,PFJ67,r2-5,TP25,P373\\
 93 & 00 43 01.10&41 13 50.8& $1.39\pm0.19$ (1.39-1.43) & $0.25\pm0.03$  &         $17$  &             & PFJ68,r2-37,P376\\
 94 & 00 43 03.06&41 15 25.1& $2.17\pm0.05$ (2.05-2.45) & $5.76 - 7.31$  &  $397 - 504$  & GCS,SV      & Bo146,TF79,PFJ70,r2-3/r2-4,P384\\
 95 & 00 43 03.12&41 10 15.5& $1.56\pm0.05$ (1.51-1.77) & $0.78 - 0.92$  &    $54 - 64$  &             & PFJ71,r3-20,P383\\
 96 & 00 43 03.31&41 21 21.6& $2.20\pm0.05$ (2.11-2.27) & $1.15 - 2.43$  &   $80 - 168$  & GCS         & Bo147,PFJ72,r3-19,TP28,P385\\
 97 & 00 43 03.88&41 18 05.2& $1.79\pm0.02$ (1.61-2.12) & $2.31 - 6.50$  &  $160 - 448$  & GCS,SV      & Bo148,TF80,PFJ73,r2-2,TP29,P386\\
 98 & 00 43 04.23&41 16 01.3& $1.80\pm0.09$ (1.73-2.03) & $0.45 - 1.17$  &    $31 - 81$  &             & r2-1,P388\\
 99 & 00 43 05.68&41 17 02.8& $3.18\pm0.06$             & $5.43\pm0.07$  &        $375$  & TR,BHC      & K124,P390\\
100 & 00 43 07.13&41 18 10.3& $1.63\pm0.08$             & $1.84\pm0.07$  &        $127$  & TR          & K125,P395\\
\hline
\end{tabular}
\end{table}

\clearpage

\begin{table}
\small
\caption{List of bright X-ray sources detected in the central part of M31 (continued). \label{source_ID_3}}
\begin{tabular}{cccccccl}
\hline
\hline
Source & R.A. & Decl.   & Photon                     & Flux$^{b}$     & $L_{\rm X}^{c}$ & Class$^{d}$ & Optical/Radio/X-ray \\
 ID    &      &         & Index$^{a}$                &                &                 &             &     ID$^{e,f}$      \\
\hline
\hline
101 & 00 43 07.51&41 20 20.1& $2.15\pm0.19$ (1.46-2.87) & $0.35 - 0.50$  &    $25 - 34$  & GCS         & Bo150,MIT246,r3-18,TP30,P396\\
102 & 00 43 08.62&41 12 48.7& $0.79\pm0.05$ (0.73-1.08) & $1.45 - 2.08$  &  $100 - 143$  & DIP,SV      & PFJ74,r3-17,P403\\
103 & 00 43 09.87&41 19 01.1& $1.77\pm0.04$ (1.77-2.81) & $0.14 - 5.16$  &               & R,SV,BKG    & B90 125,TF82,PFJ75,r3-16,P405\\
104 & 00 43 09.94&41 23 32.5& $2.27\pm0.46$             & $<0.01 - 0.44$ &    $<1 - 30$  & TR          & \\
105 & 00 43 10.62&41 14 51.5& $1.63\pm0.01$ (1.48-1.71) & $12.5 - 18.4$  & $861 - 1270$  & GCS,SV      & Bo153,MIT251,TF83,PFJ76,r3-15,TP31,P408\\
106 & 00 43 11.35&41 18 09.8& $1.72\pm0.24$ (1.68-2.29) & $0.40 - 0.77$  &    $28 - 53$  &             & TF84,r3-14,P410\\
107 & 00 43 13.30&41 18 12.8& $1.92\pm0.26$ (1.58-2.27) & $0.19 - 0.43$  &    $13 - 29$  &             & r3-13,P412\\
108 & 00 43 14.37&41 07 21.2& $0.60\pm0.01$ (0.56-0.74) & $11.2 - 26.0$  & $773 - 1798$  & GCS,DIP,SV  & Bo158,TF85,PFJ77,K132,TP32,P414\\
109 & 00 43 15.42&41 11 25.0& $2.21\pm0.27$ (2.17-2.63) & $0.22 - 0.32$  &    $15 - 22$  & GCS         & Bo161,MIT260,PFJ78,r3-10,TP34,P419\\
110 & 00 43 15.51&41 24 40.0& $3.73\pm0.40$             & $<0.02 - 1.10$ &    $<2 - 76$  & TR,BHC      & \\
111 & 00 43 16.15&41 18 41.7& $1.65\pm0.12$ (1.36-1.71) & $0.46 - 1.43$  &    $32 - 99$  &             & r3-9,P420\\
112 & 00 43 16.20&41 03 48.5& $2.22\pm0.33$             & $0.62\pm0.07$  &         $43$  &             & P422\\
113 & 00 43 18.95&41 20 17.5&                           & $<0.02 - 1.86$ &   $<2 - 129$  & REC,SSS     & TF87,r3-8,P430\\
114 & 00 43 19.47&41 17 56.9&                           & $<0.01 - 3.85$ &   $<1 - 266$  & TR,SSS      & P431\\
115 & 00 43 21.07&41 17 50.8& $2.41\pm0.14$ (1.79-3.46) & $0.45 - 0.72$  &    $31 - 50$  & SV          & PFJ79,r3-7,P437\\
116 & 00 43 24.82&41 17 26.6& $1.33\pm0.18$ (1.32-1.35) & $0.31 - 0.47$  &    $22 - 32$  &             & P443\\
117 & 00 43 27.95&41 18 30.5&                           & $0.75\pm0.01$  &         $52$  & SNR         & BA 23,B90 142,TF89,PFJ80,r3-63,P454\\
118 & 00 43 29.13&41 07 47.7& $0.89\pm0.03$ (0.69-0.97) & $8.86 - 11.5$  &  $611 - 795$  &             & PFJ81,P457\\
119 & 00 43 32.41&41 10 40.8& $0.89\pm0.04$ (0.80-1.00) & $7.06 - 10.8$  &  $487 - 745$  &             & r3-3,P463\\
120 & 00 43 34.32&41 13 23.9& $2.24\pm0.04$ (2.03-2.45) & $2.52 - 4.43$  &  $174 - 305$  & SV          & TF91,PFJ82,r3-2,P467\\
121 & 00 43 37.28&41 14 43.5& $1.43\pm0.02$ (1.41-1.55) & $8.73 - 10.9$  &  $602 - 750$  & GCS,SV      & Bo185,MIT299,TF92,PFJ83,r3-1,TP37,P471\\
122 & 00 43 43.98&41 12 31.4&                           & $0.20\pm0.01$  &         $14$  & SNR         & B90 166,PFJ84,P486\\
123 & 00 43 53.70&41 16 56.0& $2.78\pm0.05$ (2.55-2.83) & $2.22 - 5.32$  &  $154 - 368$  & BHC,SV      & TF94,PFJ86,K143,P503\\
\hline
\end{tabular}
\begin{list}{}
\item $^{a}$ -- weighted average value and the observed range of the power law model photon index.  
\item $^{b}$ -- absorbed source power law model flux in the 0.3 - 10 keV energy range in units of $10^{-13}$ erg s$^{-1}$ cm$^{-2}$
\item $^{c}$ -- estimated absorbed source luminosity in the 0.3 - 10 keV energy band in units of $10^{35}$ ergs s$^{-1}$, 
assuming the distance of 760 kpc. 
\item $^{d}$ -- source classification: SV -- spectrally variable, TR -- transient, REC -- recurrent, SSS -- supersoft 
source, SNR -- supernova remnant, BHC -- black hole candidate, Z -- Z-source candidate, GCS -- globular cluster source, 
R -- radio source, BKG -- background AGN.
\item $^{e}$ -- source identifications beginning with Bo refer to Globular Cluster candidates listed in Table IV of 
Battistini et al. (1987), MIT -- in Magnier (1993). Source identifications beginning with BA refer to the SNR 
candidates listed in Baade \& Arp (1964). ``Ford21'' refers to planetary nebula candidate from Ford \& Jacoby (1978). 
Radio source identifications beginning with 37W and B90 refer to the source lists in Walterbos, Brinks \& Shane (1985) 
and Braun (1990). 
\item $^{f}$ -- source identifications beginning with TF refer to M31 {\em Einstein} X-ray source catalog entries 
from Trinchieri \& Fabbiano (1991). Identifications beginning with PFJ refer to X-ray sources from Primini, Forman 
and Jones (1993). Identifications beginning with r and K refer to the lists of M31 X-ray sources in Kong et al. 
(2002) and Kaaret (2002). Identifications beginning with TP refer to the list of globular cluster X-ray sources 
in Trudolyubov \& Priedhorsky (2004). Identifications starting with P refer to sources from Pietsch et al. (2005). 

\end{list}
\end{table}

\clearpage

\begin{table}
\small
\caption{Model fits to the energy spectra of the persistent sources with spectral cut-off. \label{spec_par_CUTOFF}}
\tiny
\begin{tabular}{ccccccccccc}
\hline
\hline
 Obs. &Model & N$_{\rm H}$                &kT/E$_{cut}$&$R_{color}^{a}$ &Photon&Flux$^{b}$& Flux$^{c}$&$\chi^{2}$&$L_{\rm X}^{d}$&Instrument\\
  \#  &      &($\times 10^{20}$ cm$^{-2}$)&(keV)       &    (km)        &Index &          &           &(d.o.f)   &               & \\      
\hline
\multicolumn{11}{c}{SRC \# 20}\\
\hline
1 ... &  PL     &$75\pm3$             &   ...       &  ...     &$3.10\pm0.07$         &$7.88\pm0.14$& $47.25$&$161.8(109)$& 54.4 &pn\\
      &DISKBB   &$34\pm2$        &$0.81\pm0.02$&$28^{+2}_{-1}$ &  ...                 &$7.55\pm0.14$& $11.78$&$105.4(109)$& 52.2 & ... \\
2 ... &  PL     &$71\pm5$             &   ...       &  ...     &$2.99^{+0.12}_{-0.11}$&$7.56\pm0.23$& $38.48$&  $36.4(39)$& 52.3 &pn\\
      &DISKBB   &$33\pm3$        &$0.82\pm0.04$&$26^{+3}_{-2}$ &  ...                 &$7.02\pm0.21$& $10.77$&  $28.9(39)$& 48.5 & ... \\
3 ... &  PL     &$75\pm3$             &   ...       &  ...     &$3.02\pm0.06$         &$9.16\pm0.16$& $49.77$&$208.8(115)$& 63.3 &pn\\
      &DISKBB   &$35\pm2$        &$0.85\pm0.02$&$27^{+2}_{-1}$ &  ...                 &$8.88\pm0.15$& $13.56$&$133.7(115)$& 61.4 & ... \\
4 ... &  PL     &$77\pm2$             &   ...       &  ...     &$3.04\pm0.04$         &$7.17\pm0.06$& $40.04$&$645.3(436)$& 49.5 &pn+M1+M2\\
      &DISKBB   &$36\pm1$        &$0.82\pm0.01$&$26\pm1$       &  ...                 &$6.77\pm0.06$& $10.56$&$441.7(436)$& 46.8 & ... \\
7 ... &  PL     &$82\pm7$             &   ...       &  ...     &$3.14^{+0.14}_{-0.13}$&$4.15\pm0.14$& $27.46$&  $63.4(37)$& 28.7 &pn\\
      &DISKBB   &$40\pm4$        &$0.76\pm0.04$&$24\pm3$       &  ...                 &$3.88\pm0.13$&  $6.52$&  $42.7(37)$& 26.8 & ... \\
\hline
\multicolumn{11}{c}{SRC \# 25}\\
\hline
1 ... &  PL     &$27\pm3$             &   ...       &  ...     &$2.77^{+0.14}_{-0.13}$&$4.71\pm0.16$& $12.25$&  $47.7(43)$& 32.6 &M2\\
      &DISKBB   &$5\pm2$&$0.72^{+0.05}_{-0.04}$&$23\pm3$       &  ...                 &$4.19\pm0.14$&  $4.77$&  $47.8(43)$& 29.0 & ... \\
2 ... &  PL     &$24\pm2$             &   ...       &  ...     &$2.77^{+0.11}_{-0.10}$&$4.96\pm0.15$& $12.22$&  $38.5(61)$& 34.3 &pn\\
      &DISKBB   &$4\pm1$&$0.72\pm0.04$         &$24\pm3$       &  ...                 &$4.53\pm0.14$&  $5.03$&  $47.0(61)$& 31.3 & ... \\
3 ... &  PL     &$26\pm3$             &   ...       &  ...     &$3.14^{+0.16}_{-0.15}$&$2.29\pm0.07$&  $7.81$&  $33.6(46)$& 15.8 &pn\\
      &DISKBB   &$4^{+2}_{-1}$&$0.55\pm0.04$   &$28\pm4$       &  ...                 &$2.07\pm0.07$&  $2.40$&  $35.8(46)$& 14.3 & ... \\
4 ... &  PL     &$28\pm1$             &   ...       &  ...     &$3.17\pm0.07$         &$3.26\pm0.05$& $12.00$&$179.5(152)$& 22.5 &pn\\
      &DISKBB   &$5\pm1$&$0.60\pm0.02$         &$29\pm2$       &  ...                 &$3.06\pm0.05$&  $3.54$&$171.1(152)$& 21.1 & ... \\
5 ... &  PL     &$27\pm2$             &   ...       &  ...     &$2.95\pm0.10$         &$4.60\pm0.12$& $13.82$& $103.1(71)$& 31.8 &pn\\
      &DISKBB   &$4\pm1$&$0.68\pm0.03$         &$26\pm2$       &  ...                 &$4.35\pm0.11$&  $4.80$&  $82.1(71)$& 30.1 & ... \\
7 ... &  PL     &$25^{+3}_{-2}$       &   ...       &  ...     &$2.93^{+0.12}_{-0.11}$&$5.82\pm0.17$& $16.63$&  $41.2(35)$& 40.2 &pn\\
      &DISKBB   &$5\pm1$&$0.65^{+0.04}_{-0.03}$&$32^{+3}_{-5}$ &  ...                 &$5.35\pm0.16$&  $6.10$&  $28.5(35)$& 30.1 & ... \\
\hline
\multicolumn{11}{c}{SRC \# 49}\\
\hline
1 ... &  PL     &$14\pm1$             &   ...       &  ...     &$1.87\pm0.01$         &$69.95\pm0.39$&$89.78$&$787.6(628)$&483.4 &pn\\
      & CUTOFFPL&$8\pm1$          &$6.4^{+0.7}_{-0.6}$&  ...  &$1.35\pm0.05$          &$69.25\pm0.38$&$79.03$&$667.4(627)$&478.6 & ... \\
2 ... &  PL     &$11\pm1$             &   ...       &  ...     &$1.59\pm0.02$         &$74.58\pm0.72$&$86.26$&$319.1(314)$&515.4 &M1+M2\\
      & CUTOFFPL&$6\pm1$          &$5.4^{+1.2}_{-0.8}$&  ...  &$1.05\pm0.10$          &$69.40\pm0.67$&$74.61$&$287.4(313)$&479.6 & ... \\
3 ... &  PL     &$14\pm1$             &   ...       &  ...     &$1.82\pm0.01$         &$72.05\pm0.42$&$91.19$&$877.1(612)$&497.9 &pn\\
      & CUTOFFPL&$6\pm1$          &$4.2\pm0.3$        &  ...  &$1.02\pm0.05$          &$71.03\pm0.41$&$76.79$&$638.4(611)$&490.9 & ... \\
4 ... &  PL     &$13\pm1$             &   ...       &  ...     &$1.67\pm0.01$         &$51.84\pm0.28$&$61.94$&$805.4(577)$&358.3 &M1+M2\\
      & CUTOFFPL&$5\pm1$          &$4.5^{+0.4}_{-0.3}$&  ...  &$0.98\pm0.05$          &$48.87\pm0.26$&$52.17$&$624.2(576)$&337.7 & ... \\
7 ... &  PL     &$17\pm1$             &   ...       &  ...     &$1.90\pm0.02$         &$70.56\pm0.54$&$94.20$&$572.6(330)$&487.6 &pn\\
      & CUTOFFPL&$7\pm1$          &$3.9\pm0.3$        &  ...  &$1.02\pm0.07$          &$70.42\pm0.52$&$77.36$&$396.6(329)$&486.7 & ... \\
\hline
\multicolumn{11}{c}{SRC \# 80}\\
\hline
1 ... &  PL     &$10\pm1$             &   ...       &  ...     &$1.82^{+0.05}_{-0.04}$&$13.77\pm0.22$&$16.77$&$165.3(157)$& 95.2 &M1+M2\\
      & CUTOFFPL&$3\pm2$          &$4.6^{+1.9}_{-1.1}$&  ...  &$1.17\pm0.17$          &$13.11\pm0.21$&$13.92$&$153.0(156)$& 90.6 & ... \\
2 ... &  PL     &$16\pm2$             &   ...       &  ...     &$2.09\pm0.07$         &$10.34\pm0.27$&$14.79$&  $89.5(78)$& 71.5 &pn\\
      & CUTOFFPL&$6\pm3$          &$2.9^{+1.0}_{-0.6}$&  ...  &$1.09^{+0.26}_{-0.25}$ & $9.86\pm0.26$&$11.10$&  $75.6(77)$& 68.2 & ... \\
3 ... &  PL     &$12\pm2$             &   ...       &  ...     &$1.88\pm0.05$         &$11.97\pm0.22$&$15.05$&$143.5(147)$& 82.7 &M1+M2\\
      & CUTOFFPL&$10\pm3$         &$16^{+10}_{-10}$   &  ...  &$1.71^{+0.20}_{-0.23}$ &$11.73\pm0.22$&$14.17$&$143.0(146)$& 81.1 & ... \\
4 ... &  PL     &$12\pm1$             &   ...       &  ...     &$1.90\pm0.02$         &$12.37\pm0.10$&$15.78$&$772.5(720)$& 85.5 &pn+M1+M2\\
      & CUTOFFPL&$5\pm1$          &$7.0^{+1.4}_{-1.0}$&  ...  &$1.46\pm0.08$          &$12.06\pm0.09$&$13.90$&$735.9(719)$& 83.3 & ... \\
\hline
\multicolumn{11}{c}{SRC \# 118}\\
\hline
1 ... &  PL     &$30^{+10}_{-9}$      &   ...       &  ...     &$0.87^{+0.12}_{-0.11}$& $9.91\pm0.35$&$10.79$&  $37.0(40)$& 68.5 &M1+M2\\
      & CUTOFFPL&$8^{+12}_{-8}$   &$3.6^{+3.2}_{-1.3}$&  ...  &$-0.14^{+0.45}_{-0.53}$& $8.59\pm0.30$& $8.78$&  $33.0(39)$& 59.3 & ... \\
3 ... &  PL     &$12^{+4}_{-3}$       &   ...       &  ...     &$0.79\pm0.07$         &$11.57\pm0.38$&$12.05$&  $53.0(37)$& 79.9 &pn\\
      & CUTOFFPL&$<2$             &$2.8^{+0.5}_{-0.4}$&  ...  &$-0.39^{+0.16}_{-0.14}$& $9.99\pm0.33$& $9.99$&  $25.9(36)$& 69.0 & ... \\
4 ... &  PL     &$19\pm3$             &   ...       &  ...     &$0.98\pm0.05$         & $8.79\pm0.20$& $9.50$&$190.4(146)$& 60.8 &pn+M2\\
      & CUTOFFPL&$<4$             &$2.6\pm0.2$        &  ...  &$-0.34^{+0.11}_{-0.09}$& $7.67\pm0.18$& $7.67$&$142.2(145)$& 53.0 & ... \\
5 ... &  PL     &$7^{+5}_{-4}$        &   ...       &  ...     &$0.69\pm0.09$         &$11.20\pm0.46$&$11.47$&  $36.2(25)$& 77.4 &pn\\
      & CUTOFFPL&$<2$             &$3.6^{+1.1}_{-0.9}$&  ...  &$-0.20^{+0.19}_{-0.21}$& $9.84\pm0.40$& $9.84$&  $23.4(24)$& 68.0 & ... \\
\hline
\multicolumn{11}{c}{SRC \# 123}\\
\hline
1 ... &  PL     &$24\pm3$             &   ...       &  ...     &$2.71^{+0.11}_{-0.10}$&$5.32\pm0.14$& $12.56$&  $49.1(61)$& 36.8 &M1+M2\\
      &DISKBB   &$2\pm1$&$0.73\pm0.04$         &$23\pm2$       &  ...                 &$4.82\pm0.12$&  $5.12$&  $35.9(61)$& 33.3 & ... \\
2 ... &  PL     &$15^{+6}_{-5}$       &   ...       &  ...     &$2.55^{+0.40}_{-0.33}$&$2.22\pm0.16$&  $3.94$&  $15.4(14)$& 15.4 &pn\\
      &DISKBB   &$3^{+3}_{-2}$&$0.60^{+0.12}_{-0.10}$&$21^{+10}_{-7}$&  ...           &$1.72\pm0.13$&  $1.88$&  $21.8(14)$& 11.9 & ... \\
3 ... &  PL     &$23\pm3$             &   ...       &  ...     &$2.71\pm0.12$         &$4.95\pm0.14$& $11.44$&  $37.4(47)$& 34.2 &M1+M2\\
      &DISKBB   &$2\pm1$&$0.72\pm0.04$         &$22^{+3}_{-2}$ &  ...                 &$4.40\pm0.13$&  $4.59$&  $37.3(47)$& 30.4 & ... \\
4 ... &  PL     &$22\pm1$             &   ...       &  ...     &$2.80\pm0.07$         &$3.96\pm0.06$&  $9.66$&$195.4(174)$& 27.4 &pn+M1+M2\\
      &DISKBB   &$3\pm1$&$0.66\pm0.02$         &$24^{+2}_{-1}$ &  ...                 &$3.53\pm0.05$&  $3.82$&$179.4(174)$& 24.4 & ... \\
\hline
\end{tabular}

\begin{list}{}{}
\item $^{a}$ -- color radius
\item $^{b}$ -- absorbed model flux in the $0.3 - 10$ keV energy range in 
units of $10^{-13}$ erg s$^{-1}$ cm$^{-2}$
\item $^{c}$ -- unabsorbed model flux in the $0.3 - 10$ keV energy range in 
units of $10^{-13}$ erg s$^{-1}$ cm$^{-2}$
\item $^{d}$ -- absorbed luminosity in the $0.3 - 10$ keV energy range in 
units of $10^{36}$ erg s$^{-1}$, assuming the distance of 760 kpc
\end{list}
\end{table}

\clearpage

\begin{table}
\caption{M31 bright source spectral fit results, (BBODYRAD+DISKBB)*WABS and COMPTT*WABS model approximation. 
\label{spec_par_two_comp}}
\begin{tabular}{ccccccccccc}
\hline
\hline
\multicolumn{11}{c}{Model: (BBODYRAD+DISKBB)*WABS}\\
\hline
 Obs. & $N_{\rm H}$                & $kT_{BB}$ & $r_{\rm BB}$& $kT_{in}$&$r_{in} \sqrt{cos i}$&Flux$^{a}$&Flux$^{b}$&$\chi^{2}$&$L_{X}^{c}$& Instrument\\
 $\#$ &($\times 10^{20}$ cm$^{-2}$)&  (keV)    & (km)        &  (keV)   &      (km)           &          &          &   (dof)  &           &\\   
\hline
\multicolumn{11}{c}{SRC $\# 31$}\\
\hline
1 ... &$12\pm2$&$1.44^{+0.11}_{-0.10}$&$10\pm1$&$0.52\pm0.06$         &$42^{+11}_{-9}$ & $9.80\pm0.18$&$11.17$&$129.1(112)$&  67.7 &pn\\
3 ... &$11\pm1$&$1.29^{+0.05}_{-0.04}$&$20\pm2$&$0.48^{+0.04}_{-0.03}$&$84^{+14}_{-11}$&$27.63\pm0.28$&$31.79$&$282.6(295)$& 191.0 &pn\\
4 ... &$9\pm1$ &$1.34^{+0.04}_{-0.03}$&$17\pm1$&$0.57^{+0.03}_{-0.02}$&$54^{+5}_{-4}$  &$23.37\pm0.11$&$25.96$&$1279(1058)$& 161.5 &pn+M1+M2\\   
\hline
\multicolumn{11}{c}{SRC $\# 49$}\\
\hline
1 ... &$5\pm1$&$1.35^{+0.04}_{-0.03}$&$27^{+1}_{-2}$&$0.59\pm0.03$         &$89\pm7$         &$68.61\pm0.38$&$73.53$&$679.1(626)$& 474.1 &pn\\
2 ... &$4\pm1$&$1.31^{+0.09}_{-0.07}$&$29^{+4}_{-3}$&$0.63^{+0.07}_{-0.06}$&$70^{+13}_{-11}$ &$66.15\pm0.64$&$69.73$&$285.3(311)$& 457.1 &M1+M2\\
3 ... &$5\pm1$&$1.23\pm0.03$         &$34\pm2$      &$0.56\pm0.03$         &$93^{+10}_{-9}$  &$69.89\pm0.41$&$74.60$&$632.2(610)$& 483.0 &pn\\
4 ... &$3\pm1$&$1.40^{+0.08}_{-0.05}$&$21^{+1}_{-3}$&$0.75^{+0.07}_{-0.04}$&$44^{+4}_{-6}$   &$47.86\pm0.26$&$49.65$&$625.3(574)$& 330.8 &M1+M2\\
7 ... &$6\pm1$&$1.21^{+0.04}_{-0.03}$&$35^{+2}_{-3}$&$0.55\pm0.04$         &$101^{+14}_{-13}$&$69.73\pm0.54$&$75.91$&$381.6(328)$& 481.9 &pn\\
\hline
\multicolumn{11}{c}{SRC $\# 80$}\\
\hline
1 ... &$<4$   &$1.30^{+0.12}_{-0.13}$&$11^{+4}_{-2}$&$0.65^{+0.07}_{-0.10}$&$32^{+12}_{-5}$& $12.63\pm0.20$&$12.68$&$147.6(154)$& 87.3 &pn\\
3 ... &$<2$   &$1.67^{+0.67}_{-0.24}$&$8\pm2$       &$0.72^{+0.12}_{-0.08}$&$27\pm6$       & $11.61\pm0.21$&$11.83$&$142.3(134)$& 80.2 &pn\\
4 ... &$3\pm1$&$1.33\pm0.06$         &$11\pm1$      &$0.61\pm0.04$         &$35^{+4}_{-3}$ & $11.85\pm0.10$&$12.42$&$751.4(716)$& 81.9 &pn\\
\hline
\multicolumn{11}{c}{Model: Absorbed Comptonization Model (COMPTT*WABS)}\\
\hline
      & $N_{\rm H}$                & & $kT_{0}$ & $kT_{e}$ & $\tau$ & Flux$^{a}$ & Flux$^{b}$ & $\chi^{2}$&$L_{X}^{d}$& Instrument\\
      &($\times 10^{20}$ cm$^{-2}$)& &  (keV)   &  (keV)   &        &            &            & (dof)     &           & \\
\hline
\multicolumn{11}{c}{SRC $\# 31$}\\
\hline
1 ... &$9^{+5}_{-2}$ &&$0.17^{+0.03}_{-0.04}$&$2.07^{+0.30}_{-0.21}$&$18.9^{+1.4}_{-2.4}$& $9.95\pm0.19$&$10.84$&$129.6(112)$&  68.8 & pn\\
3 ... &$10^{+2}_{-3}$&&$0.14^{+0.03}_{-0.02}$&$1.83^{+0.11}_{-0.12}$&$19.0^{+1.3}_{-0.9}$&$28.06\pm0.29$&$31.61$&$271.3(295)$& 193.9 & pn\\
4 ... &$11\pm1$      &&$0.13\pm0.01$         &$1.78^{+0.05}_{-0.07}$&$19.6^{+0.7}_{-0.5}$&$23.72\pm0.12$&$27.18$&$1220(1061)$& 163.9 & pn+M1+M2\\
\hline
\multicolumn{11}{c}{SRC $\# 49$}\\
\hline
1 ... &$8\pm1$       &&$0.13\pm0.01$         &$1.75\pm0.06$         &$18.9\pm0.6$        &$68.98\pm0.38$&$77.51$&$608.8(626)$& 476.7 & pn\\
2 ... &$8\pm1$       &&$0.11^{+0.02}_{-0.04}$&$1.55^{+0.10}_{-0.08}$&$22.7\pm1.0$        &$67.06\pm0.65$&$76.79$&$285.9(312)$& 463.4 & M1+M2\\
3 ... &$9^{+2}_{-1}$ &&$0.10^{+0.02}_{-0.01}$&$1.53^{+0.04}_{-0.05}$&$21.6^{+0.9}_{-0.5}$&$70.41\pm0.41$&$79.72$&$593.5(610)$& 486.6 & pn\\
4 ... &$10\pm1$      &&$0.06^{+0.02}_{-0.06}$&$1.59^{+0.05}_{-0.04}$&$21.4\pm0.5$        &$47.68\pm0.26$&$55.26$&$621.6(575)$& 329.5 & M1+M2\\
7 ... &$9^{+1}_{-2}$ &&$0.12\pm0.02$         &$1.49\pm0.05$         &$21.7^{+1.0}_{-0.8}$&$70.15\pm0.54$&$79.63$&$357.7(328)$& 484.8 & pn\\
\hline
\end{tabular}

\begin{list}{}{}
\item $^{a}$ -- Absorbed model flux in the $0.3 - 10$ keV energy range in 
units of $10^{-13}$ erg s$^{-1}$ cm$^{-2}$
\item $^{b}$ -- unabsorbed model flux in the $0.3 - 10$ keV energy range in 
units of $10^{-13}$ erg s$^{-1}$ cm$^{-2}$
\item $^{c}$ -- Absorbed isotropic source luminosity in the $0.3 - 10$ keV 
energy range in units of $10^{36}$ erg s$^{-1}$ assuming the distance of 760 
kpc
\end{list}
\end{table}

\clearpage

\begin{table}
\small
\caption{Model fits to the energy spectra of transient sources. \label{spec_par_TR}}
\small
\begin{tabular}{ccccccccccc}
\hline
\hline
 Obs. &Model & N$_{\rm H}$                &kT/E$_{cut}$&$R_{color}^{a}$ &Photon&Flux$^{b}$& Flux$^{c}$&$\chi^{2}$&$L_{\rm X}^{d}$&Instrument\\
 $\#$ &      &($\times 10^{20}$ cm$^{-2}$)&(keV)&    (km)        &Index &       &           &(d.o.f)   &               & \\      
\hline
\multicolumn{11}{c}{SRC $\# 3$}\\
\hline
5 ... &  PL  &$53\pm5$&  ...        &  ...   &$3.17^{+0.16}_{-0.15}$&$4.71\pm0.17$& $25.12$ & $40.8(43)$& 32.6 &pn+M1+M2\\ 
      &DISKBB&$20\pm3$&$0.66\pm0.04$&$32\pm2$&  ...                 &$4.39\pm0.16$&  $6.52$ & $34.5(43)$& 30.3 & ... \\
6 ... &  PL  &$40\pm5$&  ...        &  ...   &$2.74^{+0.16}_{-0.15}$&$3.84\pm0.16$& $11.42$ & $69.0(57)$& 26.5 &pn+M1+M2\\
 &DISKBB&$14\pm3$&$0.77\pm0.05$&$19^{+4}_{-2}$&  ...           &$3.34\pm0.15$&  $4.31$ & $68.4(57)$& 23.1 & ... \\
7 ... &  PL  &$56^{+8}_{-7}$& ...   &  ...   &$3.36^{+0.24}_{-0.21}$&$4.01\pm0.18$& $27.39$ & $52.4(49)$& 27.7 &pn+M2\\         
 &DISKBB&$23\pm4$&$0.59\pm0.05$&$38^{+8}_{-7}$&  ...           &$3.65\pm0.17$&  $5.93$ & $47.1(49)$& 25.2 & ... \\
\hline
\multicolumn{11}{c}{SRC $\# 7$}\\
\hline
4 ... &  BB     &$16\pm4$ & $0.050\pm0.005$ & $19480^{+20590}_{-8630}$& ... & $0.54\pm0.03$& $5.98$& $20.7(24)$& 3.72 & M1+M2 \\
\hline
\multicolumn{11}{c}{SRC $\# 46$}\\
\hline
1 ... &  PL     &$8\pm1$             &   ...       &  ...       &$2.18\pm0.07$       & $3.33\pm0.06$& $4.32$&$155.6(138)$& 23.0&pn+M1+M2\\ 
      &DISKBB+PL&$21^{+6}_{-5}$&$0.14\pm0.02$&$830^{+775}_{-454}$&$2.01^{+0.16}_{-0.11}$& $3.63\pm0.07$&$8.33$&$121.6(134)$& 25.1& ... \\
\hline
\multicolumn{11}{c}{SRC $\# 57$}\\
\hline
5 ... &  PL  &$23\pm2$&   ...       &  ...   &$2.53\pm0.06$         &$7.85\pm0.14$& $16.01$ & $234.8(232)$& 54.3 &pn+M1+M2\\
      &DISKBB&$3\pm1$&$0.86\pm0.03$&$20^{+2}_{-3}$ &  ...           &$7.11\pm0.13$&  $7.54$ & $218.1(232)$& 49.1 & ... \\
6 ... &  PL  &$20\pm1$&   ...       & ...    &$2.38\pm0.05$         &$9.31\pm0.13$& $16.46$ & $287.2(213)$& 64.3 &pn+M1+M2\\
      &DISKBB&$2^{+2}_{-1}$&$0.97\pm0.03$&$17\pm1$&   ...           &$8.55\pm0.12$&  $8.75$ & $262.8(213)$& 59.1 & ... \\
7 ... &  PL  &$22\pm1$&   ...       & ...    &$2.33\pm0.05$        &$10.74\pm0.22$& $18.92$ & $299.7(238)$& 74.2 &pn+M1+M2\\
      &DISKBB&$2^{+2}_{-1}$&$1.00\pm0.03$&$17\pm1$&  ...            &$9.79\pm0.21$& $10.21$ & $252.3(238)$& 67.7 & ... \\
\hline
\multicolumn{11}{c}{SRC $\# 58$}\\
\hline
1 ... &  PL     &$9\pm1$          &  ...        &   ...         &$1.58\pm0.05$       &$12.49\pm0.23$&$14.09$&$157.1(148)$& 86.3 &M1+M2\\
      & CUTOFFPL&$4\pm2$          &$6.3^{+4.9}_{-1.9}$&  ...  &$1.11^{+0.21}_{-0.27}$&$11.68\pm0.21$&$12.36$&$150.4(147)$& 80.7 & ... \\
2 ... &  PL     &$7\pm1$          &  ...        &   ...         &$1.56\pm0.09$       & $6.30\pm0.27$& $6.95$&  $50.6(41)$& 43.5 &pn\\
      & CUTOFFPL&$<4$             &$4.0\pm0.9$  &   ...         &$0.76^{+0.14}_{-0.16}$& $6.11\pm0.26$& $6.11$&  $40.8(40)$& 42.2 & ... \\         
\hline
\multicolumn{11}{c}{SRC $\# 99$}\\
\hline
4 ... &  PL     &$15\pm1$         &  ...        &   ...         &$3.18\pm0.06$       & $4.93\pm0.04$&$12.80$&$593.9(435)$& 34.1 & pn+M1+M2\\
      &DISKBB+PL&$15\pm1$      &$0.16\pm0.01$&$672^{+155}_{-114}$&$2.46\pm0.09$& $5.43\pm0.04$& $12.05$& $450.0(431)$ & 37.5 & ... \\
\hline
\multicolumn{11}{c}{SRC $\# 100$}\\
\hline
4 ... &  PL     &$<2$             &   ...       &   ...         &$1.62\pm0.05$       & $1.84\pm0.07$& $1.84$&$97.6(81)$& 12.69 & M1+M2\\
\hline
\multicolumn{11}{c}{SRC $\# 104$}\\
\hline
5 ... &  PL  &$12^{+15}_{-9}$& ...    &  ...   &$2.85\pm0.50$       &$0.37\pm0.05$& $0.73$ & $11.9(8)$& 2.57 &pn\\
      &DISKBB&$4^{+11}_{-3}$&$0.32^{+0.12}_{-0.11}$&$34\pm11$& ...  &$0.28\pm0.04$& $0.35$ & $15.7(8)$& 1.91 & ... \\
6 ... &  PL  &$32^{+15}_{-11}$& ...  &  ...  &$4.17^{+0.97}_{-0.73}$&$0.44\pm0.05$& $4.17$ & $13.3(11)$& 3.04 &pn\\
      &DISKBB&$6^{+5}_{-2}$&$0.36^{+0.08}_{-0.07}$&$34^{+28}_{-14}$& ... &$0.43\pm0.05$& $5.64$ & $14.1(11)$& 2.97 & ... \\
\hline
\multicolumn{11}{c}{SRC $\# 110$}\\
\hline
5 ... &  PL  &$28\pm6$&  ...        & ...    &$3.73^{+0.42}_{-0.37}$&$1.10\pm0.06$& $6.35$ & $26.0(18)$& 7.62 &pn\\
      &DISKBB&$7\pm3$ &$0.37\pm0.05$&$49^{+20}_{-13}$&  ...         &$0.99\pm0.06$& $1.37$ & $31.6(18)$& 6.86 & ... \\
\hline
\multicolumn{11}{c}{SRC $\# 114$}\\
\hline
1 ... &  BB     &$18^{+1}_{-2}$&$0.056^{+0.002}_{-0.001}$&$34580^{+2820}_{-2680}$& ... &$3.85\pm0.09$&$44.62$&$107.5(77)$& 26.6 & pn+M1+M2 \\ 
\hline 
\end{tabular}

\begin{list}{}{}
\item $^{a}$ -- color radius
\item $^{b}$ -- absorbed model flux in the $0.3 - 10$ keV energy range in 
units of $10^{-13}$ erg s$^{-1}$ cm$^{-2}$
\item $^{c}$ -- unabsorbed model flux in the $0.3 - 10$ keV energy range in 
units of $10^{-13}$ erg s$^{-1}$ cm$^{-2}$
\item $^{d}$ -- absorbed luminosity in the $0.3 - 10$ keV energy range in 
units of $10^{36}$ erg s$^{-1}$, assuming the distance of 760 kpc
\end{list}
\end{table}

\clearpage

\begin{table}
\small
\caption{Blackbody radiation model fits to the energy spectra of supersoft sources. \label{spec_par_SSS}}
\small
\begin{tabular}{ccccccccc}
\hline
\hline
 Obs. & N$_{\rm H}$                 & kT$_{bb}$ & $R_{bb}^{a}$ & Flux$^{b}$& Flux$^{c}$ &$\chi^{2}$ & $L_{\rm X}^{d}$&Instrument\\
 $\#$ & ($\times 10^{20}$ cm$^{-2}$)& (eV) &    (km)         &           &            &(d.o.f)   &                &          \\      
\hline
\multicolumn{9}{c}{SRC $\# 83$}\\
\hline
 1 & $18^{+1}_{-2}$ & $60^{+2}_{-1}$ & $33270^{+4780}_{-5320}$ & $5.66\pm0.06$ & $63.58$ & 182.4(150) & 39.1 & pn+M1+M2 \\
 2 & $15^{+2}_{-2}$ & $60^{+3}_{-2}$ & $23000^{+6870}_{-6250}$ & $4.11\pm0.12$ & $31.80$ &   42.5(41) & 28.4 & pn \\
 3 & $22\pm2$       & $58^{+2}_{-1}$ & $42920^{+7310}_{-8680}$ & $5.30\pm0.08$ & $89.81$ &  104.5(85) & 36.6 & pn \\
 5 & $19\pm2$       & $58\pm2$       & $32080^{+7650}_{-6180}$ & $4.17\pm0.08$ & $51.15$ &  100.0(87) & 28.8 & pn \\
\hline
\multicolumn{9}{c}{SRC $\# 113$}\\
\hline
 1 & $34\pm2$       & $57\pm1$       & $43630^{+3130}_{-3100}$ & $1.69\pm0.06$ & $84.65$ &   62.7(38) & 11.7 & M1+M2 \\
 3 & $12^{+15}_{-8}$& $50^{+19}_{-17}$& $7290^{+23150}_{-7290}$& $0.11\pm0.02$ &  $0.81$ &   38.3(33) &  0.8 & pn \\
\hline 
\end{tabular}

\begin{list}{}{}
\item $^{a}$ -- color radius
\item $^{b}$ -- absorbed model flux in the $0.3 - 1.5$ keV energy range in 
units of $10^{-13}$ erg s$^{-1}$ cm$^{-2}$
\item $^{c}$ -- unabsorbed model flux in the $0.3 - 1.5$ keV energy range in 
units of $10^{-13}$ erg s$^{-1}$ cm$^{-2}$
\item $^{d}$ -- absorbed luminosity in the $0.3 - 1.5$ keV energy range in 
units of $10^{36}$ erg s$^{-1}$, assuming the distance of 760 kpc
\end{list}
\end{table}

\begin{table}
\small
\caption{Bright SNR candidate sources spectral fit results. Absorbed MEKAL and NEI models, {\em XMM}/EPIC-pn data $0.3 - 3.0$ keV energy range. 
\label{spec_par_SNR}}
\begin{tabular}{ccccccccccc}
\hline
\hline
Model    & $kT_{\rm MEKAL}/kT_{e}$&log$n_{e} t$&\multicolumn{5}{c}{Abundance$^{a}$} & N$_{\rm H}^{b}$ &Flux$^{c}$&$\chi^{2}$/(dof)\\
         & (keV)& & N & O & Ne & S & Fe &  &  &  \\
\hline
\multicolumn{11}{c}{Source $\#$85 = SNR BA 521, EPIC-pn, Obs. 1,3,4}\\
\hline
MEKAL &$0.15^{+0.01}_{-0.01}$& ... &1 & 1 & 1 & 1 & 1 & $0.65^{+0.03}_{-0.0.04}$ & $0.53\pm0.02$&111.4(60)\\
      &$0.16^{+0.01}_{-0.01}$& ... &$0.61^{d}$&$0.26^{d}$&$0.75^{d}$&$0.40^{d}$&1&$0.50^{+0.04}_{-0.02}$&$0.55\pm0.02$&109.8(60)\\
      &$0.41^{+0.03}_{-0.04}$& ... &$0.61^{d}$&$0.35^{+0.11}_{-0.10}$&$0.25^{+0.16}_{-0.15}$&$0.40^{d}$&$0.13^{+0.03}_{-0.02}$&$0.06^{+0.06}_{-0.03}$&$0.62\pm0.02$&73.4(57)\\ 
\hline
NEI&$0.72^{+0.05}_{-0.09}$&$11.0\pm0.1$& 1 & 1 & 1 & 1 & 1 &$<0.02$&$0.55\pm0.02$&133.9(59)\\
   &$4.87^{+5.54}_{-2.38}$&$9.7^{+0.2}_{-0.1}$&$0.61^{d}$&$0.26^{d}$&$0.75^{d}$&$0.40^{d}$&1&$0.20\pm0.01$&$0.66\pm0.02$&64.0(59)\\
   &$2.66^{+5.22}_{-1.19}$&$9.8\pm0.1$&$0.61^{d}$&$0.25^{+0.03}_{-0.04}$&$0.60^{+0.18}_{-0.17}$&$0.40^{d}$&$0.61^{+0.67}_{-0.19}$&$0.21^{+0.05}_{-0.04}$&$0.66\pm0.02$&63.3(56)\\
\hline
\multicolumn{11}{c}{Source $\#$117 = SNR BA 23, EPIC-pn, Obs. 1,3,4}\\
\hline
MEKAL &$0.26^{+0.01}_{-0.01}$& ... &1 & 1 & 1 & 1 & 1 & $0.02^{+0.02}_{-0.01}$ & $0.73\pm0.01$&199.8(156)\\
      &$0.18^{+0.01}_{-0.01}$& ... &$0.75^{d}$&$0.27^{d}$&$1.00^{d}$&$0.44^{d}$&1&$0.18^{+0.02}_{-0.02}$&$0.72\pm0.01$&169.4(156)\\
      &$0.25^{+0.01}_{-0.02}$& ... &$0.75^{d}$&$0.29^{+0.07}_{-0.04}$&$0.38^{+0.17}_{-0.11}$&$0.44^{d}$&$0.25^{+0.06}_{-0.05}$&$0.11^{+0.02}_{-0.03}$&$0.75\pm0.01$&141.7(153)\\ 
\hline
NEI&$1.17^{+0.34}_{-0.23}$&$10.2\pm0.1$& 1 & 1 & 1 & 1 & 1 &$0.03^{+0.03}_{-0.02}$&$0.74\pm0.01$&171.1(155)\\
   &$1.01^{+0.23}_{-0.18}$&$9.8^{+0.1}_{-0.1}$&$0.75^{d}$&$0.27^{d}$&$1.00^{d}$&$0.44^{d}$&1&$0.12^{+0.02}_{-0.03}$&$0.75\pm0.01$&158.2(155)\\
   &$0.68^{+0.25}_{-0.13}$&$10.4^{+0.2}_{-0.3}$&$0.75^{d}$&$0.41^{+0.06}_{-0.07}$&$0.68^{+0.16}_{-0.14}$&$0.44^{d}$&$0.54\pm0.09$&$0.07^{+0.04}_{-0.03}$&$0.77\pm0.01$&133.5(152)\\
\hline
\multicolumn{11}{c}{Source $\#$70, EPIC-pn, Obs. 4}\\
\hline
MEKAL &$0.44^{+0.04}_{-0.12}$& ... & 1 & 1 & 1 & 1 & 1 & $0.33^{+0.21}_{-0.14}$ & $0.18\pm0.02$&36.4(20)\\
      &$0.50^{+0.09}_{-0.15}$& ... & 1 &$0.10^{+0.40}_{-0.07}$&$<0.20$& 1 &$0.17^{+0.10}_{-0.06}$&$0.08^{+0.11}_{-0.05}$&$0.19\pm0.02$&26.3(17)\\ 
\hline
\multicolumn{11}{c}{Source $\#$122 = B90 166, EPIC-pn, Obs. 3,4}\\
\hline
MEKAL &$0.15^{+0.03}_{-0.01}$& ... &1 & 1 & 1 & 1 & 1 & $0.75^{+0.07}_{-0.10}$ & $0.18\pm0.01$&23.8(23)\\
      &$0.33^{+0.12}_{-0.08}$& ... &$1$&$0.52^{+0.52}_{-0.25}$&$0.45^{+0.51}_{-0.28}$&$1$&$0.14^{+0.10}_{-0.06}$&$0.27^{+0.18}_{-0.27}$&$0.20\pm0.01$&18.8(20)\\ 
\hline
\end{tabular}

\begin{list}{}{}
\item $^{a}$ -- Relative to the solar abundance.
\item $^{b}$ -- An equivalent hydrogen column density in units of $10^{22}$ cm$^{-2}$.
\item $^{c}$ -- Absorbed model flux in the $0.3 - 3.0$ keV energy range in units of 
$10^{-13}$ erg s$^{-1}$ cm$^{-2}$.
\item $^{d}$ -- Fixed at optical value \cite{Blair82}.
\end{list}
\end{table}

\end{document}